\documentclass[runningheads]{llncs}
\usepackage[T1]{fontenc}

\usepackage[utf8]{inputenc}
\usepackage{etoolbox}
\usepackage{url}

\usepackage{hyperref}
\usepackage{array}
\usepackage{subcaption}
\usepackage{amsmath}
\usepackage{bm}

\usepackage{amssymb}
\usepackage{graphicx}
\usepackage{booktabs}
\usepackage{multirow}
\usepackage{cleveref}
\usepackage{tablefootnote}
\usepackage{xspace}

\newtheorem{fact}{Fact}

\newcommand{\Real}{\mathsf{R}}
\newcommand{\nonnegReal}{\Real_{\geq 0}}
\newcommand{\posReal}{\Real_{> 0}}
\newcommand{\Integer}{\mathcal{N}}
\newcommand{\nonnegInteger}{\Integer_{\geq 0}}
\newcommand{\positiveInteger}{\Integer_{> 0}}

\newcommand{\rExp}[1]{\exp\left(#1\right)}

\DeclareMathOperator{\diag}{diag}

\newcommand{\multiNormal}[2]{\mathcal{N}\left(#1, #2\right)}

\newcommand{\abs}[1]{{\left| #1 \right|}}

\newcommand{\twoNorm}[1]{{\left\Vert #1 \right\Vert}_2}

\newcommand{\infNorm}[1]{{\left\Vert #1 \right\Vert}_{\infty}}

\newcommand{\rowOfMatrix}[2]{#1_{(#2)}}

\newcommand{\rX}{X}
\newcommand{\rY}{Y}
\newcommand{\rZ}{Z}

\newcommand{\rW}{W}

\def \cC {\mathcal{C}}
\def \cD {\mathcal{D}}

\def \cS {\mathcal{S}}
\def \cX {\mathcal{X}}

\def \cT {\mathcal{T}}

\def \cI {\mathcal{I}}

\newcommand{\Ex}[1]{\mathbb{E}\left[ #1 \right]}

\newcommand{\pr}[1]{\Pr\left[#1\right]}

\newcommand{\leverage}{\rowOfMatrix{p}{i}}

\newcommand{\Exf}[2]{
\mathbf{Ex}_{#1}\left[ #2 \right]
}

\newcommand{\DB}{D}
\newcommand{\Mech}{\mathcal{M}}
\newcommand{\MechNDIS}[1]{\Mech^{\mathsf{NDIS}}_{#1}}

\DeclareMathOperator*{\argmin}{arg\,min}

\newcommand{\gRP}{\mathsf{gRP}}

\newcommand{\defin}{\stackrel{\rm def}{=}}
\newcommand{\iid}{\stackrel{\mathrm{i.i.d.}}{\sim}}

\newcommand{\cancel}[1]{}

\begin{document}

\title{The Normal Distributions Indistinguishability Spectrum and its Application to Privacy-Preserving Machine Learning}
\titlerunning{NDIS and Privacy-Preserving Machine Learning}
\authorrunning{Y. Wei et al.}

\thispagestyle{plain}
\pagestyle{plain}
\sloppy

\author{Yu Wei\inst{1} \and
Yun Lu\inst{2} \and
Malik Magdon-Ismail\inst{3} \and
Vassilis Zikas\inst{1}}

\institute{Georgia Institute of Technology\\
\email{\{ywei368,vzikas\}@gatech.edu}
\and
University of Victoria\\
\email{yunlu@uvic.ca}
\and
Rensselaer Polytechnic Institute\\
\email{magdon@cs.rpi.edu}}

\maketitle

\begin{abstract}
    We investigate the privacy of {\em any} algorithm whose outputs have Gaussian distribution. This work is motivated by the prevalence of such algorithms in several useful (ML) applications, and the comparatively little research that focuses on privacy-preserving learning outside of adding Gaussian noise to the data (such as DP-SGD).

    \begin{quote} {\em What is the DP of any algorithm with multivariate Gaussian output?} \end{quote}

    We answer the above research question with a general lemma which we call {\em Normal Distributions Indistinguishability
    Spectrum} (NDIS), a closed-form analytic computation of the hockey-stick divergence $\delta$ between an arbitrary pair of multivariate Gaussians, parameterized by privacy parameter $\epsilon$.
    To show its practical implications, we prove several properties of our NDIS lemma. These properties form a {\em toolbox} of results which lead to potentially {\em easier} privacy proofs for any Gaussian-output algorithm. As an example application of our toolbox, we prove a tighter parametrisation of the privacy of {\em random projection (RP)}, and obtaining from it a more noise-frugal DP mechanism.

    Beyond random projection, NDIS can be used to lift {\em any} Gaussian-output algorithm with a `sensitivity' (which we define) to a Gaussian-output DP mechanism. The mechanism boosts the existing randomness in the algorithm, so that one can describe the mechanism's privacy as the IS between a single pair of Gaussians, which can then be analyzed via NDIS. Lastly, we leverage the connections between NDIS and the CDF of the generalized $\chi^2$ distribution (which have efficient empirical estimators) to present a tool for white-box auditing of Gaussian-output algorithms.

\end{abstract}

\section{Introduction}

Differentially privacy (DP)~\cite{TCC:DMNS06} is the most widely accepted privacy definition for the release of (queries on) sensitive data. Informally, a mechanism (randomized algorithm) is DP if slightly changing its input data(base)---adding or removing a single record/database-row---yields only somewhat {\em indistinguishable} effect on the mechanism's output. The databases that are the result of this small change are referred to as {\em neighboring}.

More concretely, an algorithm $\Mech$ is $(\varepsilon,\delta)$-DP if for all neighboring databases $\DB \sim \DB'$\footnote{Where both $\DB, \DB' \in U^*$ for some {\em universe} $U$ that defines the set of possible values for each database row.}, and all events $\cS$ over the mechanism's output space:
\begin{align*}
    & \pr{\Mech(\DB) \in \cS} \leq e^{\varepsilon}\pr{\Mech(\DB') \in \cS} + \delta
\end{align*}

We are interested in the optimal (i.e., smallest) $(\varepsilon,\delta)$ parameters (often called the mechanism's \emph{privacy profile}~\cite{TPCBalleBG20}). It is well-known that one can trade off a worse/larger parameter $\varepsilon$ for better/smaller $\delta$ and vice-versa, thus we will consider the optimal $\delta$, as a function of $\varepsilon$, written as the following:

\begin{align*}
    \delta(\varepsilon) = \sup_{\Mech(\DB)\simeq \Mech(\DB)} \sup_{\cS\in \Real^T} \Bigl(\pr{\Mech(\DB)\in \cS}-e^{\varepsilon}\pr{\Mech(\DB)\in \cS}\Bigr)_+,
\end{align*}
where we write $(x)_+ \defin{} \max\{x,0\}$, meaning  $\delta(\varepsilon) \geq 0$

Throughout the paper, we study mechanisms $\Mech$ with Gaussian output, i.e., for every input $\DB$, $\rX = \Mech(\DB)$ is a (multivariate) Gaussian random variable (r.v.). To save on notation, instead of writing $\Mech(\DB)$ and $\Mech(\DB')$, we consider Gaussian-distributed r.v.'s $\rX, \rY$ that are the output of $\Mech$ on neighbouring databases. In addition, our main result (NDIS) specifically describes $\delta(\varepsilon)$ for any fixed pair of Gaussians $\rX, \rY$ (later applied to describe $(\varepsilon, \delta)$-DP), and thus we will define:

\begin{align*}
    \delta_{\rX, \rY}(\varepsilon) \defin \sup_{\cS\in \Real^T}\Bigl(\pr{\rX\in \cS}-e^{\varepsilon}\pr{\rY\in \cS}\Bigr)_+.
\end{align*}

We will call $\delta_{\rX, \rY}(\varepsilon)$ the {\em indistinguishability} (or IS for short) between $\rX, \rY$.  When fixing an $\varepsilon \geq 0$, this is also called the \emph{hockey-stick divergence}~\cite{BalleBartheGaboardi2018}.

The investigation of the IS for a pair $(\rX,\rY)$  of \emph{Gaussian outputs} is motivated by the pervasive role of Gaussian distributions throughout private machine learning and statistics, well beyond the explicit addition of zero-mean Gaussian noise (known as the Gaussian mechanism~\cite{DworkR14} in DP literature). For instance, Gaussianity arises in sketching and random projection (RP) routines (which have applications such as randomized PCA and SVM), randomized numerical linear algebra, and classic asymptotic normality phenomena~\cite{TierneyK1986,McCN89,vanderVaart1998,RueMC09,bishop06}. In such settings, the algorithm itself produces a Gaussian (or nearly Gaussian) output. However, a complication in these algorithms is that the shape of its its output distribution may depend on the input dataset, which makes analyzing DP---the worst-case IS over {\em all} neighboring databases---a difficult task!

In more detail, in the simplest case of the standard Gaussian mechanism, Gaussian noise is added to the result of a deterministic query on the input database. The mechanism's outputs on neighboring inputs would then differ only by a \emph{mean shift} (whose magnitude is bounded by the `sensitivity' of the query, if this sensitivity exists), but otherwise share the same covariance. As such, the IS  $\delta_{\rX,\rY}(\varepsilon)$ admits a clean closed form (which, given the sensitivity bound, gives a nice form for $(\varepsilon, \delta)$-DP). In contrast, many practically important Gaussian-output algorithms induce \emph{data-dependent covariances}, so neighboring databases may yield Gaussians with \emph{different} covariance matrices. This is precisely where standard privacy analyses may fall back on loose upper bounds, indirect divergences, or worst-case sensitivity arguments, obscuring what truly governs privacy and potentially leading to unnecessary noise. One concrete example is privacy-preserving Gaussian random projection~\cite{BlockiBDS12}\footnote{First explored in~\cite{BlockiBDS12}, with follow-up works~\cite{Sheffet17,ALTSheffet19}, and those extending and applying this to private ML, e.g.,  private large graph queries~\cite{BlockiBDS12,AroraU19} and linear regression~\cite{UpadhyayU21,KharkovskiiDL20,Upadhyay14a}}: the induced pair of outputs can have the same mean but different covariances (that depend on the input), a setting not covered by standard closed-form analyses. The same phenomenon also in common second-order and Bayesian procedures, such as Bayesian Logistic Regression~\cite{bishop06} and Gaussian Process Regression~\cite{RasmussenW06}. The above challenges lead to the following core question:

\medskip
\noindent\textbf{Q1.}
\emph{Is there a general way to exactly describe the indistinguishability between \textit{any} two Gaussian distributions, and how can we use it to prove the DP of  (data-dependent) Gaussian-output algorithms?}
\medskip

Answering the above is not only of theoretical interest, but has several practical implications which we explore in this work (discussed in Contributions subsection). We prove a general lemma (which we call NDIS, for \textbf{n}ormal \textbf{d}istributions \textbf{i}ndistinguishability \textbf{s}pectrum) for the indistinguishability (IS) between any two Gaussians. Then, we present a general toolbox for proving the privacy of Gaussian-output algorithms, demonstrating it for random projection and expose the \emph{actual} quantities that govern its privacy loss.

Beyond the task of analyzing specific mechanisms, a major task in DP literature is creating new DP mechanisms. This raises a natural design question, which we address in this work:

\medskip
\noindent\textbf{Q2.}
\emph{Can a tight characterization of IS be used to design new DP mechanisms?}

\subsection{Our Contributions.}

In this paper, we answer both research questions in the affirmative. \\

\noindent
\textbf{NDIS lemma and toolbox} We present NDIS, which exactly describes the IS between two Gaussian r.v.'s $\rX, \rY$. Specifically, it describes the IS as the expectation of a quadratic expression where the constants depend on the means/covariances of $\rX, \rY$, and the variable is a standard normal r.v..

To aid the reader in applying NDIS, we construct a toolbox which helps simplify the IS in various cases. The first tool is a connection between the IS and the CDF of the generalized-$\chi^2$ distribution, which can be easier to work with. Since there exist widely-used algorithms~\cite{Imhof1961,Davies1973,DuchesneDeMicheaux2010} (and actively going research~\cite{CompQuadForm2025,Das2025} on how) to numerically compute the CDF, another nice implication is then the easy numerical computation of IS. One application of this is white-box auditing, discussed later in this subsection.

The second tool simplifies $\delta_{\rX,\rY}(\varepsilon)$ into \emph{clean analytic forms} (exact expressions in special cases and explicit upper bounds in more general regimes). These analytic characterizations make it possible to prove monotonicity and to identify the worst-case  neighboring pair for a mechanism (as demonstrated in our analyses of random projection and of the general NDIS-calibrated wrapper that lifts Gaussian-output algorithms to $(\varepsilon,\delta)$-DP mechanisms).

In the easiest case, we give an equivalent but simpler IS expression for $\rX, \rY$ when they have the same covariance matrices, and possibly different means. This captures the well-known Gaussian mechanism (and its linear query variant~\cite{LiMHMR15matrix}). In the case when $\rX, \rY$ share the same mean but have differing covariances, with covariances comparable by Loewner order (`Loewner-comparable'), we provide simpler upper-bound expressions for the IS of  $\rX, \rY$. We show and use a property of IS that when the covariance of $\rX$ is `larger' than $\rY$ (by Loewner order), the IS of $(\rX, \rY)$ is always larger than the IS of $(\rY, \rX)$. For the most complex case of any two Gaussians $\rX, \rY$, we decompose this case into two pairs of simpler distributions, where one pair has the same mean  and the other has the same covariance, then combine the IS expressions of these two pairs based on the triangle inequality property of IS.

We apply the first tool (connections between NDIS with CDF of generalized $\chi^2$) in conjunction with the property of IS when one covariance in the pair of distributions being larger than the other, to exactly  characterize the  $(\varepsilon, \delta)$-DP of random projection, discussed below.\\

\noindent
\textbf{Applications to privacy-preserving random projection (RP).}
Gaussian RP projects a high-dimensional database $\DB$ onto a lower-dimensional space using a Gaussian matrix. The NDIS toolbox can be applied to achieve an exact characterization of the differential privacy of RP on neighboring inputs (and thus tighter than previous analysis from~\cite{BlockiBDS12,Sheffet17,ALTSheffet19}). This in turn creates a more noise-frugal privacy-preserving RP mechanism that achieves any desired DP parameters.

Informally, for RP (which is well-studied) we know the mean and covariance of its output as a function of the input database. Applying the NDIS lemma, we can determine a `worst-case' neighbor pair that maximizes IS. We then use the NDIS toolbox to obtain, and then simplify the IS expression to show that the privacy of RP is best described by a quantity of the input databases called the {\em leverage} (denoted by $\leverage$, where higher leverage implies worse privacy). In order to achieve $(\varepsilon,\delta)$-DP, our mechanism regularizes (via ridge regularization) the input database to lower its leverage down to a target value, that the NDIS lemma identifies as corresponding to the desired $(\varepsilon,\delta)$.
As a further step, we show that the NDIS-based analysis composes cleanly with the standard DP toolkit from the literature, enabling \emph{database-agnostic} privacy guarantees while delivering \emph{database-dependent} utility improvements. Specifically, we apply {\em subsampling} and {\em Propose-Test-Release (PTR)} techniques, which always preserve DP but may help achieve better utility for some inputs. We discuss under what conditions on the input database can our optimized mechanisms achieve better utility than our baseline RP mechanism.\\

\noindent
\textbf{Lifting any Gaussian-output algorithm to a DP mechanism with Gaussian Output.} We present a definition of `sensitivity' that generalizes the notion of sensitivity from the standard Gaussian mechanism. Here, we consider a Gaussian-output algorithm whose output's mean/covariance may depend on the input. Assuming this sensitivity exists (and we show  examples of such sensitivity in useful Gaussian algorithms), it allows us to define a smallest sufficient boost of the algorithm's randomness, to achieve a desired level of $(\varepsilon, \delta)$-DP. Inside this more general Gaussian mechanism is the NDIS theorem---by introducing sufficient additional randomness, the mechanism's privacy can be described as the IS ($\delta_{\rX, \rY}(\varepsilon)$) for a known pair of Gaussians $\rX, \rY$. NDIS can then be applied to analyze this IS to compute the privacy parameters $(\varepsilon, \delta)$-DP of the mechanism.\\

\noindent
\textbf{White-box auditing.} One neat consequence of NDIS is a simple tool to white-box audit a Gaussian-output mechanism. {\em Auditing} refers to attempting to find violations of  privacy in a mechanism claiming to be $(\varepsilon,\delta)$-DP, and `white-box' refers to having certain knowledge about the algorithm (such as its code). One way to find such violations is to compute/estimate the IS of the mechanism outputs for a specific pair of neighboring databases, and treating this IS as a {\em lower-bound} of the true privacy of the mechanism. When the lower-bound is higher than the claimed privacy, we have found a violation. Computing IS for a specific pair of Gaussian distributions is exactly in our wheelhouse, and using the connection between IS and the CDF of the generalized-$\chi^2$ distribution (which can be estimated efficiently), we obtain an efficient audit/lower-bound while requiring only for the mean and covariance of algorithm output to be known for the tested input databases.

\section{Related Work}

The distance between pairs of multivariate Gaussian is a fundamental problem, with widespread applications in ML, e.g., ~\cite{LanMHWYL22,SohoniDAGR20}, and in security and privacy, e.g.,~\cite{AlabiKTVZ23,ZhangGFCY23}. For {\em statistical distance} (different from the divergence used in DP), Devroye et al.~\cite{Devroye2018} established a lower and an upper bound on the distance between two multivariate Gaussian with the same mean; more recently, Arbas et al.~\cite{ArbasAL23}  generalized this result and proved a tight bound on the distance between two arbitrary multivariate Gaussians. In contrast, our work centers on (closed-form expressions for the) {\em DP distance} between any two Gaussians, instead of statistical distance. Moreover, while~\cite{CCS:BCSV25} also considers the DP between two Gaussians, their work constructs an empirical verifier (with some error) verifying the DP-distance of a specific pair of one-dimensional Gaussians, which does not readily extend to proofs of DP guarantees.

\vspace{1mm}
\noindent
\textbf{The Gaussian Mechanism.} A standard DP mechanism called the {\em Gaussian mechanism}, introduced by~\cite{DworkMN06,Dwork2014} and further analyzed in subsequent works, such as~\cite{BalleW18}, adds a zero-mean Gaussian vector to the output of a (deterministic) query. Building on this concept, Chanyaswad et al.~\cite{ChanyaswadDPM18} proposed the Matrix-Variate Gaussian (MVG) mechanism, which incorporated matrix noise from a matrix-variate Gaussian distribution. Liu~\cite{liu19} presented a generalized Gaussian mechanism, which adds noise whose scale is computed from a more generalized sensitivity function. In this work, our goal is to solve a strictly more general problem: the indistinguishability between arbitrary multivariate Gaussians, whose mean and covariance can both vary (unlike for Gaussian mechanisms).

\smallskip
\noindent
\textbf{Random Projection (RP).}
RP projects a high-dimensional database $\DB$ onto a lower-dimensional space, with various applications such as randomized PCA and SVM.
The seminal work of Blocki et al.~\cite{BlockiBDS12} showed that RP achieves DP, if the input database's least singular value is sufficiently large and each record's norm is bounded. Recently Sheffet~\cite{Sheffet17,ALTSheffet19} studied an $r$-fold composition of RP and improved on previous works' DP analysis. Concurrently, another line of works~\cite{BlumR13,Upadhyay14b,upadhyay2014randomness} studied the DP of random projection using random matrices other than standard Gaussian matrices. Beyond the study of RP itself, it has been explored as a pre-processing step to achieve DP in many applications~\cite{AroraU19,BartanP20,KharkovskiiDL20,Upadhyay14a,UpadhyayU21}. Our objective aligns with~\cite{BlockiBDS12,Sheffet17,ALTSheffet19}, focusing on Gaussian RP, and our results improve upon the analyses of previous work. The NDIS lemma derives a {\em tight} privacy statement for Gaussian RP, as a function of the (global) {\em leverage score}.

To put our leverage-based characterization in context, prior work showed that Gaussian RP can be made differentially private by calibrating its randomness using a worst-case lower bound on the least singular value (LSV) of datasets. In the same spirit, our RP mechanism $\Mech_{RP}$ takes a target privacy budget $(\varepsilon,\delta)$ and computes a \emph{maximum allowable} leverage threshold $\leverage$ and calibrate the noise using this threshold. Since larger leverage corresponds to weaker privacy (by~\Cref{lemma: rp inherent,prop:grp-monotone}), $\leverage$ directly captures the privacy-critical regime. Moreover, leverage can be upper bounded from LSV and norm of records. Consequently, LSV-based calibration is inherently more conservative: it controls leverage only indirectly (via an upper bound) and thus can yield strictly looser privacy/utility tradeoffs than working with leverage directly, while the reverse implication does not hold in general. We refer~\Cref{fig:leverage-ratio} in \Cref{subsec: utility} for further illustration.

\smallskip
\noindent
\textbf{Using an Algorithm's Inherent Randomness for Privacy.} At the heart of the concept of `lifting' an already randomized algorithm to a DP mechanism (in our work, we show as examples Gaussian RP, Bayesian Logistic Regression, and Gaussian Process Regression), is the idea of using the inherent randomness in queries or randomization operations for privacy. This has been seen in previous work, a notable example being ``privacy amplification by subsampling"~\cite{BalleBG18,ParkFCW20,Steinke22}, i.e., running a DP mechanism on a random subsample of a dataset offers stronger privacy guarantees compared to applying it to the entire dataset. This subsampling technique has been successfully employed in the privacy analysis of DP algorithms, such as DP-SGD~\cite{AbadiCGMMT016}. In addition, there is a growing body of work investigating the inherent privacy of specific randomized algorithms, such as the Flajolet-Martin Sketch~\cite{Hu0LGWGLD21,Smith0T20}, Thompson Sampling~\cite{OuCA24}, and random projection~\cite{BlockiBDS12,KharkovskiiDL20,ALTSheffet19,Upadhyay14a}.

\medskip

\noindent
\textbf{Terminology of {\em indistinguishability} (IS).}
We use the term {\em indistinguishability} (IS) to denote the optimal $\delta$ as a function of $\epsilon$ for a {\em specific pair} of distributions $(\rX, \rY)$. We do so to distinguish it from the notion of the optimal $(\varepsilon, \delta)$-DP parameters of a {\em mechanism}---the worst-case $\delta(\epsilon)$, over {\em all pairs} of neighboring databases.
The optimal DP parameters of a mechanism has been introduced and studied in prior work~\cite{TPCBalleBG20,BartheO13,DworkR16,abs-2405-20769,SommerMM19,WangMWJM23,0005DW22} under names such as privacy profile/privacy curve.

\medskip
\noindent
\textbf{Studies on optimal $(\varepsilon, \delta)$-DP; DP Variants.}
The study of a optimal $(\varepsilon, \delta)$-DP parameters via the privacy profile (or, the hockey-stick divergence of the mechanism's {\em privacy loss random variable}) has been extensively developed for subsampled Gaussian mechanisms~\cite{Steinke22,0005DW22} and for the standard Gaussian mechanism (i.e., two normal distributions with the same covariance matrix but different means)~\cite{BalleW18,SommerMM19}. To the best of our knowledge, there has been no prior work analyzing the optimal $(\varepsilon, \delta)$ pairs for arbitrary pairs of Gaussian distributions, particularly in the case where the covariance matrices differ. The NDIS lemma can also be viewed as a more explicit (in fact, quadratic polynomial) form of the privacy loss random variable, which can then be further simplified (using our toolbox) to prove $(\varepsilon,\delta)$-DP.

The study of tighter composition bounds and more tractable privacy analysis via the privacy loss random variable has led to several variants of differential privacy, such as (Zero-)Concentrated Differential Privacy (zCDP)~\cite{DworkR16,BunS16} and Rényi Differential Privacy (RDP)~\cite{Mironov17}. These frameworks enable refined analysis of iterative algorithms like stochastic gradient descent, particularly in the context of subsampled Gaussian mechanisms~\cite{abs-1908-10530}. Unlike the NDIS lemma, however, zCDP and RDP measure privacy using Rényi divergence rather than the standard DP distance. While conversions from RDP and zCDP to $(\varepsilon, \delta)$-DP are well established, it remains unclear how to obtain closed-form expressions for the optimal $(\varepsilon, \delta)$ parameters in the DP setting, especially for arbitrary Gaussian distributions, from the existing results under RDP or zCDP. Lastly, Gaussian ($f$-)differential privacy (GDP)~\cite{Dong2022} is a recently proposed variant of differential privacy that emphasizes measuring privacy using a function (specifically, a tradeoff function) rather than a fixed parameter pair. In particular, GDP uses the tradeoff function between two Gaussian distributions with the same (unit) variance and different means as the privacy measure. Although conversions from the GDP tradeoff function to the optimal $(\varepsilon, \delta)$-DP curve are known, it remains an open question how to derive closed-form expressions for the optimal $(\varepsilon, \delta)$ parameters---especially for arbitrary Gaussian distributions---based on the existing results in GDP.

\medskip

\noindent
\textbf{(White-box) auditing of DP.}
{\em Auditing} a (DP) mechanism refers to attempting to find counterexample(s) to a claim that a given algorithm satisfies $(\varepsilon, \delta)$-DP. In general, an auditor can be either {\em black-box} (where the auditor can only feed inputs to the audited algorithm and observe outputs before deciding if a mechanism passes the audit), or {\em white-box/`semi'-white-box}\footnote{A sample of the many works on white-box auditing:~\cite{POPL:ZhaKif17,PLDI:WDWKZ19,CCS:WDKZ20,OOPSLA:ZRHPR20}}, where the auditor knows everything (or some partial information) about the algorithm. While black-box auditing has advantages (e.g., allows auditing without access to the actual code), white-box auditing can be more accurate~\cite{wang2024curator}, and in some restricted cases, even {\em prove} DP through analysing the algorithm's code (e.g., when the algorithm's code is differentiable~\cite{CCS:BGDTV18}).
Nevertheless, we are not aware of auditors (white- or black-box) that are specialized for an arbitrary pair of Gaussian distributions (that can differ in mean and covariance)---the work~\cite{CCS:BCSV25} verifies algorithms that use one-dimensional Gaussians through computing/estimating its distribution. Thus we believe our tool can be a useful addition to the arsenal of any white-box auditor.
While our auditing tool still requires empirically estimating the IS and treating this as the lower-bound on privacy parameters, we remark that its white-box nature gives advantages over black-box counterparts, such as~\cite{LuWMZ22} which also produces tight estimation of the IS for a pair of (neighbouring) databases (\cite{NeurIPS:LiuOh19} is also tight, but requires enumerating the mechanism's output space, making it inapplicable for Gaussians). In contrast to our white-box tool, black-box works produce probabilistic statements, i.e., with some probability, the estimate of IS is close to the true IS. In contrast, assuming a generalized $\chi^2$ CDF oracle, we guarantee our estimates are close to the real IS. Lastly, while a general-purpose $f$-DP (and thus Gaussian DP) auditor exists (e.g.,~\cite{fDPestimator}), Gaussian DP considers a pair of Gaussian distributions with the same variance, which makes it unclear how to apply it audit arbitrary Gaussians.

\section{Preliminaries}

\begin{definition}[Generalized-$\chi^2$ distribution]
\label{def:genchi2}
A real-valued random variable $W$ is said to have a \emph{generalized-$\chi^2$ distribution} if there exist an integer $r\geq 1$, coefficients $\lambda_1,\dots,\lambda_r\in\Real\setminus\{0\}$, degrees of freedom $\nu_1,\dots,\nu_r\in\positiveInteger$, noncentrality parameters $\rho_1,\dots,\rho_r\geq 0$, a Gaussian coefficient $\sigma\ge 0$, and a shift $\kappa\in\Real$, such that
\begin{align*}
    W \overset{d}{=} \kappa + \sum_{j=1}^r \lambda_j X_j + \sigma G,
\end{align*}
where $X_1,\dots,X_r$ are independent noncentral chi-square random variables $X_j\sim \chi'^2_{\nu_j}(\rho_j)$ and $G\sim \multiNormal{0}{1}$ is independent of $\{X_j\}_{j\in[r]}$.
\end{definition}

\subsection{Gaussian random projection and leverage score.}

We recall the Gaussian random projection algorithm and the definition of a leverage score.

\begin{definition}[Gaussian random projection]
\label{def:grp}
Fix an integer $r\ge 1$. Let $G\in\Real^{n\times r}$ have i.i.d. $\multiNormal{0}{1}$ entries. For a database $\DB\in\Real^{n\times d}$, define the \emph{Gaussian random projection} as a $\Real^{d\times r}$ random matrix
\begin{align*}
    \gRP_r(\DB) \defin \DB^\top G.
\end{align*}
Equivalently, if we write $\gRP_r(\DB) = [X^{(1)},\dots,X^{(r)}]$ by columns, then
$X^{(1)},\dots,X^{(r)}$ are i.i.d., and for each $j\in[r],$
\begin{align*}
    X^{(j)} \sim \multiNormal{0}{\DB^\top \DB}.
\end{align*}
\end{definition}

Assume $\DB$ has full column rank, and let $v^\top\in\Real^{1\times d}$ denote the $i$-th row of $\DB$. The \emph{leverage score} (or simply \emph{leverage}) of record $i$ with respect to database $\DB$ is
\begin{align}
    \label{def: leverage}
    \leverage \defin v^\top(\DB^\top\DB)^{-1}v.
\end{align}
Equivalently,
\begin{align*}
    \leverage = \twoNorm{(\DB^\top\DB)^{-1/2}v}^2
\end{align*}
so $\leverage$ is the squared norm of the record $v$ after \emph{whitening} by the covariance matrix $\DB^\top\DB$. Intuitively, $\leverage$ is large when the record has a large $\ell_2$ norm, or when it is aligned with directions in which $\DB^\top\DB$ is poorly conditioned (small eigenvalues). We will use the following standard fact.

\begin{fact}
\label{fact:leverage-basic}
For every $i\in[n]$,
\begin{align*}
    0 \leq \leverage \leq 1, \qquad\text{and}\qquad
    \sum_{i=1}^n \leverage = \mathrm{tr}(P) = d.
\end{align*}
Moreover, $\leverage=1$ if and only if $P e_i = e_i$ (equivalently, $e_i\in \mathrm{col}(\DB)$).
\end{fact}

\section{The NDIS Lemma and Toolbox}

The Normal-Distributions Indistinguishability (NDIS) lemma (\Cref{lemma: ndis}) provides a closed-form characterization of the indistinguishability between any two multivariate Gaussians $\rX$ and $\rY$ with means $\mu_1,\mu_2$ and covariances $\Sigma_1,\Sigma_2$, respectively.
The characterization is an expectation over a \emph{standard} Gaussian $Z\sim\mathcal{N}(0,I_d)$, where the expression's dependence on $(\mu_1,\Sigma_1)$ and $(\mu_2,\Sigma_2)$ is captured by a simple quadratic polynomial $g_\varepsilon(Z)$, thereby making explicit how privacy is governed by the relative geometry of the two Gaussians. Intuitively, $g_\varepsilon(Z)$ can be viewed as the $\varepsilon$-shifted privacy loss (log-likelihood ratio) expressed in a whitened/diagonalized coordinate system: the region ${g_\varepsilon(Z)\leq 0}$ corresponds exactly to where the likelihood ratio exceeds $e^\varepsilon$, and the hinge form $\Ex{1-e^{g_\varepsilon(Z)})_+}$ captures both the probability mass and the severity of this privacy loss. This expression connects $\delta$ to the classical theory of (generalized) quadratic forms in Gaussian random variables, enabling efficient evaluation and further analysis (as we do in the next sections).

\begin{lemma}[NDIS Lemma; Proof is in~\Cref{proof for lemma: ndis}]
\label{lemma: ndis}
Let $\rX \sim \multiNormal{\mu_1}{\Sigma_1}$ and $\rY \sim \multiNormal{\mu_2}{\Sigma_2}$ on $\Real^d$, where $\Sigma_1,\Sigma_2\succ 0$. Let $\Delta\mu\defin \mu_1-\mu_2$ and $M = \Sigma_1^{1/2}\Sigma_2^{-1}\Sigma_1^{1/2}.$ Let $U\diag(\tau_1,\dots,\tau_d)U^T$ be an Eigenvalue Decomposition (EVD) of $M$, and define
\begin{align*}
    a_i \defin{}& 1-\tau_i ~\text{for}~i \in [d], \quad b \defin -U^T\Sigma_1^{1/2}\Sigma_2^{-1}\Delta\mu \\
    c(\varepsilon)\defin{}& \varepsilon+\frac12\log\frac{\det\Sigma_1}{\det\Sigma_2}-\frac12\Delta\mu^T\Sigma_2^{-1}\Delta\mu.
\end{align*}
Then for $Z\sim \multiNormal{0}{I_d}$,
\begin{align*}
    \delta_{X,Y}(\varepsilon) ={}& \Ex{\Bigl(1-\rExp{g_\varepsilon(Z)}\Bigr)_{+}},
\end{align*}
where $g_\varepsilon(Z) \defin{} c(\varepsilon)+b^T Z+\frac12\sum_{i=1}^d a_i Z_i^2$.
\end{lemma}

To prove~\Cref{lemma: ndis}, we work directly from the standard expression $\sup_{\cS}\bigl(\pr{\rX\in\cS}-e^\varepsilon\pr{\rY\in\cS}\bigr)$.
At a high level, the goal is to rewrite this set-optimization problem as an expectation under a \emph{standard} Gaussian measure, so that the expression's dependence on $(\mu_1,\Sigma_1)$ and $(\mu_2,\Sigma_2)$ becomes explicit. We first expand $\pr{\rX\in\cS}$ and $\pr{\rY\in\cS}$ using their Gaussian density formulas and combine them inside a single integral over $\cS$. Next, we re-express $\rX$ via the change of variables $x=\mu_1+\Sigma_1^{1/2}z$, which turns the $\rX$-density into the standard Gaussian density $\varphi(z)$ and turns $\cS$ into $\cS_z=\Sigma_1^{-1/2}(\cS-\mu_1)$. After this standardization, the integrand becomes $\varphi(z)\bigl(1-e^{\psi_\varepsilon(z)}\bigr)$, where $\psi_\varepsilon(z)$ is an explicit quadratic function obtained by expanding the exponent of the $\rY$ density and rearranging the expression; the quadratic term is characterized by $M=\Sigma_1^{1/2}\Sigma_2^{-1}\Sigma_1^{1/2}$.

We then diagonalize $M=U\diag(\tau_1,\ldots,\tau_d)U^T$ and rotate coordinates via $v=U^Tz$. This preserves the standard Gaussian measure but separates out the quadratic part, yielding $\psi_\varepsilon(z)=g_\varepsilon(v)=c(\varepsilon)+b^Tv+\frac12\sum_{i=1}^d a_i v_i^2$. Finally, the set-optimisation: since $\varphi(v)\ge 0$, the integral is maximized by choosing the region where the bracket is positive, i.e., where $1-e^{g_\varepsilon(v)}>0$ or equivalently $g_\varepsilon(v)\leq 0$. This gives the claimed representation in~\Cref{lemma: ndis}.

\subsection{Tool 1: Efficient Computation via Generalized-$\chi^2$ CDFs}
Building on the NDIS lemma, we next highlight a key \emph{computational} consequence. \Cref{thm:ndis-two-genchi2-cdf} shows that, by pushing the NDIS representation one step further, $\delta_{\rX,\rY}(\varepsilon)$ can be written as the difference of \emph{two} tail probabilities of quadratic forms,
namely $\pr{g_\varepsilon(Z)\leq 0}$ and $\pr{g_\varepsilon(\widetilde Z)\leq 0}$. Crucially, both $g_\varepsilon(Z)$ and $g_\varepsilon(\widetilde Z)$ are generalized-$\chi^2$ random variables (cf.~\Cref{def:genchi2}). Thus, computing indistinguishability between arbitrary multivariate Gaussians reduces to a well-studied numerical task: evaluating generalized-$\chi^2$ CDFs, for which there exist mature and widely used algorithms based on characteristic-function inversion and related techniques (e.g., Imhof-type and Davies-type methods).

Two applications of this tool include the analysis of random projection (Sec.~\ref{sec:rpmech}) and white-box auditing (Sec.~\ref{sec:whitebox}).

\begin{theorem}[Connecting IS and CDF of generalized-$\chi^2$ r.v.s; Proof is in~\Cref{proof for thm:ndis-two-genchi2-cdf}]
\label{thm:ndis-two-genchi2-cdf}
Let $\varepsilon,\rX,\rY$ and the quantities $g_\varepsilon(\cdot),\,b,$ and $\tau_1,\dots,\tau_d$ be as in~\Cref{lemma: ndis}. Let $D = \diag(\tau_1,\dots,\tau_d)$ and $m = D^{-1}b$. Let $Z\sim \multiNormal{0}{I_d}$ and $\widetilde Z\sim \multiNormal{m}{D^{-1}}$.
Then both $g_\varepsilon(Z)$ and $g_\varepsilon(\widetilde Z)$ are generalized-$\chi^2$ random variables (cf.~\Cref{def:genchi2}), and
\begin{align*}
\delta_{\rX,\rY}(\varepsilon) = \pr{g_\varepsilon(Z)\leq 0} - e^{\varepsilon}\,\pr{g_\varepsilon(\widetilde Z)\leq 0}.
\end{align*}
Consequently, computing $\delta_{\rX,\rY}(\varepsilon)$ reduces to evaluating two generalized-$\chi^2$ CDFs.
\end{theorem}

\subsection{Tool 2: Simplifying NDIS to Cleaner Analytic Forms, Beyond CDF Evaluation}
While \Cref{thm:ndis-two-genchi2-cdf} shows that $\delta_{\rX,\rY}(\varepsilon)$ for an \emph{arbitrary} Gaussian pair can be computed efficiently via generalized-$\chi^2$ CDF evaluation, for several purposes it is also valuable to have \emph{explicit analytic} expressions or bounds. An analytic form is useful when performing further analysis,
for example: proving monotonicity properties of this expression, or deriving sufficient conditions for achieving $(\varepsilon,\delta)$-DP. Motivated by this, we next develop analytic characterizations of $\delta_{\rX,\rY}(\varepsilon)$ in progressively more general settings (using the NDIS lemma as the starting point of their proofs).

We begin with the `mean-shift' case, where $\rX$ and $\rY$ share the same covariance but have different means (a well-known example of this case is the Gaussian mechanism). We then turn to the `covariance-shift' case, where $\rX$ and $\rY$ share the same mean but have different covariances (a well-known example is the random projection-based mechanisms). Finally, we combine these ingredients to obtain a simple analytic upper bound for the fully general case of simultaneous mean and covariance changes when the two covariances are comparable in the Loewner order (\Cref{thm:ndis-analytic-form}).

\medskip
\noindent\textbf{Equal covariance (mean shift).}
In the simplest nontrivial regime, $\rX$ and $\rY$ differ only in their means while sharing a common covariance. In this case, \Cref{prop:ndis-mean-analytic-form} shows that $\delta_{\rX,\rY}(\varepsilon)$ admits a closed form that depends \emph{only} on the Mahalanobis norm of the difference of the mean $\Delta\mu$ under the common covariance, namely $\tau=\sqrt{\Delta\mu^\top\Sigma^{-1}\Delta\mu}$. This leads to the well-known privacy profile of the Gaussian mechanism~\cite{BalleW18} (and its linear-query variants~\cite{LiMHMR15matrix,kairouz21b,DenisovMRST22optimalPLO}) as a special case.

\begin{proposition}[IS for mean shift; Proof is in~\Cref{proof for prop:ndis-mean-analytic-form}]
\label{prop:ndis-mean-analytic-form}
Let $\varepsilon,\rX \sim \multiNormal{\mu_1}{\Sigma_1}, \rY \sim \multiNormal{\mu_2}{\Sigma_2}$, and $\Delta\mu$ be as in~\Cref{lemma: ndis}. Define the Mahalanobis distance
\begin{align*}
    \tau \defin \sqrt{\Delta\mu^\top \Sigma^{-1}\Delta\mu}.
\end{align*}
Assume $\Sigma_2 = \Sigma_1.$ Then,
\begin{align*}
    \delta_{\rX,\rY}(\varepsilon) = \Phi\left(-\frac{\varepsilon}{\tau}+\frac{\tau}{2}\right) -  e^{\varepsilon}\Phi\left(-\frac{\varepsilon}{\tau}-\frac{\tau}{2}\right),
\end{align*}
where $\Phi$ is the CDF of $\multiNormal{0}{1}$.
\end{proposition}

\medskip
\noindent\textbf{Equal mean (covariance shift).}
We next consider the case where $\rX$ and $\rY$ share the same mean but differ in their covariances. In this case, \Cref{lem:ndis-cov-analytic-form} shows that when the covariances are comparable in the Loewner order (denoted $\Sigma_2\succeq \Sigma_1$), $\delta_{\rX,\rY}(\varepsilon)$ admits an analytic \emph{upper bound} controlled by the relative eigen-geometry of $(\Sigma_1,\Sigma_2)$ through the eigenvalues $\tau_1,\ldots,\tau_d$ from \Cref{lemma: ndis}. In particular, the lemma identifies a \emph{zero-leakage} threshold: if $\varepsilon \geq \tfrac12\log\frac{\det\Sigma_2}{\det\Sigma_1}$ then $\delta_{\rX,\rY}(\varepsilon)=0$, i.e., the pair satisfies \emph{pure} $\varepsilon$-DP (no $\delta$ slack). Otherwise, \Cref{lem:ndis-cov-analytic-form} provides a closed-form Markov-type bound that can be optimized over two scalar parameters $(s,u)$. We note in passing that this covariance-shift setting also covers random-projection-based mechanisms, where neighboring datasets can induce different (but ordered) post-projection covariances.

\begin{lemma}[Upper bound IS for covariance shift; Proof is in~\Cref{proof for lem:ndis-cov-analytic-form}]
\label{lem:ndis-cov-analytic-form}
Let $\varepsilon,\rX \sim \multiNormal{\mu_1}{\Sigma_1}, \rY \sim \multiNormal{\mu_2}{\Sigma_2}$, and $\tau_1,\dots,\tau_d$ be as in~\Cref{lemma: ndis}. Define
\begin{align*}
    f(\varepsilon)\defin \log\frac{\det\Sigma_2}{\det\Sigma_1}-2\varepsilon, \qquad
    k \defin \Bigl(2\max_{i\in[d]}\frac{1-\tau_i}{\tau_i}\Bigr)^{-1}\in(0,\infty],
\end{align*}
(with the convention $k=\infty$ if $\max_i\frac{1-\tau_i}{\tau_i}=0$). Assume $\Sigma_2\succeq \Sigma_1$ and $\mu_1 = \mu_2$.

If $\varepsilon\geq \frac12\log\frac{\det\Sigma_2}{\det\Sigma_1}$, then $\delta_{\rX,\rY}(\varepsilon)=0$; otherwise
\begin{align*}
    \delta_{\rX,\rY}(\varepsilon) \leq  \inf_{\substack{s\geq 0\\ u\in(0,k)}}
    &\rExp{ sf(\varepsilon) - \frac12\sum_{i=1}^d \log\!\bigl(1+2s(1-\tau_i)\bigr)} \\
    &- e^\varepsilon\Bigg(1 - \rExp{-uf(\varepsilon) - \frac12\sum_{i=1}^d \log\Bigl(1-2u \, \frac{1-\tau_i}{\tau_i}\Bigr)}\Bigg).
\end{align*}
\end{lemma}

To see how~\Cref{lem:ndis-cov-analytic-form} is proved, we specialize the NDIS lemma (\Cref{lemma: ndis}) to the covariance-only setting $\mu_1=\mu_2$. In this case, the NDIS polynomial simplifies substantially: the linear term disappears ($b=0$) and $g_\varepsilon(Z)$ becomes a shifted quadratic form. Consequently, the privacy-loss region $\{g_\varepsilon(Z)\leq 0\}$ can be rewritten as an one-sided linear constraint on a weighted $\chi^2$ sum,
\begin{align*}
    \sum_i (1-\tau_i)Z_i^2 \leq f(\varepsilon),
\end{align*}
where $f(\varepsilon)=\log\frac{\det\Sigma_2}{\det\Sigma_1}-2\varepsilon$ measures the remaining “slack’’ after accounting for the determinant gap between the two covariances. This reformulation immediately gives a threshold: if $f(\varepsilon)\leq 0$ (equivalently $\varepsilon\geq \tfrac12\log\frac{\det\Sigma_2}{\det\Sigma_1}$), the inequality cannot hold since the left-hand side is nonnegative, and thus $\delta_{\rX,\rY}(\varepsilon)=0$.

In the nontrivial regime $f(\varepsilon)>0$, we invoke the two-CDF decomposition from \Cref{thm:ndis-two-genchi2-cdf} and control the resulting quadratic-form probabilities using exponential (Markov/Chernoff) bounds. At a high level, we upper bound the first term by applying Markov to an exponentially tilted version of the weighted $\chi^2$ sum, and we lower bound the second term by upper bounding the complementary tail of an analogous exponentially tilted $\chi^2$ sum. The closed-form $\chi^2$ moment generating functions then produce the explicit log-product expressions in $(s,u)$, and optimizing over these $(s,u)$ yields the stated analytic upper bound.

\medskip
$\delta_{\rX,\rY}(\varepsilon)$ is inherently asymmetric in $(\rX,\rY)$, which means in general $\delta_{\rX,\rY}(\varepsilon) \neq \delta_{\rY, \rX}(\varepsilon)$. In the covariance-only setting, however, this asymmetry becomes \emph{structured}. \Cref{lem: asymetric order for covariance} shows that if $\Sigma_1\succeq \Sigma_2$ (so $\rX$ is the “more spread’’ Gaussian), then $\delta_{\rX,\rY}(\varepsilon)\geq \delta_{\rY,\rX}(\varepsilon)$ for all $\varepsilon\geq 0$. Thus, in the comparable-covariance regime there is a “hard’’ direction.~\footnote{This asymmetry property is useful beyond saving a second computation: in the hard direction $\delta_{\rX,\rY}(\varepsilon)$, the NDIS coefficients satisfy $1-\tau_i\geq 0$, so the privacy-loss event reduces to a one-sided constraint on a \emph{nonnegative weighted} $\chi^2$ sum. This favorable sign structure is what enables stable evaluation and the clean Markov-type product bounds used in \Cref{lem:ndis-cov-analytic-form}.} Consequently, it suffices to upper bound $\delta_{\rX,\rY}(\varepsilon)$ in this direction to upper bound the covariance-induced leakage, which is exactly what \Cref{lem:ndis-cov-analytic-form} provides.

\begin{lemma}[$\delta_{\rX,\rY}$ upper bounds $\delta_{\rY,\rX}$ when covariances in Loewner order; Proof is in~\Cref{proof for lem: asymetric order for covariance}]
\label{lem: asymetric order for covariance}
Let $\rX \sim \multiNormal{0}{\Sigma_1}$ and $\rY \sim \multiNormal{0}{\Sigma_2}$ on $\Real^d$, where $\Sigma_1,\Sigma_2\succ 0$ and $\Sigma_1 \succeq \Sigma_2$. Then, for all $\varepsilon \geq 0$ and $d \in \nonnegInteger$,
\begin{align*}
    \delta_{\rX,\rY}(\varepsilon) \geq \delta_{\rY,\rX}(\varepsilon).
\end{align*}
\end{lemma}

To prove~\Cref{lem: asymetric order for covariance}, we first whiten by $\Sigma_2$ to reduce the comparison to the pair $(M,I_d)$ with $M=\Sigma_2^{-1/2}\Sigma_1\Sigma_2^{-1/2}\succeq I_d$, so it suffices to show $\delta_{M,I_d}(\varepsilon)\geq \delta_{I_d,M}(\varepsilon)$. Using the NDIS lemma together with the radial--angular expression of the standard multi-normal random variable $\rZ=RU$, both divergences can be written as expectations over the sphere: conditional on a direction $U$, they reduce to one-dimensional “radial’’ expressions of the form $g_d(\cdot,\alpha(U))$ and $h_d(\cdot,\beta(U))$, where $\alpha(U)=U^\top(M-I)U$ and $\beta(U)=U^\top(I-M^{-1})U$. The eigenvalue dominance $M\succeq I$ implies a pointwise coupling $\beta(U)\leq \alpha(U)/(1+\alpha(U))$. Combining this coupling with \Cref{prop:radial-comparison} below (and monotonicity properties of the resulting radial functions) gives a \emph{direction-wise dominance}: for every fixed direction $U$, the contribution to $\delta_{I_d,M}(\varepsilon)$ is bounded by the corresponding contribution to $\delta_{M,I_d}(\varepsilon)$. Averaging over $U$ then proves the claimed ordering.

\begin{proposition}[Proof is in~\Cref{proof for prop:radial-comparison}]
\label{prop:radial-comparison}
Let $d\ge 2$, $t\ge 0$, and $a\in\Real$. Define function, for $R^2\sim\chi^2_d$,
\begin{align*}
    g_d(a,t) \defin \Ex{\Bigl(1-\rExp{a-\tfrac{t}{2}R^2}\Bigr)_{+}},
    \quad
    h_d(a,t) \defin \Ex{\Bigl(1-\rExp{a+\tfrac{t}{2}R^2}\Bigr)_{+}}.
\end{align*}
Then, for every $t\ge 0$ and every $a\in\bigl[-\tfrac12\log(1+t),0\bigr]$, we have
\begin{align*}
    h_d\Bigl(a,\tfrac{t}{1+t}\Bigr)\;\le\; g_d\bigl(a+\log(1+t),t\bigr).
\end{align*}
\end{proposition}

\medskip
\noindent\textbf{General case (mean \emph{and} covariance shift).}
We now consider the general setting where $\rX$ and $\rY$ may differ in both mean and covariance. When the two covariances are comparable in Loewner order, \Cref{thm:ndis-analytic-form} provides a simple analytic upper bound on $\delta_{\rX,\rY}(\varepsilon)$ by decomposing the privacy leakage into two parts: (i) a \emph{mean-shift} component, corresponding to the case of equal covariance but different means, and (ii) a \emph{covariance-shift} component, corresponding to the case of equal mean but different covariances. The theorem then gives an explicit analytic bound by combining these two special-cases, whose results were established above, via triangle inequality.

\begin{theorem}[IS for mean and covariance shift; Proof is in~\Cref{proof for thm:ndis-analytic-form}]
\label{thm:ndis-analytic-form}
Let $\varepsilon\geq 0$ and $\rX \sim \multiNormal{\mu_1}{\Sigma_1}, \rY \sim \multiNormal{\mu_2}{\Sigma_2}$ on $\Real^d$, where $\Sigma_1,\Sigma_2\succ 0$. Assume either $\Sigma_1\succeq \Sigma_2$ or $\Sigma_2\succeq \Sigma_1$. Let $\Sigma_{\min},\Sigma_{\max}\in\{\Sigma_1,\Sigma_2\}$ be such that $\Sigma_{\max}\succeq \Sigma_{\min}$.
For $\varepsilon\geq 0$, define the two upper bounds
\begin{align*}
    \overline{\delta}^{(1)}(\varepsilon) \defin \inf_{\substack{\varepsilon_1,\varepsilon_2\geq 0\\ \varepsilon_1+\varepsilon_2=\varepsilon}}
        \Bigl[\delta_{\multiNormal{\mu_1}{\Sigma_{\min}},\,\multiNormal{\mu_2}{\Sigma_{\min}}}(\varepsilon_1)
            + e^{\varepsilon_1} \delta_{\multiNormal{\mu_2}{\Sigma_{\max}},\,\multiNormal{\mu_2}{\Sigma_{\min}}}(\varepsilon_2) \Bigr],\\
    \overline{\delta}^{(2)}(\varepsilon) \defin \inf_{\substack{\varepsilon_1,\varepsilon_2\ge 0\\ \varepsilon_1+\varepsilon_2=\varepsilon}}
        \Bigl[\delta_{\multiNormal{\mu_1}{\Sigma_{\max}},\,\multiNormal{\mu_1}{\Sigma_{\min}}}(\varepsilon_1)
            + e^{\varepsilon_1} \delta_{\multiNormal{\mu_1}{\Sigma_{\min}},\,\multiNormal{\mu_2}{\Sigma_{\min}}}(\varepsilon_2) \Bigr].
\end{align*}

Then,
\begin{align*}
    \delta_{\rX,\rY}(\varepsilon) \leq \min\bigl\{\overline{\delta}^{(1)}(\varepsilon),\,\overline{\delta}^{(2)}(\varepsilon)\bigr\}.
\end{align*}

Moreover, each component admits a simple analytic form: the mean-shift terms admits an analytic form in \Cref{prop:ndis-mean-analytic-form}, and the covariance-shift terms admit an analytic upper bound in \Cref{lem:ndis-cov-analytic-form}.
\end{theorem}

At a high level, \Cref{thm:ndis-analytic-form} bounds the $\delta_{\rX,\rY}$ by \emph{separating} the change from $\rX$ to $\rY$ into two simpler steps. We introduce an intermediate Gaussian $\rW$ so that one step changes only the mean (holding the covariance fixed) and the other step changes only the covariance (holding the mean fixed). The triangle inequality for the DP divergence (\Cref{prop: triangle for DP divergence}) then upper bounds $\delta_{\rX,\rY}(\varepsilon)$ as the sum of the divergences incurred on these two steps, with a privacy-budget split $\varepsilon=\varepsilon_1+\varepsilon_2$ that allocates how much “privacy cost’’ we pay on each step.

There are two natural ways to choose the intermediate distribution, corresponding to whether we perform the mean change first or the covariance change first. These two choices lead to the bounds $\overline{\delta}^{(1)}(\varepsilon)$ and $\overline{\delta}^{(2)}(\varepsilon)$, respectively. Optimizing over the budget split and taking the better of the two paths yields
\begin{align*}
    \delta_{\rX,\rY}(\varepsilon)\le \min\bigl\{\overline{\delta}^{(1)}(\varepsilon),\overline{\delta}^{(2)}(\varepsilon)\bigr\}.
\end{align*}

\begin{proposition}[Triangle inequality for DP divergence; Proof is in~\Cref{proof for prop: triangle for DP divergence}]
\label{prop: triangle for DP divergence}
Let $X,Y,Z$ be random variables, then for all $\varepsilon_1,\varepsilon_2 \geq 0$,
\begin{align*}
  \delta_{X,Z}(\varepsilon_1 + \varepsilon_2) \leq  \delta_{X,Y}(\varepsilon_1) + e^{\varepsilon_1}\, \delta_{Y,Z}(\varepsilon_2).
\end{align*}
\end{proposition}

\section{Random-Projection DP Mechanisms via NDIS}\label{sec:rpmech}

In this section, we present an application of NDIS (and associated tools) to prove the $(\varepsilon,\delta)$-DP of random projection. We treat Gaussian random projection $\gRP_r(\DB)$ as a \emph{Gaussian-output algorithm} and use NDIS (cf.~\Cref{lemma: ndis,thm:ndis-two-genchi2-cdf}) to characterize the indistinguishability $\delta_{\rX,\rY}(\varepsilon)$ between neighboring outputs $\gRP_r(\DB)$ and $\gRP_r(\DB')$. This leads to an explicit privacy characterization, which we then use to ``lift'' random projection into a mechanism that can meet desired privacy  $(\varepsilon,\delta)$. Building on this, we prove the privacy and utility guarantees of the resulting Random Projection mechanism $\Mech_{RP}$.
We also directly show that using leverage to calibrate noise compares favourably with LSV-based differentially-private RP mechanisms from previous work (Fig.~\ref{fig:leverage-ratio}).
Finally, we show how to further optimize $\Mech_{RP}$ by combining it with standard DP techniques such as subsampling and the \emph{Propose--Test--Release (PTR)} paradigm, demonstrating that NDIS-based analyses fit cleanly into existing tools in DP literature.

\subsection{Construction}

Our RP mechanism $\Mech_{RP}$ is found in~\Cref{alg::RP mechanism}. At a high level, its design is guided by~\Cref{prop:grp-monotone}, which formalizes the principle that \emph{larger leverage yields weaker privacy}, and builds on the pairwise indistinguishability analysis in \Cref{lemma: rp inherent}. Concretely, the mechanism requires finding a sufficiently small leverage threshold $\leverage$ corresponding to the desired $\delta(\epsilon)$, and achieving (at most) such leverage through regularizing the input database. We note that the `amount' of regularization depends only on the leverage threshold and input space, and not on the input database itself.

We begin by~\Cref{lemma: rp inherent} studying the \emph{inherent} privacy of Gaussian random projection under the row-deletion neighboring relation, i.e., one database is a neighbor of another through deleting a row.
Concretely, fix the $i$-th record and let $\DB'$ be obtained by removing row $i$ from $\DB$. We analyze the indistinguishability $\delta_{\rX,\rY}(\varepsilon)$ between the neighboring outputs $\rX=\gRP_r(\DB)$ and $\rY=\gRP_r(\DB')$.~\footnote{Deletion let the RP output covariance have $\Sigma(\DB)\preceq \Sigma(\DB')$, and by the covariance-order asymmetry in \Cref{lem: asymetric order for covariance}, the deletion orientation is the worst case in the sense that $\delta_{\rX, \rY} \geq \delta_{\rY, \rX}$. Hence it suffices to analyze the row-deletion direction; the add-one direction follows automatically (details refer to the proof of \Cref{thm:RP-mech-DP}).} \Cref{lemma: rp inherent} expresses $\delta_{\rX,\rY}(\varepsilon)$ as an explicit analytic function of the leverage $\leverage$, showing that the privacy impact of record $i$ is entirely captured by its leverage score in $\DB$. Moreover, the target dimension $r$ governs the concentration of the RP output, and hence the tail behavior that determines $\delta$.

\begin{lemma}[Proof is in~\Cref{proof for lemma: rp inherent}]
\label{lemma: rp inherent}
Let $n, r, d \in \positiveInteger,$ and fix $i \in [n]$. Let $\DB\in\Real^{n\times d}$ such that $\DB^\top\DB \succ 0$, and let $\DB'$ be obtained from $\DB$ by removing row $i$. Let $\leverage$ be the leverage score of row $i$ in $\DB$ (cf.~\Cref{def: leverage}), and let
\begin{align*}
    \rho = \frac{1}{1-\leverage}\geq 1, \qquad t_0 = \frac{2\bigl(\varepsilon+\tfrac{r}{2}\log\rho\bigr)}{\rho-1}.
\end{align*}

Then, for random matrices $\rX \defin \gRP_r(\DB), \rY \defin \gRP_r(\DB_{-i})$ ($\gRP_r(\cdot)$ is defined in~\Cref{def:grp}), and for every $\varepsilon\geq 0$,
\begin{align*}
    \delta_{\rX,\rY}(\varepsilon) = \frac{1}{\Gamma(\tfrac{r}{2})} \Bigl(\Gamma\Bigl(\tfrac{r}{2}, \tfrac{t_0}{2}\Bigr) - e^{\varepsilon} \Gamma\Bigl(\tfrac{r}{2}, \tfrac{\rho t_0}{2}\Bigr)\Bigr),
\end{align*}
where $\Gamma(\cdot,\cdot)$ denotes the upper incomplete gamma function.
\end{lemma}

To prove~\Cref{lemma: rp inherent}, we treat it as a direct application of the NDIS lemma specialized to the \emph{zero-mean, covariance-shift} setting. Indeed, $\gRP_r(\DB)$ produces $r$ i.i.d.\ Gaussian columns with covariance $\Sigma_1=\DB^\top\DB$, whereas for the row-deleted database $\DB'$ the output has the same form with covariance $\Sigma_2=\DB'^\top\DB'=\Sigma_1-vv^\top$, where $v^\top$ is the removed row. After vectorizing the $d\times r$ output, the neighboring pair becomes two centered Gaussians in $\Real^{dr}$ with covariances $I_r\otimes\Sigma_1$ and $I_r\otimes\Sigma_2$, placing us exactly in the zero-mean covariance-shift regime.

We then inspect the resulting NDIS quadratic form $g_\varepsilon(Z)$ and observe that all dependence on the database enters through two quantities: (i) the determinant ratio in the constant term $c(\varepsilon)$ and (ii) the spectrum of $\Sigma_1^{1/2}\Sigma_2^{-1}\Sigma_1^{1/2}$. Both are controlled by the leverage of the removed record $v^\top$.  By the matrix determinant lemma,
\begin{align*}
    \frac{\det(\Sigma_1)}{\det(\Sigma_2)} =\bigl(1-v^\top\Sigma_1^{-1}v\bigr)^{-1} = (1-\leverage)^{-1} =\rho,
\end{align*}
and writing $u=\Sigma_1^{-1/2}v$ yields
\begin{align*}
    \Sigma_1^{1/2}\Sigma_2^{-1}\Sigma_1^{1/2} = (I-uu^\top)^{-1},
\end{align*}
whose eigenvalues are $\rho$ in the direction $u$ and $1$ on $u^\perp$. Consequently, the privacy impact, $\delta_{\rX,\rY}(\varepsilon)$, is fully captured by the single leverage-induced scalar $\rho=(1-\leverage)^{-1}$.

Finally, the same rank-one structure explains why the resulting expression is \emph{clean}. While NDIS in general reduces $\delta_{\rX,\rY}(\varepsilon)$ to generalized-$\chi^2$ CDF computations, here the rank-one update $\Sigma_2=\Sigma_1-vv^\top$ leaves only $r$ nontrivial directions in the quadratic form (one per RP column), with all remaining directions canceling. Consequently, $g_\varepsilon(Z)$ depends only on the simple statistic $T=\sum_{k=1}^r Z_k^2\sim\chi_r^2$, so the $dr$-dimensional expectation collapses to a one-dimensional $\chi^2$ tail integral. Evaluating this tail yields upper incomplete gamma functions, giving the stated analytic formula.

\medskip
Building on \Cref{lemma: rp inherent}, \Cref{prop:grp-monotone} formalizes the intuition that \emph{larger leverage implies weaker privacy} by showing that, for every fixed $\varepsilon\geq 0$, the resulting smallest $\delta$ value of the RP mechanism is nondecreasing in the leverage parameter. This monotone ordering is central to our subsequent DP analysis: once we enforce a uniform leverage bound for all records (e.g., via regularization), the worst-case neighboring pair is realized at the maximal leverage.
Consequently, proving that a random-projection mechanism is $(\varepsilon,\delta)$-DP reduces to analyzing the indistinguishability $\delta_{\rX,\rY}(\varepsilon)$ for a \emph{single} neighboring pair $(\rX,\rY)=(\gRP_r(\DB),\gRP_r(\DB'))$ --- precisely the type of Gaussian indistinguishability problem that the NDIS toolbox is designed to handle.

\begin{proposition}[Proof is in~\Cref{proof for prop:grp-monotone}]
\label{prop:grp-monotone}
Let $r\in\positiveInteger$ and $\varepsilon\geq 0$. For $p\in[0,1)$ define
\begin{align*}
    \rho(p) \defin \frac{1}{1-p}\geq 1, \qquad
    t_0(p) \defin \frac{2\bigl(\varepsilon+\tfrac{r}{2}\log\rho(p)\bigr)}{\rho(p)-1}.
\end{align*}
Define the \emph{RP $\delta$-curve} (as a function of leverage) as
\begin{align*}
    \delta^{\gRP}_{r}(\varepsilon; p) \defin \frac{1}{\Gamma(\tfrac{r}{2})}
    \Bigl( \Gamma\Bigl(\tfrac{r}{2}, \tfrac{t_0(p)}{2}\Bigr) - e^{\varepsilon} \Gamma\Bigl(\tfrac{r}{2}, \tfrac{\rho(p) t_0(p)}{2}\Bigr) \Bigr).
\end{align*}
Then $p\mapsto \delta^{\gRP}_{r}(\varepsilon;p)$ is nondecreasing on $[0,1)$.
\end{proposition}

\begin{small}
\begin{figure}[htp]
\captionsetup{skip=2pt}
\centering
\setlength{\fboxrule}{0pt}
\fbox{
\setlength{\fboxrule}{0.5pt}
\framebox{
\begin{minipage}[t]{0.94\linewidth}

\begin{footnotesize}
\textbf{Input:} A database $\DB\in\Real^{n\times d}$; target dimension $r\geq 1$; privacy parameters $\varepsilon\in\nonnegReal$, $\delta\in[0,1)$; and a public parameter $l\in\posReal$ that upper bounds the row norm in the record universe.

\textbf{Output:} A matrix $\widetilde M \in\Real^{d\times r}$.

\begin{enumerate}
    \item Use binary search to find $p^\star\in(0,1)$ such that
    \begin{align*}
        \delta^{\gRP}_{r}(\varepsilon; p^\star) \leq \delta,
    \end{align*}
    where $\delta^{\gRP}_{r}(\varepsilon; p)$ is defined in Prop.~\ref{prop:grp-monotone} ($p\mapsto \delta^{\gRP}_{r}(\varepsilon;p)$ is monotone in $p$)

    \item Set the ridge level $\lambda = \frac{l^2}{p^\star}$ and define the augmented (regularized) database
    \begin{align*}
        \overline \DB \defin
        \begin{bmatrix}
            \DB \\ \sqrt{\lambda}\, I_d
        \end{bmatrix}
        \in \Real^{(n+d)\times d}.
    \end{align*}

    \item Return $\widetilde M = \gRP_r(\overline \DB).$
\end{enumerate}
\end{footnotesize}

\end{minipage}
}}
\caption{Random Projection (RP) mechanism $\Mech_{RP}$.}
\label{alg::RP mechanism}
\end{figure}
\end{small}

Equipped with \Cref{lemma: rp inherent,prop:grp-monotone}, we can now `upgrade'/`lift' Gaussian RP into an $(\varepsilon,\delta)$-DP mechanism ($\Mech_{RP}$ in~\Cref{alg::RP mechanism}), where two databases are neighbors/adjacent when they are the same, except one row is added or removed. Informally, the privacy of Gaussian random projection over a given input domain is governed by how large a leverage score any single record can attain within that domain: higher maximal leverage leads to larger $\delta_{\rX,\rY}(\varepsilon)$, while uniformly small leverage yields strong privacy. Using this connection, $\Mech_{RP}$ performs two key tasks: (1) it converts the desired $(\varepsilon, \delta)$ privacy parameters into the maximum allowable leverage that a desired input domain could have to ensure the privacy, and (2) it regularizes the input domain, so that the regularized input domain's maximum leverage is at desired level and applies RP over the the regularized input domain to achieve $(\varepsilon, \delta)$-DP. We recall that \Cref{lemma: rp inherent} is stated for the row-deletion neighboring relation; the discussion above for the add-one-row direction follows automatically from the covariance-order asymmetry in~\Cref{lem: asymetric order for covariance}.

\subsection{Privacy and Utility analysis of $\Mech_{RP}$}
\label{subsec: utility}

We present the concrete mechanism $\Mech_{RP}$ in \Cref{alg::RP mechanism}, and \Cref{thm:RP-mech-DP} states that $\Mech_{RP}$ is $(\varepsilon,\delta)$-DP.

\begin{theorem}[Privacy of $\Mech_{RP}$; Proof is in~\Cref{proof for thm:RP-mech-DP}]
\label{thm:RP-mech-DP}
Let $\varepsilon \geq 0, \delta \in (0,1)$. The Random Projection (RP) mechanism $\Mech_{RP}$ in \Cref{alg::RP mechanism} is $(\varepsilon,\delta)$-DP under the standard add/remove (unbounded) notion of neighboring databases.
\end{theorem}

At a high level, $\Mech_{RP}(\DB)$ first applies a \emph{public} ridge regularization, producing the augmented matrix $\overline\DB$. Because this regularization is data-independent, the same transformation applies to the deletion neighbor $\DB'$, yielding the paired augmented matrices $\overline\DB$ and $\overline\DB'$. Distinguishing $\rX=\gRP_r(\overline\DB)$ and $\rY=\gRP_r(\overline\DB')$ then places us exactly in the rank-one covariance-shift regime of \Cref{lemma: rp inherent}.

The role of the ridge is to enforce a uniform leverage bound. Writing $\Sigma_1=\overline\DB^\top\overline\DB=\DB^\top\DB+\lambda I_d$, we have $\Sigma_1\succeq \lambda I_d$, and hence the leverage of the deleted row $v^\top$ in the regularized design satisfies
\begin{align*}
    \overline{\leverage} = v^\top(\DB^\top\DB+\lambda I_d)^{-1}v \leq \frac{1}{\lambda}\twoNorm{v}^2 \leq \frac{l^2}{\lambda} = p^\star.
\end{align*}
With this leverage control in hand, \Cref{lemma: rp inherent} identifies the deletion-direction indistinguishability as $\delta_{\rX,\rY}(\varepsilon)=\delta^{\gRP}_r(\varepsilon;\overline\leverage)$, and \Cref{prop:grp-monotone} turns the leverage bound $\overline\leverage\le p^\star$ into the worst-case bound $\delta_{\rX,\rY}(\varepsilon)\le \delta^{\gRP}_r(\varepsilon;p^\star)\le \delta$ by the choice of $p^\star$ in the mechanism (cf.~\Cref{alg::RP mechanism}). Finally, the add-one direction follows from the covariance-order asymmetry (\Cref{lem: asymetric order for covariance}), so controlling deletions suffices to obtain $(\varepsilon,\delta)$-DP under add/remove adjacency.

\Cref{thm:mech-rp-utility} studies the utility of $\Mech_{RP}$ by examining how accurately the released sketch preserves the Gram matrix $\Sigma=\DB^\top\DB$. This is a natural criterion as Gaussian random projection depends on the data only through $\Sigma$: each released column of $\widetilde M=\gRP_r(\overline\DB)$ is an independent draw from $\multiNormal{0}{\Sigma+\lambda I_d}$.\footnote{This “sketch-as-second-moment” viewpoint is standard in randomized numerical linear algebra and sketching~\cite{HalkoMT11,Woodruff14}.} Accordingly, the estimator $\widehat\Sigma=\frac1r \widetilde M\widetilde M^\top-\lambda I_d$ is {\em unbiased} for the original Gram matrix, i.e., $\Ex{\widehat\Sigma}=\Sigma$.

\Cref{thm:mech-rp-utility} also shows that $\widehat\Sigma$ concentrates to $\DB^\top\DB$ in {\em spectral norm} at the rate $\tilde O(\lambda\sqrt{d/r})$, making the utility--privacy tradeoff explicit. In particular, accuracy improves with the sketch dimension $r$, while privacy is enforced through the ridge level $\lambda=\lambda(\varepsilon,\delta,r)$ chosen via the monotone calibration in Step~1 of \Cref{alg::RP mechanism}. It highlights that tighter privacy targets $(\varepsilon,\delta)$ translate into stronger regularization and hence larger estimation error.

\begin{theorem}[Utility of $\Mech_{RP}$; Proof is in~\Cref{proof for thm:mech-rp-utility}]
\label{thm:mech-rp-utility}
Given a database $\DB\in\Real^{n\times d}$, and let $\widetilde M=\Mech_{RP}(\DB)$, where $\Mech_{RP}$ is defined in~\Cref{alg::RP mechanism}. Let
\begin{align*}
    \widehat\Sigma \defin \frac{1}{r}\,\widetilde M\widetilde M^\top-\lambda I_d, \qquad
    \Sigma \defin \DB^\top\DB,
\end{align*}
where $\lambda=l^2/p^\star$ and $p^\star=p^\star(\varepsilon,\delta,r)\in(0,1)$ is the value selected in Step~1 of~\Cref{alg::RP mechanism}. Then:
\begin{enumerate}
    \item \textbf{Unbiasedness.} $\Ex{\widehat\Sigma}=\Sigma$.
    \item \textbf{High-probability spectral-norm error.} There exists an absolute constant $C>0$ such that for every $\beta\in(0,1)$,
    with probability at least $1-\beta$,
    \begin{align*}
        \twoNorm{\widehat\Sigma-\Sigma} \leq C\left(\sqrt{\frac{d+\log(2/\beta)}{r}}+\frac{d+\log(2/\beta)}{r}\right)\Bigl(\twoNorm{\DB}^2+\lambda\Bigr).
    \end{align*}
\end{enumerate}
\end{theorem}

\noindent\textbf{Leverage Calibration: NDIS vs. LSV-Based Bounds.}
We compare the leverage thresholds implied by our NDIS-based calibration and by the LSV-based calibration of~\cite{Sheffet17,ALTSheffet19}. We fix $\delta=10^{-6}$ and a row-norm bound $l=1$, and vary the target privacy parameter $\varepsilon$ over a grid (here, $\varepsilon\in[0.1,5]$). We report results for several sketch dimensions $r\in\{50,100,200,500\}$.

\begin{figure}[t]
    \centering
    \includegraphics[width=0.8\linewidth]{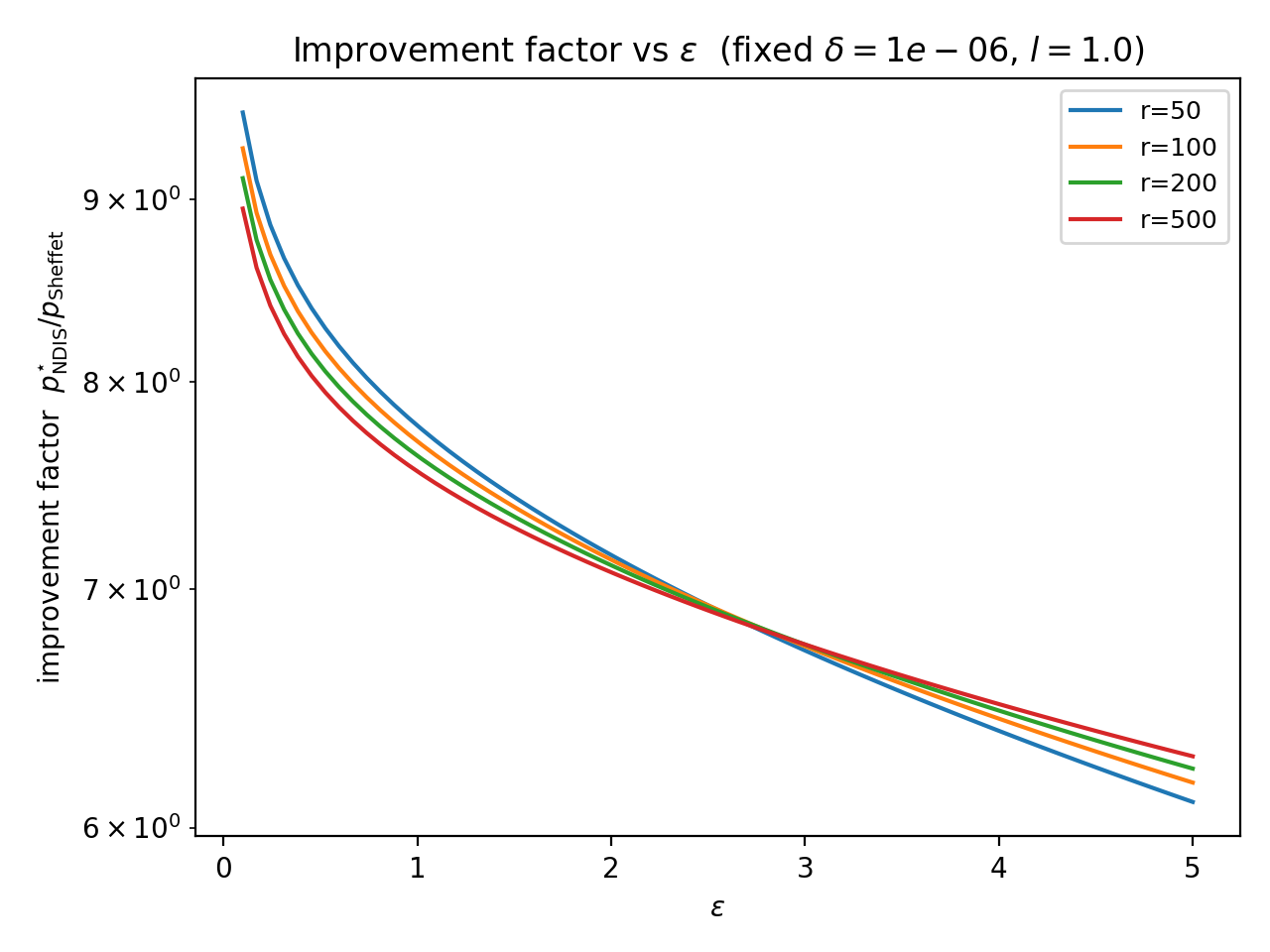}
    \caption{Improvement factor in allowable leverage:
    $p^\star_{\mathrm{NDIS}}/p_{\mathrm{LSV}}$ versus $\varepsilon$ (log scale), with $\delta=10^{-6}$, $l=1$, and $r\in\{50,100,200,500\}$.
    Here $p^\star_{\mathrm{NDIS}}$ is the maximal leverage threshold computed by our NDIS-based RP analysis, while $p_{\mathrm{LSV}}$ is the leverage bound implied by the LSV-based calibration of~\cite{ALTSheffet19,Sheffet17}. Ratios above $1$ indicate that LSV-based calibration is more conservative, forcing smaller allowable leverage and thus larger ridge regularization.}
    \label{fig:leverage-ratio}
\end{figure}

For each $(\varepsilon,\delta,r)$, our method computes the maximal leverage threshold
\begin{align*}
    p^\star_{\mathrm{NDIS}}(\varepsilon,\delta;r) = \sup\{p\in[0,1): \delta^{\gRP}_{r}(\varepsilon; p)\leq \delta\},
\end{align*}
obtained using binary search. As a baseline, we compute the leverage threshold \emph{implicitly} enforced by the LSV-based analysis: Theorem~2 of~\cite{ALTSheffet19} calibrates a ridge parameter $w$ (equivalently $w^2 I$ added to $\DB^\top \DB$), and the condition $\sigma_{\min}(A)\geq w$ together with $\twoNorm{x_i} \leq l$ implies
\begin{align*}
    p_{\mathrm{LSV}}(\varepsilon,\delta;r) \defin \ell_{\max}(A)\le \frac{l^2}{w^2} = \frac{\varepsilon}{4\bigl(\sqrt{2r\log(4/\delta)}+\log(4/\delta)\bigr)}.
\end{align*}

\Cref{fig:leverage-ratio} plots the improvement factor $\frac{p^\star_{\mathrm{NDIS}}(\varepsilon,\delta;r)}{p_{\mathrm{LSV}}(\varepsilon,\delta;r)}$ as a function of $\varepsilon$, for several values of $r$.
A ratio larger than $1$ means that our analysis guarantees $(\varepsilon,\delta)$-DP while allowing a strictly larger worst-case leverage, which translates into less required regularization. Across the full range of $\varepsilon$ shown, the ratios are consistently above $1$, demonstrating that the LSV-based condition controls privacy only through a conservative upper bound on leverage and can substantially over-regularize relative to an analysis that characterizes privacy directly in terms of leverage.

\subsection{Optimizing $\Mech_{RP}$ via Subsampling and Propose--Test--Release}
We have shown that the NDIS lemma gives an \emph{exact} characterization of the IS for random projection given neighboring databases, which in turn yields a method to calibrate $\Mech_{RP}$ to satisfy $(\varepsilon,\delta)$-DP. Such \emph{global} calibration, however, is inherently conservative: it injects noise uniformly over \emph{all} databases in the universe, and thus can over-regularize well-conditioned instances. In this section, we show that the NDIS-based analysis composes cleanly with today’s standard DP toolkit, enabling \emph{database-agnostic} privacy guarantees while delivering \emph{database-dependent} utility improvements.

\begin{small}
\begin{figure}[htp]
\captionsetup{skip=2pt}
\centering
\setlength{\fboxrule}{0pt}
\fbox{
\setlength{\fboxrule}{0.5pt}
\framebox{
\begin{minipage}[t]{0.94\linewidth}

\begin{footnotesize}
\textbf{Input:} A database $\DB\in\Real^{n\times d}$; target dimension $r\ge 1$; privacy parameters $\varepsilon\in\nonnegReal$, $\delta\in[0,1)$; a Poisson subsampling rate $1 \geq q > \delta$; and a public parameter $l\in\posReal$ that upper bounds the row norm in the record universe.

\textbf{Output:} A matrix $\widetilde M\in\Real^{d\times r}$.

\begin{enumerate}
    \item Construct a sub-database $\DB_S$ by including each row of $\DB$ independently with probability $q$.

    \medskip
    \item Set $\varepsilon_0 = \ln\Bigl(1+\frac{e^{\varepsilon}-1}{q}\Bigr),$ and $\delta_0 = \frac{\delta}{q}.$

    \medskip
    \item Return $\widetilde M \gets \Mech_{RP}\bigl(\DB_S; r,\varepsilon_0,\delta_0,l\bigr),$ where $\Mech_{RP}$ is defined in~\Cref{alg::RP mechanism}.
\end{enumerate}
\end{footnotesize}

\end{minipage}
}}
\caption{Poisson-subsampled wrapper $\Mech^{\mathsf{Pois}}_{RP}$ for $\Mech_{RP}$.}
\label{alg:subsampled-wrapper-rp}
\end{figure}
\end{small}

\Cref{alg:subsampled-wrapper-rp} presents a Poisson-subsampled wrapper, denoted $\Mech^{\mathsf{Pois}}_{RP}$. On input $\DB$, the $\Mech^{\mathsf{Pois}}_{RP}$ first forms a random sub-database $\DB_S$ by including each row independently with probability $q$, and then runs the baseline mechanism $\Mech_{RP}$ on $\DB_S$ with \emph{weaker} privacy parameters $(\varepsilon_0,\delta_0)$ chosen so that the overall output remains $(\varepsilon,\delta)$-DP. \Cref{thm:subsampling RP-mech-DP} states that $\Mech^{\mathsf{Pois}}_{RP}$ is $(\varepsilon,\delta)$-DP.

Subsampling improves utility in two complementary ways. From a privacy perspective, subsampling making the privacy ``event'' that distinguishes neighboring databases occurs less often, so that we can afford a weaker level of regularization inside $\Mech_{RP}$. From a utility perspective, when the dataset is sufficiently redundant so that each record is, in aggregate, representable by the others, dropping a random $(1-q)$ fraction of rows has limited effect on the same geometric signal ($\DB^\top\DB$) that drives the random projection.

\begin{theorem}[Proof is in~\Cref{proof for thm:subsampling RP-mech-DP}]
\label{thm:subsampling RP-mech-DP}
Let $\varepsilon \geq 0, \delta \in (0,1)$. The mechanism $\Mech^{\mathsf{Pois}}_{RP}$ in \Cref{alg:subsampled-wrapper-rp} is $(\varepsilon,\delta)$-DP under the standard add/remove neighboring notion.
\end{theorem}

\begin{small}
\begin{figure}[htp]
\captionsetup{skip=2pt}
\centering
\setlength{\fboxrule}{0pt}
\fbox{
\setlength{\fboxrule}{0.5pt}
\framebox{
\begin{minipage}[t]{0.94\linewidth}

\begin{footnotesize}
\textbf{Input:} A database $\DB\in\Real^{n\times d}$; target dimension $r\ge 1$; privacy parameters $\varepsilon\in\nonnegReal$, $\delta\in[0,1)$; and a public parameter $l\in\posReal$ that upper bounds the row norm in the record universe.

\textbf{Parameters:} A privacy split $\delta=\delta_R+\delta_T+\delta_{\mathrm{ptr}}$ and an eigenvalue-noise scale $\tau\in\posReal$.

\textbf{Output:} A matrix $\widetilde M\in\Real^{d\times r}$.

\begin{enumerate}
    \item Use binary search to find $\varepsilon_T \geq 0$ such that
    \begin{align*}
        \Phi\Bigl(\frac{l^2}{2\tau}-\frac{\varepsilon_T\cdot\tau}{l^2}\Bigr) -e^{\varepsilon_T}\Phi\Bigl(-\frac{l^2}{2\tau}-\frac{\varepsilon_T\cdot\tau}{l^2}\Bigr) \leq \delta_T
    \end{align*}
    Set $\varepsilon_R = \max\{\varepsilon-\varepsilon_T,\,0\}$.

    \medskip
    \item Use binary search to find $p^\star\in(0,1)$ such that
    \begin{align*}
        \delta^{\gRP}_{r}(\varepsilon_R; p^\star) \leq \delta_R,
    \end{align*}
    where $\delta^{\gRP}_{r}(\cdot;\cdot)$ is defined in~\Cref{prop:grp-monotone}.

    \medskip
    \item Let $\lambda = \lambda_{\min}(\DB^\top\DB),$ and $\alpha = \tau\cdot \Phi^{-1}(1-\delta_{\mathrm{ptr}})$. Set
    \begin{align*}
        \lambda_{\mathrm{lb}} = \max\{\lambda+\eta-\alpha, 0\}, \qquad \text{where} \quad \eta \sim \multiNormal{0}{\tau^2}.
    \end{align*}

    \medskip
    \item Set the ridge level $\lambda = \max\Bigl\{\frac{l^2}{p^\star}-\lambda_{\mathrm{lb}},0\Bigr\}$, and define the augmented (regularized) database
    \begin{align*}
        \overline \DB \defin
        \begin{bmatrix}
            \DB \\ \sqrt{\lambda}\, I_d
        \end{bmatrix}
        \in \Real^{(n+d)\times d}.
    \end{align*}

    \item Return $\widetilde M = \gRP_r(\overline \DB).$
\end{enumerate}
\end{footnotesize}

\end{minipage}
}}
\caption{PTR-based wrapper $\Mech^{\mathsf{PTR}}_{RP}$ for $\Mech_{RP}$.}
\label{alg:ptr-rp}
\end{figure}
\end{small}

\Cref{alg:ptr-rp} presents a PTR-based wrapper, denoted as $\Mech^{\mathsf{PTR}}_{RP}$. It follows classical \emph{Propose--Test--Release} (PTR) paradigm, and the idea is: i) \emph{proposing} a data-dependent condition under which random projection is inherently more stable when removing a record; here a larger minimum eigenvalue $\lambda_{\min}(\DB^\top\DB)$ implies smaller leverage scores ii) \emph{testing} this condition privately by releasing a conservative lower bound $\lambda_{\mathrm{lb}}$ on $\lambda_{\min}(\DB^\top\DB)$ with failure probability $\delta_{\mathrm{ptr}}$; the lower bound is derived from a differentially private query on $\lambda_{\min}$ iii) \emph{releasing} a projection computed on an augmented database $\overline{\DB}=[\DB;\sqrt{\lambda}I_d]$ with the ridge level $\lambda$ chosen using the lower bound $\lambda_{\mathrm{lb}}$. In particular, the ridge is set to satisfy the same NDIS-based indistinguishability constraint as used in~\Cref{alg::RP mechanism}, while the privacy budget is split between the test and release steps. \Cref{thm:ptr RP-mech-DP} states that $\Mech^{\mathsf{PTR}}_{RP}$ is $(\varepsilon,\delta)$-DP.

\begin{theorem}[Proof is in~\Cref{proof for thm:ptr RP-mech-DP}]
\label{thm:ptr RP-mech-DP}
Let $\varepsilon \geq 0, \delta \in (0,1)$. The mechanism $\Mech^{\mathsf{PTR}}_{RP}$ in \Cref{alg:ptr-rp} is $(\varepsilon,\delta)$-DP under the standard add/remove neighboring notion.
\end{theorem}

\Cref{prop:ptr-profitability} states that, $\Mech^{\mathsf{PTR}}_{RP}$ improves utility on databases that has a large enough $\lambda_{\min}(\DB^\top\DB)$. Informally, whenever the data geometry is sufficiently favorable, the (small) privacy cost of privately estimating a lower bound on $\lambda_{\min}(\DB^\top\DB)$ is repaid by a strictly less conservative release, i.e., the strictly less regularization in random projection. More precisely, if $\lambda_{\min}(\DB^\top\DB)$ exceeds a concrete threshold (capturing the ``slack'' needed to pay for the noisy lower-bound test, via $\alpha$, and the NDIS-calibrated noise scale, via $l^2/p^\star$), then the PTR-selected ridge level $\lambda_{\mathsf{ptr}}$ is strictly smaller than the baseline ridge $\lambda_{\mathsf{rp}}$ used by $\Mech_{RP}$.

\begin{proposition}
\label{prop:ptr-profitability}
Let $\lambda_{\mathsf{rp}}$ denote the ridge level used by $\Mech_{RP}$ (cf.~Step~2 of \Cref{alg::RP mechanism}). Let $(\varepsilon_R,p^\star,\alpha,\lambda_{\mathrm{lb}})$ be the variables defined in $\Mech^{\mathsf{PTR}}_{RP}$ (cf.~Steps~1--3 of \Cref{alg:ptr-rp}. If a database $\DB$ satisfies
\begin{align*}
    \lambda_{\min}(\DB^\top \DB) > \alpha + \frac{l^2}{p^\star} - \lambda_{\mathsf{rp}},
\end{align*}
then the PTR ridge level $\lambda_{\mathsf{ptr}} = \max\Bigl\{\frac{l^2}{p^\star}-\lambda_{\mathrm{lb}},0\Bigr\}$ is strictly smaller than $\lambda_{\mathsf{rp}}$.
\end{proposition}

\section{Creating DP Mechanisms from Gaussian-output Algorithms via NDIS}

Many algorithms in modern statistics and machine learning naturally output a Gaussian distribution whose \emph{mean and covariance both depend on the dataset}---for example, posterior or asymptotic normal approximations~\cite{TierneyK1986,RueMC09}, randomized estimators~\cite{McCN89}, and second-order procedures~\cite{vanderVaart1998,bishop06}. Thus, releasing Gaussian outputs with differential privacy is central to private learning. However, DP mechanisms (with Gaussian output) for learning tasks use various forms of the Gaussian mechanism (see related work), or the privacy of specific algorithms (e.g., RP).

In this section, we present a `wrapper' which calibrates a general Gaussian-output algorithm's output distribution so that it is turned/`lifted' into a Gaussian-output DP mechanism. Importantly, this wrapper does not depend on how the algorithm is implemented, only on the `sensitivity' of this algorithm's output on neighboring databases.  Here, our notion of sensitivity generalizes from that of the standard Gaussian mechanism, which describes the maximal shift in the output distribution's mean on neighboring inputs. A Gaussian-output algorithm has a NDIS sensitivity if there is a bound in the shift in mean and covariance for the outputs of neighboring inputs, and the covariances satisfy a mild \emph{Loewner-comparability} condition (required to apply our NDIS tools). Given the NDIS sensitivity exists (and we show examples for useful Gaussian-output algorithms), this framework for creating DP mechanisms leverages NDIS to summarize the worst-case change in the output distribution. Then, it \emph{automatically calibrate} the amount of additional noise needed to achieve $(\varepsilon,\delta)$-DP.

More precisely, we first introduce an abstraction of \emph{Gaussian-output algorithms} (cf.~\Cref{def:gaussian-output-alg}) via an underlying deterministic map $q(\DB)=(\mu(\DB),\Sigma(\DB))$. We then describe a family of  \emph{additive-noise mechanisms} (cf.~\Cref{def:additive-noise-family}) that increases/inflates the covariance of the output of these algorithms (which preserves Gaussianity). The core challenge is that privacy depends on how the \emph{entire} Gaussian distribution changes across neighboring datasets; when the covariance is data-dependent, proving DP would in principle require controlling indistinguishability over \emph{all} neighboring pairs, which is typically intractable.

NDIS makes this problem tractable via a \emph{pairwise reduction}: as shown in~\Cref{thm:ndis-two-genchi2-cdf}, the IS for any fixed neighboring pair can be written as the scalar functional $\delta_\varepsilon(m,R)$ of an invariant pair $(m,R)$ derived from  two Gaussians (cf. Eq.~\ref{def: nids function}). We then define a NDIS-based sensitivity model (Def.~\ref{def:ndis-sensitivity-loewner}), and use it to upper-bound the collection of invariants (over pairs of neighboring databases) by an \emph{envelope} that is tractable to work with. Finally, leveraging our NDIS tool for the bound on IS when covariances are Loewner-comparable  (\Cref{thm:ndis-analytic-form}) allows us to simplify and upper bound $\delta_\varepsilon(m,R)$ via a small number of meaningful parameters. This gives us an efficient upper bound of the envelope’s worst-case privacy and an explicit $(\varepsilon,\delta)$-DP guarantee. The resulting NDIS-Calibrated Gaussian Mechanism is shown in~\Cref{alg::ndis wrapper}.

\subsection{Our NDIS-Calibrated Gaussian Mechanism}

\Cref{def:gaussian-output-alg} formalizes the class of algorithms studied in this section: randomized procedures whose output distribution is multivariate Gaussian, with both the mean and covariance could be data-dependent.

\begin{definition}[Gaussian-output algorithm]
\label{def:gaussian-output-alg}
Let $\cD$ be a dataset space equipped with the remove/add adjacency relation. Let $m \in \positiveInteger.$ A randomized algorithm $\Mech_q$ is called a \emph{Gaussian-output algorithm} if there exists a deterministic map
\begin{align*}
    q:\cD \to \Real^m\times \mathbb S_{++}^m, \qquad
    q(\DB)=(\mu(\DB),\Sigma(\DB)),
\end{align*}
such that given any input $\DB\in \cD$, $\Mech_q$ outputs $\Mech_q(\DB) \sim \multiNormal{\mu(D)}{\Sigma(\DB)}$.
\end{definition}

A natural way to privatize such outputs while preserving their Gaussianity is to add independent Gaussian noise. \Cref{def:additive-noise-family} models this by a family of \emph{additive-noise} mechanisms parameterized by a positive semi-definite (PSD) matrix $N\succeq 0$, which inflates the output covariance from $\Sigma(\DB)$ to $\Sigma(\DB)+N$. This family include the standard isotropic Gaussian noise, when $N=\sigma^2 I_m$.

\begin{definition}[Family of Additive-noise Mechanisms for Gaussian-output Algorithms]
\label{def:additive-noise-family}
Let $\Mech_q$ be a Gaussian-output algorithm (cf.~\Cref{def:gaussian-output-alg}). For any \emph{positive-semi-definite} (PSD) matrix $N\succeq 0$, define the \emph{additive-noise mechanism} $\Mech_{N,q}$ by
\begin{align*}
    \Mech_{N,q}(\DB) ~\sim~ \multiNormal{\mu(\DB)}{\Sigma(\DB)+N}.
\end{align*}
\end{definition}

Our next step is to connect the privacy of $\Mech_{N,q}$ to the NDIS toolbox. Concretely, \Cref{thm:ndis-two-genchi2-cdf} gives a representation of Gaussian indistinguishability, and \Cref{prop:mr-reduction} is an immediate specialization of that result to the two Gaussian distributions induced by a neighboring pair under additive Gaussian noise. \Cref{prop:mr-reduction} states that, for any ordered neighbors $\DB\sim \DB'$, the $\delta$-curve of $\Mech_{N,q}$ is fully captured by a pair of invariants $(m_N(\DB,\DB'),R_N(\DB,\DB'))$ (cf.~Equation\ref{def: m,R tuple}) through the scalar functional $\delta_\varepsilon(m,R)$ (cf.~\Cref{def: nids function}).

\begin{proposition}
\label{prop:mr-reduction}
Let $q(\DB)=(\mu(\DB),\Sigma(\DB))$ be a Gaussian-output map and $N\succeq 0$, and let $\Mech_{N,q}$ (cf.~\Cref{def:additive-noise-family}) be the additive-noise mechanism. For any ordered neighboring pair $\DB \sim \DB'$, define
\begin{align}
\label{def: m,R tuple}
    m_{N}(\DB,\DB') \defin{}& (\Sigma(\DB')+N)^{-1/2}\bigl(\mu(\DB)-\mu(\DB')\bigr), \nonumber\\
    R_{N}(\DB,\DB') \defin{}& (\Sigma(\DB')+N)^{-1/2}\bigl(\Sigma(\DB)+N\bigr)(\Sigma(\DB')+N)^{-1/2}.
\end{align}
Define the NDIS function
\begin{align}
\label{def: nids function}
    \delta_\varepsilon(m,R) \defin \delta_{\rX,\rY}(\varepsilon)
    \quad\text{where}\quad \rX\sim \multiNormal{m}{R}, \rY\sim \multiNormal{0}{I_d}.
\end{align}
Then, for every such $(\DB,\DB')$,
\begin{align*}
\delta_{\Mech_{N,q}(\DB),\Mech_{N,q}(\DB')}(\varepsilon) =  \delta_\varepsilon\Bigl(m_N(\DB,\DB'),R_N(\DB,\DB')\Bigr).
\end{align*}
\end{proposition}

To turn the pairwise reduction into a DP guarantee, we must control the invariants $(m_N(\DB,\DB'),R_N(\DB,\DB'))$ \emph{uniformly} over all neighboring pairs. Let
\begin{align}
\label{def: m,R tuple set}
    \cS_{q,N}\defin\{(m_N(\DB,\DB'),R_N(\DB,\DB')):\DB\sim \DB'\}
\end{align}
denote the set of feasible invariant pairs induced by $(q,N)$. By \Cref{prop:mr-reduction}, calibrating $N$ so that $\Mech_{N,q}$ is $(\varepsilon,\delta)$-DP is equivalent to enforcing
\begin{align*}
    \sup_{(m,R)\in \cS_{q,N}} \delta_\varepsilon(m,R) \leq \delta.
\end{align*}
which is, however, usually intractable because the feasible set $\cS_{q,N}$ can be highly complex for an arbitrary data-dependent Gaussian map $q$. We address this issue by replacing the exact worst-case condition with a single conservative check obtained from two relaxations: (i) we upper bound the unknown feasible set by an \emph{envelope} $\cS_{q,N}\subseteq \cC(q,N)$ via an NDIS-based sensitivity model (cf.~\Cref{def:ndis-sensitivity-loewner}) which is tractable to work with, and (ii) we upper bound the NDIS functional itself by an analytic expression $\overline{\delta}_{\mathrm{LC}}$ (cf.~\Cref{def:delta-LC}) on that envelope. Combining the two reduces privacy calibration to verifying
\begin{align*}
    \sup_{(m,R)\in \cC(q,N)} \overline{\delta}_{\mathrm{LC}}(\varepsilon;\cdot) \leq \delta,
\end{align*}
which is efficiently computable and directly supports automatic noise selection.

\Cref{def:ndis-sensitivity-loewner} introduces a model for the  sensitivity of general Gaussian-output algorithms, which we call Loewner-comparable NDIS sensitivity, that we use to induce an envelope $\cC(q,N)$ of $\cS_{q,N}$, which we can work with. \Cref{def:ndis-sensitivity-loewner} is tailored to \emph{Gaussian-output} queries in that it constrains exactly the NDIS invariants $(m_N,R_N)$ appearing in the pairwise reduction. The \emph{mean sensitivity} $\twoNorm{m_N(\DB,\DB')}\leq \Delta$ bounds the largest standardized mean difference (i.e., a Mahalanobis distance under the noised covariance). The \emph{Loewner-comparability} condition requires the two noised covariances to be ordered, which is precisely the structural regime in which our analytic NDIS bounds apply. Finally, the \emph{covariance sensitivity} constraints $\infNorm{\boldsymbol{\ell}_N} \leq \rho_\infty$ and $\bigl|\sum_i \ell_{N,i}\bigr|\leq \nu$ bound, respectively, the worst per-direction multiplicative distortion and the aggregate log-determinant change.

We note that the Loewner-comparable NDIS sensitivity is ``tight'' as a modeling abstraction: since $\delta_\varepsilon$ depends on the output distribution only through $(m_N,R_N)$, the sensitivity conditions are stated directly in terms of these invariants. The only conservatism comes from replacing $\cS_{q,N}$ by an explicit envelope $\cC(q,N)$, which is chosen to be as sharp as possible while still being verifiable.

\begin{definition}[Loewner-comparable NDIS sensitivity]
\label{def:ndis-sensitivity-loewner}
Let $q(\DB)=(\mu(\DB),\Sigma(\DB))$ (corresponding to the output distribution of Gaussian-output algorithm) be a map with $\Sigma(\DB)\succ 0$ for all $\DB\in\cD$, and fix a PSD matrix $N\succeq 0$. Let $(m_N(\DB,\DB'), R_N(\DB,\DB'))$ be as defined in~Equation~\ref{def: m,R tuple} for an ordered neighboring pair $\DB\sim \DB'$.

\medskip
\noindent We say that $q$ has \emph{$(\Delta,\rho_\infty,\nu, N)$-Loewner-comparable NDIS sensitivity} if for every ordered neighboring pair $\DB\sim \DB'$:
\begin{enumerate}
    \item \textbf{Mean sensitivity:} $\twoNorm{m_N(\DB,\DB')} \leq \Delta$.

    \item \textbf{Loewner comparability:} $R_N(\DB,\DB')\succeq I$ or $R_N(\DB,\DB')\preceq I$.

    \item \textbf{Covariance sensitivity:}
    Let $\tau_1,\dots,\tau_d$ be the eigenvalues of $R_N(\DB,\DB')$ and define $\boldsymbol{\ell}_N(\DB,\DB')\defin(\log\tau_1,\dots,\log\tau_d)$.
    Then
    \begin{align*}
        \infNorm{\boldsymbol{\ell}_N(\DB,\DB')} \leq \rho_\infty,
        \qquad
        \Bigl|\sum_{i=1}^d \boldsymbol{\ell}_N(\DB,\DB')_i\Bigr| \leq \nu.
    \end{align*}
\end{enumerate}
\end{definition}

\Cref{def:delta-LC} defines an analytic upper bound on $\delta_\varepsilon(m, R)$ over the envelope induced by \Cref{def:ndis-sensitivity-loewner}. Concretely, \Cref{thm:ndis-analytic-form} shows that under Loewner comparability, the indistinguishability between two Gaussians can be upper bounded by splitting the privacy budget $\varepsilon=\varepsilon_1+\varepsilon_2$ and separating the effect of (i) a \emph{mean shift} under a common covariance and (ii) a \emph{covariance shift} under a common mean. Definition~\ref{def:delta-LC} is a direct instantiation of this template in terms of the sensitivity parameters: the term $\overline{\delta}_{\mathrm{mean}}(\varepsilon_1;\Delta)$ is exactly the analytic form of the mean-shift bound (cf.~\Cref{prop:ndis-mean-analytic-form}) when the standardized mean difference is bounded by $\Delta$, while $\overline{\delta}_{\mathrm{cov}}(\varepsilon_2;\rho_\infty,\nu)$ upper bounds the covariance-shift contribution (cf.~\Cref{lem:ndis-cov-analytic-form}) when the log-eigenvalue statistics of the relative covariance satisfy $\|\boldsymbol{\ell}\|_\infty\le \rho_\infty$ and $|\sum_i \ell_i|\le \nu$.

As a result, once map $q$ satisfies Loewner-comparable NDIS sensitivity with parameters $(\Delta,\rho_\infty,\nu)$, the worst-case privacy over the entire envelope $\cC(q,\sigma I_m)$ can be controlled by the single scalar quantity $\overline{\delta}_{\mathrm{LC}}(\varepsilon;\Delta,\rho_\infty,\nu)$, enabling an efficient calibration rule for $\sigma I_m$ in the mechanism of \Cref{alg::ndis wrapper}.

\begin{definition}
\label{def:delta-LC}
Let $\varepsilon\geq 0, \Delta\geq 0$, $\rho_\infty\geq 0$ and $\nu\geq 0$. Define
\begin{align*}
    \overline{\delta}_{\mathrm{LC}}(\varepsilon;\Delta,\rho_\infty,\nu) \defin \inf_{\substack{\varepsilon_1,\varepsilon_2\ge 0\\ \varepsilon_1+\varepsilon_2=\varepsilon}}
    \Bigl[\overline{\delta}_{\mathrm{mean}}(\varepsilon_1;\Delta) + e^{\varepsilon_1}\,\overline{\delta}_{\mathrm{cov}}(\varepsilon_2;\rho_\infty,\nu)\Bigr],
\end{align*}
where $\overline{\delta}_{\mathrm{mean}}$ and $\overline{\delta}_{\mathrm{cov}}$ are defined as follows.

\medskip
\noindent\textbf{Mean-shift upper bound.} Define, for $\Delta\geq 0$,
\begin{align*}
\overline{\delta}_{\mathrm{mean}}(\varepsilon;\Delta) \defin
\begin{cases}
    \Phi\left(\frac{\Delta}{2}-\frac{\varepsilon}{\Delta}\right)- e^{\varepsilon}\Phi\left(-\frac{\Delta}{2}-\frac{\varepsilon}{\Delta}\right), & \Delta>0,\\[3pt]
    0,& \Delta=0.
\end{cases}
\end{align*}

\noindent\textbf{Covariance-shift upper bound.} Define
\begin{align*}
    c(\rho_\infty) &\defin e^{\rho_\infty}-1, \qquad
    k \defin \frac{1}{2\,c(\rho_\infty)}\in(0,\infty], \tag{$k=\infty$ when $\rho_\infty=0$},
\end{align*}
Next define the envelope functionals
\begin{align*}
    A_\star(s;\rho_\infty,\nu)
    &\defin
    \sup\Biggl\{
        \sum_{i=1}^{d}\phi_s(\ell_i)
        :
        0\le \ell_i\le \rho_\infty \ \text{for all } i,\
        \sum_{i=1}^{d}\ell_i \le \nu
    \Biggr\},\\
    B_\star(u;\rho_\infty,\nu)
    &\defin
    \sup\Biggl\{
        \sum_{i=1}^{d}\psi_u(\ell_i)
        :
        0\le \ell_i\le \rho_\infty \ \text{for all } i,\
        \sum_{i=1}^{d}\ell_i \le \nu
    \Biggr\}.
\end{align*}
Then define
{\small
\begin{align*}
\overline{\delta}_{\mathrm{cov}}(\varepsilon;\rho_\infty,\nu)
\defin
\begin{cases}
    0, & \varepsilon\ge \nu/2,\\[4pt]
    \displaystyle
    \inf_{\substack{s\ge 0\\ u\in(0,k(\rho_\infty))}}
    \Bigg[
        \rExp{-2\varepsilon s + A_\star(s;\rho_\infty,\nu)}
        -
        e^{\varepsilon}
        \Bigl(
            1-\rExp{2\varepsilon u + B_\star(u;\rho_\infty,\nu)}
        \Bigr)
    \Bigg],
    & \varepsilon<\nu/2.
\end{cases}
\end{align*}
}
\end{definition}

\begin{small}
\begin{figure}[htp]
\captionsetup{skip=2pt}
\centering
\setlength{\fboxrule}{0pt}
\fbox{
\setlength{\fboxrule}{0.5pt}
\framebox{
\begin{minipage}[t]{0.94\linewidth}

\begin{footnotesize}
\textbf{Input:} A database $\DB\in\Real^{n\times d}$; privacy parameters $\varepsilon\in\nonnegReal$, $\delta\in[0,1)$.

\textbf{Output:} A vector $\widetilde M \in \Real^m$

\begin{enumerate}
    \item Use binary search to find $\sigma^\star$ such that
    \begin{align*}
        \overline{\delta}_{\mathrm{LC}}\bigl(\varepsilon;\Delta(\sigma^\star),\rho_\infty(\sigma^\star),\nu(\sigma^\star)\bigr) \leq \delta.
    \end{align*}

    \item Return $\Mech_{\sigma I_m,q}(\DB)$
\end{enumerate}
\end{footnotesize}

\end{minipage}
}}
\caption{An NDIS-Calibrated Gaussian Mechanism for Loewner-comparable Gaussian-output algorithms $\MechNDIS{q}$}

\label{alg::ndis wrapper}
\end{figure}
\end{small}

We are now ready to state the NDIS-Calibrated Gaussian Mechanism (cf.~\Cref{alg::ndis wrapper}). The mechanism takes as input the target privacy parameters $(\varepsilon,\delta)$ and adds \emph{isotropic} Gaussian noise $N=\sigma I_m$ to the the underlying Gaussian-output algorithm. The key step is to choose the smallest noise level $\sigma^\star$ that makes the worst-case privacy bound fall below $\delta$. Concretely, for each candidate $\sigma$, the query-dependent sensitivity bounds produce parameters $(\Delta(\sigma),\rho_\infty(\sigma),\nu(\sigma))$, and \Cref{def:delta-LC} yields the upper bound
\begin{align*}
    \overline{\delta}_{\mathrm{LC}}\bigl(\varepsilon;\Delta(\sigma),\rho_\infty(\sigma),\nu(\sigma)\bigr)
\end{align*}
on the worst-case indistinguishability over all neighboring pairs. Since increasing $\sigma$ only increases regularization (thereby shrinking mean shifts and bringing covariances closer), this bound is monotone in $\sigma$, which allows us to locate $\sigma^\star$ efficiently by binary search. Finally, the mechanism outputs $\Mech_{\sigma^\star I_m,q}(\DB)$, i.e., a Gaussian draw with the same mean as the original algorithm and covariance inflated by $\sigma^\star I_m$.

\begin{theorem}[Proof is in~\Cref{proof for thm:NDIS-mech-DP}]
\label{thm:NDIS-mech-DP} Let $\varepsilon \geq 0, \delta \in (0,1)$.
If $q$ has $(\Delta(\sigma),\rho_\infty(\sigma),\nu(\sigma),\sigma I_d)$-Loewner-comparable NDIS sensitivity then the mechanism $\MechNDIS{q}$ in \Cref{alg::ndis wrapper} is $(\varepsilon,\delta)$-DP under the standard add/remove neighboring notion.
\end{theorem}

\subsection{Application}

Below, we give two examples of commonly-used Gaussian-output algorithms and their NDIS sensitivities (Def.~\ref{def:ndis-sensitivity-loewner}).

\begin{definition}[Bayesian Logistic Regression surrogate as a Gaussian-output algorithm]
\label{def:blr-gaussian-output}
Let $\lambda > 0$ be a regularization parameter and $\cD = (\Real^d \times \{-1,1\})^n$ be the dataset space. The \emph{B\"ohning-style Bayesian Logistic Regression surrogate}, denoted $\Mech_{q_{\mathrm{BLR}}}$, is a Gaussian-output algorithm (cf.~\Cref{def:gaussian-output-alg}) determined by the map $q_{\mathrm{BLR}}$.
\noindent Given a dataset $\DB = \{(x_i, y_i)\}_{i=1}^n \in \cD$, define the regularized negative log-likelihood
\begin{align*}
    \mathcal{L}(\theta; \DB) \defin \sum_{i=1}^n \ln\!\bigl(1 + \exp(-y_i x_i^\top \theta)\bigr) + \frac{\lambda}{2}\|\theta\|_2^2.
\end{align*}
The map $q_{\text{BLR}}: \cD \to \Real^d \times \mathbb S_{++}^d$ is defined as $q_{\text{BLR}}(\DB) = (\hat{\theta}, \Sigma)$, where:
\begin{align*}
    \hat{\theta} ={} \argmin_{\theta \in \Real^d} \mathcal{L}(\theta; \DB), \qquad \Sigma ={} \left(\lambda I_d + \frac{1}{4}\sum_{i=1}^n x_i x_i^\top\right)^{-1}.
\end{align*}
\end{definition}

\begin{proposition}[Proof is in~\Cref{proof for prop:blr-ndis-sens}]
\label{prop:blr-ndis-sens}
Let $l \in \posReal$ be the row norm upper bound, i.e., for every $(x,y) \in \Real^d \times \{-1,1\}$, $\twoNorm{x} \leq l$.
Then $q_{\textnormal{BLR}}$ has $(\Delta(\sigma), \rho_\infty(\sigma), \nu(\sigma), \sigma I_d)$-Loewner-comparable NDIS sensitivity, where:
\begin{align*}
    \Delta(\sigma) = \frac{l/\lambda}{\sqrt{\sigma+\frac{1}{\lambda+\frac{n l^2}{4}}}}, \qquad
    \rho_\infty(\sigma) = \log\Bigl(\frac{\sigma+\frac{1}{\lambda}}{\sigma+\frac{1}{\lambda+\frac{n l^2}{4}}}\Bigr)  \qquad
    \nu(\sigma) = d \rho_\infty(\sigma).
\end{align*}
\end{proposition}

\begin{definition}[Gaussian Process Regression~\cite{RasmussenW06} as a Gaussian-output algorithm]
\label{def:gpr-gaussian-output}
Let $\cX\subseteq \Real^d$ be the input domain. Fix a positive definite kernel $k:\cX\times \cX\to\Real$ and a noise variance $\sigma_n^2>0$. Fix a public query point $x_\star\in\cX$. The \emph{Gaussian Process Regression (GPR)} mechanism, denoted $\Mech_{q_{\textnormal{GPR}}}$, is a Gaussian-output algorithm (cf.~\Cref{def:gaussian-output-alg}) determined by the map $q_{\textnormal{GPR}}$.

\medskip
\noindent Given a dataset $\DB=\{(x_i,y_i)\}_{i=1}^n\in(\cX\times\Real)^n$, define the Gram matrix $K\in\Real^{n\times n}$ and the kernel vector $k_\star\in\Real^n$ by
\begin{align*}
    K_{ij}\defin k(x_i,x_j), \qquad (k_\star)_i \defin k(x_\star,x_i), \qquad k_{\star\star} \defin k(x_\star,x_\star).
\end{align*}
Let $A \defin K+\sigma_n^2 I_n$. The map $q_{\textnormal{GPR}}:(\cX\times\Real)^n\to \Real\times \Real$ is defined as $q_{\textnormal{GPR}}(\DB)=(\mu_\star,\Sigma_\star)$, where
\begin{align*}
    \mu_\star    \defin{}& k_\star^\top A^{-1} y,\\
    \Sigma_\star \defin{}& k_{\star\star} - k_\star^\top A^{-1} k_\star \in \Real.
\end{align*}
\end{definition}

\begin{proposition}[Proof is in~\Cref{proof for prop:gpr-ndis-sens}]
\label{prop:gpr-ndis-sens}
Let $B \in \posReal$ be a public upper bound on the label magnitude, i.e., for every $(x,y)\in \cX\times\Real$, we have $\abs{y} \leq B$. Let $l \in \posReal$ be a public upper bound on the kernel diagonal, i.e., for every $x \in \cX$, we have $k(x,x)\leq l^2$. Then $q_{\textnormal{GPR}}$ has $(\Delta(\sigma),\rho_\infty(\sigma),\nu(\sigma),\sigma)$-Loewner-comparable NDIS sensitivity, where
\begin{align*}
    \Delta(\sigma) = \frac{l B \sqrt{n}}{\sigma_n\,\sqrt{\sigma}}, \qquad
    \rho_\infty(\sigma) = \log\Bigl(\frac{\sigma+l^2}{\sigma}\Bigr), \qquad
    \nu(\sigma) = \rho_\infty(\sigma).
\end{align*}
\end{proposition}

\section{Whitebox Auditing Tool Via NDIS}\label{sec:whitebox}
\label{sec: auditing}
In this section, we use NDIS to construct a tool for {\em white-box auditing} of mechanisms with Gaussian outputs. This tool does not require any understanding of NDIS and its tools---as long as we know the mean/covariance of the mechanism's output on the tested inputs, and have access to a CDF oracle (e.g., Davies method)---which makes it a simple addition to any white-box auditor's toolbox.

Our new tool empirically but accurately estimates the $\delta_{X,Y}(\varepsilon)$ for any normal distributions $X,Y$ and any $\epsilon>0$. In other words, for any mechanism $M$ we can estimate a {\em lower bound} on $\delta$ such that $M$ is $(\varepsilon, \delta)$-DP, by picking $X = M(D)$ and $Y = M(D')$ for any neighbouring databases $D, D'$, then estimating $\delta_{X,Y}(\varepsilon)$. If the claimed $\delta$ is lower than the estimated lower-bound (taking into account some estimation error), then this is a counterexample/refutation to the claim that $M$ is $(\epsilon, \delta)$-DP. Importantly, our estimate $\delta_{M(D),M(D')}$ is tight, and given a CDF oracle, is guaranteed/non-probabilistic.

\subsection{Estimating $\delta_{X,Y}(\varepsilon)$ through connection between NDIS and generalized-$\chi^2$ distribution}

We define a CDF oracle (with some error $\eta_0$) as the following.
\begin{definition}[CDF oracle for the generalized-$\chi^2$ distribution]\label{def:genchi2-cdf-oracle}
    A CDF oracle for the generalized-$\chi^2$ distribution with error $\eta_0$ is one which, on any  generalized-$\chi^2$ distribution $\rX$, outputs $\hat{p}$ such that
    $$
    |\hat{p} - \Pr(X \leq 0)| \leq \eta_0
    $$
\end{definition}

Given a CDF oracle described above, we can construct an IS estimator, which can be purposed into a white-box auditor by treating the estimator's output as a lower-bound on the tested mechanism's DP.

\begin{theorem}[Proof is in~\Cref{proof for lem:ndis-upper-oracle}]
    \label{lem:ndis-upper-oracle}
    Suppose there exists a CDF oracle for the generalized-$\chi^2$ distribution with error $\eta_0 = \frac{\eta}{2(1+e^\varepsilon)}.$ Then we can construct a deterministic algorithm for estimating IS $\delta_{\rX,\rY}(\epsilon)$ for any $\rX,\rY, \epsilon$, which outputs $\hat{\delta}$ such that:
$$
|\hat{\delta} - \delta_{\rX,\rY}(\epsilon)| \leq \eta
$$
\end{theorem}

\bibliographystyle{splncs04}
\bibliography{abbrev3, crypto, ref, additional}

\appendix
\section{Missing Proofs for NDIS toolbox}
\subsection{Proof for~\Cref{lemma: ndis} }
\begin{proof}[Proof of \Cref{lemma: ndis}]
\label{proof for lemma: ndis}

Fix a set $\cS\subseteq \Real^d$. Let $p_{\rX}$ and $p_{\rY}$ denote the densities of $\rX$ and $\rY$. We start from the density formulas.
{\small
    \begin{align*}
        \pr{\rX\in \cS} ={}& \int_{\cS} \frac{1}{(2\pi)^{d/2}\det(\Sigma_1)^{1/2}} \rExp{-\frac12 (x-\mu_1)^T\Sigma_1^{-1}(x-\mu_1)} \mathrm{d}x,\\
        \pr{\rY\in \cS} ={}& \int_{\cS} \frac{1}{(2\pi)^{d/2}\det(\Sigma_2)^{1/2}} \rExp{-\frac12 (x-\mu_2)^T\Sigma_2^{-1}(x-\mu_2)} \mathrm{d}x.
    \end{align*}
}
Plugging the above into $\pr{\rX\in \cS}-e^\varepsilon\pr{\rY\in \cS}$, we have that
\begin{align*}
    & \pr{\rX\in \cS}-e^\varepsilon\pr{\rY\in \cS}\\
    ={}& \int_{\cS}\Bigg[ \frac{1}{(2\pi)^{d/2}\det(\Sigma_1)^{1/2}} \rExp{-\frac12 (x-\mu_1)^T\Sigma_1^{-1}(x-\mu_1)}\\
    &\hspace{1.5cm}- \frac{e^\varepsilon}{(2\pi)^{d/2}\det(\Sigma_2)^{1/2}} \rExp{-\frac12 (x-\mu_2)^T\Sigma_2^{-1}(x-\mu_2)} \Bigg]\mathrm{d}x.
\end{align*}
Performing a change of variable,
\begin{align*}
    x=\mu_1+\Sigma_1^{1/2}z, \qquad z=\Sigma_1^{-1/2}(x-\mu_1), \qquad \mathrm{d}x=\det(\Sigma_1)^{1/2} \mathrm{d}z.
\end{align*}
and let
\begin{align*}
    \cS_z \defin \Sigma_1^{-1/2}(\cS-\mu_1) = \bigl\{z\in\Real^d: \mu_1+\Sigma_1^{1/2}z\in \cS\bigr\}.
\end{align*}
gives
\begin{align*}
    & \pr{\rX\in \cS}-e^\varepsilon\pr{\rY\in \cS}\\
    ={}& \int_{\cS_z}\Bigg[\frac{1}{(2\pi)^{d/2}\det(\Sigma_1)^{1/2}} \rExp{-\frac12 z^Tz} -\frac{e^\varepsilon}{(2\pi)^{d/2}\det(\Sigma_2)^{1/2}}\\
    &\hspace{1.5cm} \cdot \rExp{-\frac12(\Sigma_1^{1/2}z+\Delta\mu)^T\Sigma_2^{-1}(\Sigma_1^{1/2}z+\Delta\mu)} \Bigg]\det(\Sigma_1)^{1/2} \mathrm{d}z\\
    ={}& \int_{\cS_z}\frac{\exp\!\left(-\frac12 z^Tz\right)}{(2\pi)^{d/2}} \Bigg[1-\frac{e^\varepsilon\det(\Sigma_1)^{1/2}}{\det(\Sigma_2)^{1/2}}\\
    &\hspace{1.5cm} \cdot \rExp{-\frac12(\Sigma_1^{1/2}z+\Delta\mu)^T\Sigma_2^{-1}(\Sigma_1^{1/2}z+\Delta\mu)+\frac12 z^Tz}\Bigg] \mathrm{d}z.
\end{align*}

Define $\psi_\varepsilon(z)$ as the log of the exponential factor inside the brackets:
\begin{align*}
    \psi_\varepsilon(z) \defin& \varepsilon+\frac12\log\frac{\det\Sigma_1}{\det\Sigma_2} -\frac12(\Sigma_1^{1/2}z+\Delta\mu)^T\Sigma_2^{-1}(\Sigma_1^{1/2}z+\Delta\mu) +\frac12 z^Tz\\
    ={}& c(\varepsilon) - \Delta\mu^T\Sigma_2^{-1}\Sigma_1^{1/2}z +\frac12 z^T\Bigl(I_d-\Sigma_1^{1/2}\Sigma_2^{-1}\Sigma_1^{1/2}\Bigr)z\\
    ={}& c(\varepsilon) - \Delta\mu^T\Sigma_2^{-1}\Sigma_1^{1/2}z + \frac12 z^T(I_d-M)z.
\end{align*}
Then the bracket becomes $1-\exp(\psi_\varepsilon(z))$, and we obtain
\begin{align*}
\pr{\rX\in \cS}-e^\varepsilon\pr{\rY\in \cS} &= \int_{\cS_z} \frac{\rExp{-\frac12 z^Tz}}{(2\pi)^{d/2}} \left(1-\rExp{\psi_\varepsilon(z)}\right) \mathrm{d}z.
\end{align*}

Now diagonalize $M=U\diag(\tau_1,\dots,\tau_d)U^T$ and let $v = U^Tz$. Since $U$ is orthogonal, $\mathrm{d}z=\mathrm{d}v$ and $z^Tz=v^Tv$, and moreover
\begin{align*}
    z^T(I_d-M)z = v^T\Bigl(I_d-\diag(\tau_1,\dots,\tau_d)\Bigr)v = \sum_{i=1}^d a_i v_i^2.
\end{align*}
Also,
\begin{align*}
   -\Delta\mu^T\Sigma_2^{-1}\Sigma_1^{1/2}z = -\Delta\mu^T\Sigma_2^{-1}\Sigma_1^{1/2}Uv = \bigl(-U^T\Sigma_1^{1/2}\Sigma_2^{-1}\Delta\mu\bigr)^T v = b^T v.
\end{align*}
Therefore $\psi_\varepsilon(z)=g_\varepsilon(v)$, where
\begin{align*}
    g_\varepsilon(v)=c(\varepsilon)+b^Tv+\frac12\sum_{i=1}^d a_i v_i^2.
\end{align*}
Let $\cT\defin U^T\cS_z$. Then
\begin{align*}
\pr{\rX\in \cS}-e^\varepsilon\pr{\rY\in \cS} = \int_{\cT}\frac{\rExp{-\frac12 v^Tv}}{(2\pi)^{d/2}} \left(1-\rExp{g_\varepsilon(v)}\right)\,dv.
\end{align*}

To maximize this expression over $\cS$ (equivalently, over $\cT$), observe that the Gaussian density factor is nonnegative, hence the integral is maximized by taking $\cT$ to include exactly the points where the integrand $\left(1-\rExp{g_\varepsilon(v)}\right)$ is positive, i.e., where $g_\varepsilon(v)\le 0$. Thus the optimal value is
\begin{align*}
\delta_{\rX,\rY}(\varepsilon) ={}& \int_{\Real^d}\frac{\rExp{-\frac12 v^Tv}}{(2\pi)^{d/2}} \Bigl(1-\rExp{g_\varepsilon(v)}\Bigr)_+ \mathrm{d}v\\
={}& \Ex{\Bigl(1-\rExp{g_\varepsilon(Z)}\Bigr)_+},
\end{align*}
where $Z\sim \multiNormal{0}{I_d}$. This concludes the proof.
\end{proof}

\subsection{Proof for~\Cref{thm:ndis-two-genchi2-cdf}}
\begin{proof}[Proof of~\Cref{thm:ndis-two-genchi2-cdf}]
\label{proof for thm:ndis-two-genchi2-cdf}
Let $a\in\Real^d$, $b\in\Real^d$, and $c(\varepsilon)\in\Real$ be as in \Cref{lemma: ndis}, and recall that for $\rZ\sim \multiNormal{0}{I_d}$,
\begin{align*}
    \delta_{\rX,\rY}(\varepsilon) = \Ex{\bigl(1-\rExp{g_\varepsilon(\rZ)}\bigr)_+}, \quad g_\varepsilon(z) = c(\varepsilon)+b^T z+\frac12\sum_{i=1}^d a_i z_i^2 .
\end{align*}
Then, we have that
\begin{align*}
    & \delta_{\rX,\rY}(\varepsilon)\\
    ={}& \Ex{\bigl(1-\rExp{g_\varepsilon(\rZ)}\bigr)_+}\\
    ={}& \Ex{\bigl(1-\rExp{g_\varepsilon(\rZ)}\bigr) \mathbf 1_{\{g_\varepsilon(Z)\leq 0\}}}\\
    ={}& \pr{g_\varepsilon(Z)\leq 0}-   \Ex{e^{g_\varepsilon(Z)}\mathbf 1_{\{g_\varepsilon(Z)\leq 0\}}}.
\end{align*}
Our first step is to rewrite $\delta_{\rX,\rY}(\varepsilon)$ in terms of two quadratic-form CDF values. Denote $D = \diag(\tau_1,\dots,\tau_d)$, where $a_i = 1-\tau_i$, and define
\begin{align*}
    m \defin{} D^{-1}b, \qquad \widetilde Z \sim \multiNormal{m}{D^{-1}}.
\end{align*}
Let $\varphi(z) = (2\pi)^{-d/2}\exp(-\twoNorm{z}^2/2)$ be the standard Gaussian density on $\Real^d$. Express $g_\varepsilon(z)$ in quadratic form:
\begin{align*}
    g_\varepsilon(z) = c(\varepsilon) + b^\top z + \frac12 z^\top (I-D) z .
\end{align*}
Indeed, performing change of variable gives
\begin{align*}
    &\Ex{e^{g_\varepsilon(Z)}\mathbf 1_{\{g_\varepsilon(Z)\leq 0\}}}\\
    ={}&\int_{g_\varepsilon(z)\leq 0} e^{g_\varepsilon(z)} \varphi(z) \mathrm{d}z\\
    ={}& \int_{g_\varepsilon(z)\le 0} (2\pi)^{-d/2} \rExp{c(\varepsilon)+b^\top z-\tfrac12 z^\top D z}\mathrm{d}z\\
    ={}& \int_{g_\varepsilon(z)\le 0} (2\pi)^{-d/2} \rExp{c(\varepsilon)+\tfrac12 m^\top Dm} \rExp{-\tfrac12 (z-m)^\top D(z-m)}\mathrm{d}z\\
    ={}& \int_{g_\varepsilon(z)\le 0} e^\varepsilon \widetilde\varphi(z)\mathrm{d}z\\
    ={}& e^\varepsilon \pr{g_\varepsilon(\widetilde Z)\leq 0},
\end{align*}
where $\widetilde\varphi(z)$ is the density of $\widetilde Z\sim\multiNormal{m}{D^{-1}}$.
Plugging this back yields
\begin{align*}
\delta_{\rX,\rY}(\varepsilon) = \pr{g_\varepsilon(Z)\leq 0} - e^{\varepsilon}\pr{g_\varepsilon(\widetilde Z)\leq 0},
\end{align*}
with $\widetilde Z\sim\multiNormal{m}{D^{-1}}$ as defined above.

The important observation is that both $g_\varepsilon(Z)$ and $g_\varepsilon(\widetilde Z)$ are (possibly indefinite) quadratic polynomials in independent Gaussian coordinates, and hence are generalized-$\chi^2$ random variables. To fit it in the characterization given in~\Cref{def:genchi2}, we rewrite both $g_\varepsilon(Z)$ and $g_\varepsilon(\widetilde Z)$ as a weighted sum of independent (noncentral) $\chi^2$ variables with an Gaussian linear term if needed.

\paragraph{Term 1: $g_\varepsilon(\rZ)$.} Recall that
\begin{align*}
    g_\varepsilon(\rZ)=c(\varepsilon)+b^\top Z+\frac12\sum_{i=1}^d a_i \rZ_i^2, \quad \rZ\sim\multiNormal{0}{I_d}.
\end{align*}
Define the set $\cI = \{i\in[d]: a_i\neq 0\},$ and for each $i \in I$, define
\begin{align*}
    \lambda_i = \frac{a_i}{2},\quad \mu_i = \frac{b_i}{a_i}, \quad \rX_i \sim \chi'^2_{1}(\mu_i^2).
\end{align*}
Then we can rewrite $g_\varepsilon(\rZ)$ as
\begin{align*}
    g_\varepsilon(\rZ) = c(\varepsilon) - \sum_{i\in I}\frac{b_i^2}{2a_i} + \sum_{i\in I}\lambda_i \rX_i + \sum_{i\notin I} b_i \rZ_i .
\end{align*}
In particular, if $\abs{\cI} \neq d,$ then $\sum_{i\notin I} b_i \rZ_i$ is Gaussian. Therefore, $g_\varepsilon(\rZ)$ is a generalized-$\chi^2$ random variable (cf.~\Cref{def:genchi2}) with parameters
\begin{align*}
    r=|\cI|,\quad (\lambda_i,\nu_i,\rho_i) = \Big(\frac{a_i}{2},\,1,\,\big(\tfrac{b_i}{a_i}\big)^2\Big)\ \text{for } i\in \cI,
    \quad \sigma = \Big(\sum_{i \in [d] \setminus \cI} b_i^2\Big)^{1/2},
\end{align*}
and threshold
\begin{align*}
    c \defin{} -\Big(c(\varepsilon)-\sum_{i\in I}\frac{b_i^2}{2a_i}\Big),
\end{align*}
since $\pr{g_\varepsilon(Z)\leq 0}= \pr{\sum_{i\in I}\lambda_i X_i + \sigma G \leq c}$.

\paragraph{Term 2: $g_\varepsilon(\widetilde \rZ)$.} For $i\in[d]$, write
\begin{align*}
    \widetilde \rZ_i = m_i + \frac{1}{\sqrt{\tau_i}} G_i,
\end{align*}
where $G_1,\dots,G_d \iid \multiNormal{0}{1}$ are independent. Recall that
\begin{align*}
    g_\varepsilon(\widetilde \rZ) = c(\varepsilon)+b^\top \widetilde \rZ + \frac12\sum_{i=1}^d a_i \widetilde \rZ_i^2 .
\end{align*}
Recall that $\cI = \{i\in[d]: a_i\neq 0\}$. For each $i\in\cI$, complete the square in $\widetilde \rZ_i$:
\begin{align*}
    b_i \widetilde \rZ_i + \frac{a_i}{2}\widetilde \rZ_i^2 = \frac{a_i}{2}\Bigl(\widetilde \rZ_i + \frac{b_i}{a_i}\Bigr)^2 - \frac{b_i^2}{2a_i}.
\end{align*}
Moreover,
\begin{align*}
    \widetilde \rZ_i + \frac{b_i}{a_i} = \frac{1}{\sqrt{\tau_i}}\Bigl(G_i + \widetilde\mu_i\Bigr), \quad \widetilde\mu_i = \sqrt{\tau_i}\Bigl(m_i+\frac{b_i}{a_i}\Bigr)
    = \frac{b_i}{a_i\sqrt{\tau_i}},
\end{align*}
where the last equality uses $m_i=b_i/\tau_i$ and $a_i+\tau_i=1$.
\begin{align*}
    \widetilde\lambda_i = \frac{a_i}{2\tau_i}, \qquad \widetilde \rX_i  \sim \chi'^2_{1}(\widetilde\mu_i^2)
    \quad\text{for } i\in\cI.
\end{align*}
Then we can rewrite
\begin{align*}
    g_\varepsilon(\widetilde \rZ) ={}& c(\varepsilon) - \sum_{i\in\cI}\frac{b_i^2}{2a_i} + \sum_{i\in\cI}\widetilde\lambda_i\,\widetilde X_i
                                        + \sum_{i\in[d]\setminus\cI} b_i \widetilde \rZ_i.
\end{align*}
In particular, for $i\notin\cI$ we have $a_i=0$ and hence $\tau_i=1$, so $\widetilde \rZ_i = m_i + G_i = b_i + G_i$, and thus
\begin{align*}
    \sum_{i\in[d]\setminus\cI} b_i \widetilde \rZ_i = \sum_{i\in[d]\setminus\cI} b_i^2 + \sum_{i\in[d]\setminus\cI} b_i G_i
    \overset{d}{=} \sum_{i\in[d]\setminus\cI} b_i^2 + \sigma \widetilde G,
\end{align*}
where $\sigma = \big(\sum_{i\in[d]\setminus\cI} b_i^2\big)^{1/2}$ and $\widetilde G\sim\multiNormal{0}{1}$.
Therefore, $g_\varepsilon(\widetilde \rZ)$ is a generalized-$\chi^2$ random variable (cf.~\Cref{def:genchi2}) with parameters
\begin{align*}
    r=|\cI|,\quad
    (\lambda_i,\nu_i,\rho_i)=\Big(\frac{a_i}{2\tau_i},\,1,\,\big(\tfrac{b_i}{a_i\sqrt{\tau_i}}\big)^2\Big)\ \text{for } i\in\cI,
    \quad \sigma=\Big(\sum_{i\in[d]\setminus\cI} b_i^2\Big)^{1/2}.
\end{align*}
Let $\widetilde c(\varepsilon)\defin c(\varepsilon)-\sum_{i\in\cI}\frac{b_i^2}{2a_i}+\sum_{i\in[d]\setminus\cI} b_i^2,$ and threshold $c \defin{} -\widetilde c(\varepsilon)$ since $\pr{g_\varepsilon(\rZ)\leq 0}= \pr{\sum_{i\in I}\widetilde \lambda_i \rX_i + \sigma \widetilde G \leq c}$.

\end{proof}

\subsection{Proof for \Cref{thm:ndis-analytic-form}}

\medskip
\noindent\textbf{Proof for~\Cref{prop:ndis-mean-analytic-form}}
\begin{proof}[Proof of \Cref{prop:ndis-mean-analytic-form}]
\label{proof for prop:ndis-mean-analytic-form}
Let $\Sigma \defin \Sigma_1=\Sigma_2\succ 0$ and $\Delta\mu\defin \mu_1-\mu_2$. If $\Delta\mu=0$ (equivalently $\tau=0$), then $\rX$ and $\rY$ are identical and hence $\delta_{\rX,\rY}(\varepsilon)=0$ for all $\varepsilon\ge 0$. Assume henceforth $\tau>0$.

We specialize \Cref{lemma: ndis} and \Cref{thm:ndis-two-genchi2-cdf} to the equal-covariance case. Since
\begin{align*}
    M=\Sigma^{1/2}\Sigma^{-1}\Sigma^{1/2}=I_d,
\end{align*}
we may take $U=I_d$ and $\tau_i=1$ for all $i\in[d]$, hence $a_i=1-\tau_i=0$. Moreover,
\begin{align*}
    b=-U^\top \Sigma^{1/2}\Sigma^{-1}\Delta\mu = -\Sigma^{-1/2}\Delta\mu, \qquad
    c(\varepsilon)=\varepsilon-\frac12\Delta\mu^\top \Sigma^{-1}\Delta\mu =\varepsilon-\frac{\tau^2}{2}.
\end{align*}
Therefore, for $Z\sim\multiNormal{0}{I_d}$,
\begin{align*}
    g_\varepsilon(Z) = c(\varepsilon)+b^\top Z = \varepsilon-\frac{\tau^2}{2}+b^\top Z.
\end{align*}
Since $\twoNorm{b}^2=\Delta\mu^\top \Sigma^{-1}\Delta\mu=\tau^2$, let the scalar $W =  \frac{b^\top Z}{\tau}\sim \multiNormal{0}{1},$ and thus
\begin{align*}
    \pr{g_\varepsilon(Z)\le 0} =\pr{\varepsilon-\frac{\tau^2}{2}+\tau W\le 0}
    =\pr{W\le -\frac{\varepsilon}{\tau}+\frac{\tau}{2}}
    =\Phi\left(-\frac{\varepsilon}{\tau}+\frac{\tau}{2}\right).
\end{align*}

Next, in~\Cref{thm:ndis-two-genchi2-cdf} we have $D=I_d$ and $m=D^{-1}b=b$, so $\widetilde Z\sim\multiNormal{b}{I_d}$; write $\widetilde Z=b+Z'$ with $Z'\sim\multiNormal{0}{I_d}$ independent. Then
\begin{align*}
    g_\varepsilon(\widetilde Z) = \varepsilon-\frac{\tau^2}{2}+b^\top(b+Z') =\varepsilon+\frac{\tau^2}{2}+b^\top Z'.
\end{align*}
Let $W' = \frac{b^\top Z'}{\tau}\sim\multiNormal{0}{1}$. Hence
\begin{align*}
    \pr{g_\varepsilon(\widetilde Z)\leq 0} ={}& \pr{\varepsilon+\frac{\tau^2}{2}+\tau W'\leq 0}
    =\pr{W'\le -\frac{\varepsilon}{\tau}-\frac{\tau}{2}} =\Phi\left(-\frac{\varepsilon}{\tau}-\frac{\tau}{2}\right).
\end{align*}

Finally, applying \Cref{thm:ndis-two-genchi2-cdf} gives
\begin{align*}
    \delta_{\rX,\rY}(\varepsilon) ={}& \pr{g_\varepsilon(Z)\leq 0}-e^\varepsilon\pr{g_\varepsilon(\widetilde Z)\leq 0}\\
    ={}&\Phi\left(-\frac{\varepsilon}{\tau}+\frac{\tau}{2}\right) -e^\varepsilon\Phi\left(-\frac{\varepsilon}{\tau}-\frac{\tau}{2}\right),
\end{align*}
as claimed.
\end{proof}

\medskip
\noindent\textbf{Proof for~\Cref{lem:ndis-cov-analytic-form}}

\begin{proof}[Proof of~\Cref{lem:ndis-cov-analytic-form}]
\label{proof for lem:ndis-cov-analytic-form}
In this lemma, we compute the indistinguishablity $\delta_{\rX, \rY}$, where the leakage is only from the covariance (i.e., $\mu_1=\mu_2$, hence $\Delta\mu=0$). In this case, $b=0$ in~\Cref{lemma: ndis}, and therefore
\begin{align*}
    c(\varepsilon) = \varepsilon+\frac12\log\frac{\det\Sigma_1}{\det\Sigma_2} = -\frac12 f(\varepsilon), \qquad
    g_\varepsilon(Z)=c(\varepsilon)+\frac12\sum_{i=1}^d (1-\tau_i)\rZ_i^2.
\end{align*}
Moreover, by~\Cref{thm:ndis-two-genchi2-cdf} (with $b=0$ and $m=0$),
\begin{align}
    \label{eq:delta-two-cdf-cov}
    \delta_{\rX,\rY}(\varepsilon) = \pr{g_\varepsilon(Z)\leq 0} - e^\varepsilon \pr{g_\varepsilon(\widetilde Z)\leq 0},
\end{align}
where $Z\sim\multiNormal{0}{I_d}$ and $\widetilde Z\sim \multiNormal{0}{D^{-1}}$.

By the condition, $\Sigma_2\succeq \Sigma_1$, we have that for all $i \in [d]$, $\tau_i\in(0,1]$. We then define
\begin{align*}
    w_i\defin 1-\tau_i\geq 0 \qquad\text{and}\qquad w_i' \defin \frac{1-\tau_i}{\tau_i}\geq 0.
\end{align*}
Since $g_\varepsilon(Z)\leq 0$ is equivalent to $\sum_i w_i Z_i^2 \leq f(\varepsilon)$, we have that
\begin{align*}
    \pr{g_\varepsilon(Z)\leq 0}=\pr{\sum_{i=1}^d w_i Z_i^2 \leq f(\varepsilon)}.
\end{align*}

For the case that $\varepsilon\geq \frac12\log\frac{\det\Sigma_2}{\det\Sigma_1}$, $f(\varepsilon)\leq 0$ while $\sum_i w_i Z_i^2\geq 0$ almost surely; hence $\pr{g_\varepsilon(Z)\le 0}=0.$ Plugging into~\Cref{eq:delta-two-cdf-cov} yields $\delta_{\rX,\rY}(\varepsilon)\leq 0$, and since $\delta_{\rX,\rY}(\varepsilon)\geq 0$ always, therefore $\delta_{\rX,\rY}(\varepsilon)=0$.

We then consider case that $f(\varepsilon)>0$. We upper bound the two probabilities in~\Cref{eq:delta-two-cdf-cov} by Markov's bounds.

\noindent\textbf{Upper bound $\quad \pr{g_\varepsilon(Z)\leq 0}$.}
Let $\chi_i^2\defin \rZ_i^2\sim\chi^2_1$ be i.i.d. For any $s\geq 0$,
\begin{align*}
    & \pr{\sum_{i=1}^d w_i \chi_i^2 \le f(\varepsilon)}\\
    ={}& \pr{\rExp{-s\sum_{i=1}^d w_i \chi_i^2}\geq e^{-s f(\varepsilon)}} \\
    \leq{}& e^{s f(\varepsilon)} \Ex{\rExp{-s\sum_{i=1}^d w_i \chi_i^2}} \\
    ={}& e^{s f(\varepsilon)}\prod_{i=1}^d \Ex{e^{-s w_i \chi^2_1}} = e^{s f(\varepsilon)}\prod_{i=1}^d (1+2s w_i)^{-1/2} \\
    ={}& \rExp{s f(\varepsilon)-\frac12\sum_{i=1}^d \log(1+2s w_i)},
\end{align*}
where we used $\Ex{e^{-t\chi^2_1}}=(1+2t)^{-1/2}$ for $t\geq 0$.

\noindent\textbf{Lower bound $\quad \pr{g_\varepsilon(\widetilde Z)\le 0}$.}
Since $\widetilde \rZ\sim \mathcal N(0,D^{-1})$, we can write $\widetilde \rZ_i=\tau_i^{-1/2}\rZ_i'$ with $\rZ'\sim \multiNormal{0}{I_d}$, hence
\begin{align*}
    g_\varepsilon(\widetilde \rZ)=c(\varepsilon)+\frac12\sum_{i=1}^d w_i' (\rZ_i')^2, \qquad
    \pr{g_\varepsilon(\widetilde Z)\leq 0} =\pr{\sum_{i=1}^d w_i' (Z_i')^2 \leq f(\varepsilon)}.
\end{align*}
Let $\chi_i'^2\defin (Z_i')^2\sim\chi^2_1$ be i.i.d. Then
\begin{align*}
    \pr{\sum_{i=1}^d w_i' \chi_i'^2 \leq f(\varepsilon)} =1-\pr{\sum_{i=1}^d w_i' \chi_i'^2 > f(\varepsilon)}.
\end{align*}
For any $u\in(0,k)$, where $k=(2\max_i w_i')^{-1}$, we have $2u w_i'<1$ for all $i$, and thus $\Ex{e^{u w_i'\chi_1^2}}=(1-2u w_i')^{-1/2}$. By Markov's inequality,
\begin{align*}
    & \pr{\sum_{i=1}^d w_i' \chi_i'^2 > f(\varepsilon)}\\
    ={}& \pr{\rExp{u\sum_{i=1}^d w_i' \chi_i'^2} \geq e^{u f(\varepsilon)}} \\
    \leq{}& e^{-u f(\varepsilon)}\Ex{\rExp{u\sum_{i=1}^d w_i' \chi_i'^2}} \\
    ={}& e^{-u f(\varepsilon)}\prod_{i=1}^d \Ex{e^{u w_i'\chi^2_1}} = e^{-u f(\varepsilon)}\prod_{i=1}^d (1-2u w_i')^{-1/2} \\
    ={}& \rExp{-u f(\varepsilon)-\frac12\sum_{i=1}^d \log(1-2u w_i')}.
\end{align*}
Therefore,
\begin{align*}
    \pr{g_\varepsilon(\widetilde Z)\leq 0} \geq 1-\rExp{-u f(\varepsilon)-\frac12\sum_{i=1}^d \log(1-2u w_i')}.
\end{align*}

Plugging the bounds above into~\Cref{eq:delta-two-cdf-cov} gives, for all $s\geq 0$ and $u\in(0,k)$,
\begin{align*}
    \delta_{\rX,\rY}(\varepsilon) \leq &\rExp{ sf(\varepsilon) - \frac12\sum_{i=1}^d \log\!\bigl(1+2s(1-\tau_i)\bigr)} \\
    &- e^\varepsilon\Bigg(1 - \rExp{-uf(\varepsilon) - \frac12\sum_{i=1}^d \log\Bigl(1-2u \, \frac{1-\tau_i}{\tau_i}\Bigr)}\Bigg).
\end{align*}
Taking the infimum over $(s,u)$ completes the proof.
\end{proof}

\medskip
\noindent\textbf{Proof for~\Cref{prop:radial-comparison}}

\begin{proof}[Proof of~\Cref{prop:radial-comparison}]
\label{proof for prop:radial-comparison}
If $t=0$, then both sides equal $(1-e^{a})_{+}$ and the claim holds with equality. Assume henceforth $t>0$.

Let $Y\sim\chi^2_d$ and set $k\defin d/2\ge 1$. Equivalently, $Y/2\sim\Gamma(k,1)$. Let $P_k$ and $Q_k$ denote the CDF and survival function of $\Gamma(k,1)$, i.e., $P_k(x)=\pr{G\le x}$ and $Q_k(x)=\pr{G\ge x}$ for $G\sim\Gamma(k,1)$.
A direct change-of-variables calculation gives the following truncated exponential moments:
for every $y_0\ge 0$,
\begin{align*}
    \Ex{\rExp{-\tfrac{u}{2}Y} \mathbf 1\{Y\ge y_0\}} &=(1+u)^{-k} Q_k\Bigl(\tfrac{1+u}{2}y_0\Bigr), && u\geq 0,\\
    \Ex{\rExp{\tfrac{u}{2}Y}\,\mathbf 1\{Y\le y_0\}} &=(1-u)^{-k} P_k\Bigl(\tfrac{1-u}{2}y_0\Bigr), && u\in[0,1).
\end{align*}

Our first step is to get closed forms for $g_d$ and $h_d$. Since $a\in[-\tfrac12\log(1+t),0]$, we have $a+\log(1+t)\ge 0$, hence the positivity region
$\{1-\exp(a+\log(1+t)-\tfrac{t}{2}Y)>0\}$ is $\{Y\ge y_g\}$ with
\begin{align*}
    y_g \defin \frac{2(a+\log(1+t))}{t}\geq 0, \qquad
    x_g \defin \frac{y_g}{2}=\frac{a+\log(1+t)}{t}.
\end{align*}
Therefore,
\begin{align*}
    g_d\bigl(a+\log(1+t),t\bigr)
    &= \Ex{\Bigl(1-\rExp{a+\log(1+t)-\tfrac{t}{2}Y}\Bigr)\mathbf 1\{Y\ge y_g\}}\\
    &= \pr{Y\ge y_g} - \rExp{a+\log(1+t)}\Ex{\rExp{-\tfrac{t}{2}Y}\mathbf 1\{Y\ge y_g\}}\\
    &= Q_k(x_g) - \rExp{a}(1+t)^{1-k} Q_k\bigl((1+t)x_g\bigr),
\end{align*}
where in the last line we used the first truncated-moment identity with $u=t$.

Similarly, since $a\leq 0$, the positivity region $\{1-\exp(a+\tfrac{t}{2(1+t)}Y)>0\}$ is $\{Y\le y_h\}$ with
\begin{align*}
    y_h \defin \frac{-2a(1+t)}{t}\geq 0, \qquad
    x_h \defin \frac{y_h}{2}=\frac{-a(1+t)}{t}.
\end{align*}
Using the second truncated-moment identity with $u=t/(1+t)\in[0,1)$ yields
\begin{align*}
    h_d\Bigl(a,\frac{t}{1+t}\Bigr)
    &= \Ex{\Bigl(1-\rExp{a+\tfrac{t}{2(1+t)}Y}\Bigr)\mathbf 1\{Y\le y_h\}}\\
    &= \pr{Y\le y_h} - \rExp{a}\Ex{\rExp{\tfrac{t}{2(1+t)}Y}\mathbf 1\{Y\le y_h\}}\\
    &= P_k(x_h) - \rExp{a}(1+t)^{k}\,P_k\!\Bigl(\frac{x_h}{1+t}\Bigr).
\end{align*}

Subtracting and regrouping, we have that
\begin{align*}
    &g_d\bigl(a+\log(1+t),t\bigr) - h_d\Bigl(a,\frac{t}{1+t}\Bigr)\\
    ={}&\Bigl[Q_k(x_g)-\rExp{a}(1+t)^{1-k}Q_k\bigl((1+t)x_g\bigr)\Bigr] +\Bigl[\rExp{a}(1+t)^k P_k\!\Bigl(\tfrac{x_h}{1+t}\Bigr)-P_k(x_h)\Bigr].
\end{align*}
Using that $Q_k$ is nonincreasing and $P_k$ is nondecreasing, we have $Q_k((1+t)x_g)\leq Q_k(x_g)$ and $P_k(x_h)\geq P_k(x_h/(1+t))$, hence
\begin{align*}
    &g_d\bigl(a+\log(1+t),t\bigr) - h_d\Bigl(a,\frac{t}{1+t}\Bigr)\\
    \ge{}& Q_k(x_g)\Bigl[1-\rExp{a}(1+t)^{1-k}\Bigr] + P_k\!\Bigl(\tfrac{x_h}{1+t}\Bigr)\Bigl[\rExp{a}(1+t)^k-1\Bigr].
\end{align*}

It remains to check that both bracketed coefficients are nonnegative under $a\in[-\tfrac12\log(1+t),0]$ and $k\geq 1$. Indeed, since $\rExp{a}\leq 1$ and $(1+t)^{1-k}\le 1$, we get $\rExp{a}(1+t)^{1-k}\le 1$, so the first bracket is $\geq 0$. Also, $\rExp{a}\geq (1+t)^{-1/2}$, hence
\begin{align*}
    \rExp{a}(1+t)^k \geq (1+t)^{k-\frac12} \geq 1,
\end{align*}
so the second bracket is nonnegative as well. Since $Q_k(\cdot)\geq 0$ and $P_k(\cdot)\geq 0$, the RHS is nonnegative, proving the desired inequality.
\end{proof}

\medskip
\noindent\textbf{Proof for~\Cref{lem: asymetric order for covariance}}

\begin{proof}[Proof of \Cref{lem: asymetric order for covariance}]
\label{proof for lem: asymetric order for covariance}
Recall the definition of the matrix $M = \Sigma_2^{-1/2}\Sigma_1\Sigma_2^{-1/2}$. Since $\Sigma_1 \succeq \Sigma_2$, we know that $M\succeq I_d$, hence all eigenvalues $\tau_1,\dots,\tau_d$ of $M$ satisfy $\tau_i\geq 1$. Write $L = \log\det M=\sum_{i=1}^d\log\tau_i\geq 0$.

\medskip
\noindent\textbf{Reduction to $(M,I_d)$.} Applying the NDIS lemma (\Cref{lemma: ndis}) to the pair $\bigl(\multiNormal{0}{\Sigma_1},\multiNormal{0}{\Sigma_2}\bigr)$ and using the change of variables $z=\Sigma_2^{-1/2}x$, we obtain the invariance
\begin{align*}
    \delta_{\multiNormal{0}{\Sigma_1},\multiNormal{0}{\Sigma_2}}(\varepsilon) ={}& \delta_{\multiNormal{0}{M},\multiNormal{0}{I_d}}(\varepsilon) \\
    \delta_{\multiNormal{0}{\Sigma_2},\multiNormal{0}{\Sigma_1}}(\varepsilon) ={}& \delta_{\multiNormal{0}{I_d},\multiNormal{0}{M}}(\varepsilon).
\end{align*}
Therefore it suffices to prove that, for all $\varepsilon\geq 0$,
\begin{align*}
    \delta_{M,I_d}(\varepsilon) \geq \delta_{I_d,M}(\varepsilon).
\end{align*}

\medskip
\noindent\textbf{Radial--Angular decomposition.}
Let $Z\sim\multiNormal{0}{I_d}$ and write $Z=RU$, where $R\geq 0$ and $U\in\mathbb S^{d-1}$ are independent, $R^2\sim\chi^2_d$, and $U$ is uniform on the unit sphere. Define the Rayleigh quotients
\begin{align*}
    \alpha(U)\defin U^\top(M-I_d)U \geq 0, \qquad \beta(U)\defin U^\top(I_d-M^{-1})U \geq 0.
\end{align*}

Applying~\Cref{lemma: ndis} to the zero-mean case gives
\begin{align*}
    \delta_{M,I_d}(\varepsilon)
    ={}& \Ex{1-\rExp{\varepsilon+\tfrac12\log\det M+\tfrac12 Z^\top(I_d-M)Z}}_{+}\\
    ={}& \Exf{U}{\Exf{R^2}{\Bigl(1-\rExp{\varepsilon+\tfrac{L}{2}-\tfrac{\alpha(U)}{2}R^2}\Bigr)_{+}}}\\
    ={}& \Exf{U}{g_d\bigl(\varepsilon+\tfrac{L}{2},\alpha(U)\bigr)},
\end{align*}
and similarly
\begin{align*}
    &\delta_{I_d,M}(\varepsilon) \\
    ={}& \Ex{1-\rExp{\varepsilon-\tfrac12\log\det M+\tfrac12 Z^\top(I_d-M^{-1})Z}}_{+}\\
    ={}& \Exf{U}{\Exf{R^2}{\Bigl(1-\rExp{\varepsilon-\tfrac{L}{2}+\tfrac{\beta(U)}{2}R^2}\Bigr)_{+}}}
    = \Exf{U}{h_d\bigl(\varepsilon-\tfrac{L}{2},\,\beta(U)\bigr)},
\end{align*}
where $g_d,h_d$ are exactly the functions in \Cref{prop:radial-comparison}.

\medskip
\noindent\textbf{Coupling $\beta(U)$ and $\alpha(U)$.} Recall that $M=Q^\top \diag(\tau_1,\dots,\tau_d)Q$ with $\tau_i\geq 1$. Let $V = QU$, note that $V$ is also uniform on the sphere and $\twoNorm{V} = 1$. Then, we have that
\begin{align*}
    \alpha(U)=\sum_{i=1}^d(\tau_i-1)V_i^2, \qquad
    \beta(U)=\sum_{i=1}^d\Bigl(1-\frac1{\tau_i}\Bigr)V_i^2 =\sum_{i=1}^d \frac{\tau_i-1}{\tau_i}V_i^2.
\end{align*}
With $f(x)\defin \frac{x}{1+x}$ (concave on $\nonnegReal$), Jensen's inequality gives
\begin{align*}
\beta(U) = \sum_i f(\lambda_i-1)V_i^2 \leq f\Bigl(\sum_i(\lambda_i-1)V_i^2\Bigr) =\frac{\alpha(U)}{1+\alpha(U)}.
\end{align*}

Fix $U$ and set $t = \alpha(U)\geq 0$. The above inequality and the fact that $h_d(a,\cdot)$ is nondecreasing in its second argument give
\begin{align*}
    h_d\bigl(\varepsilon-\tfrac{L}{2},\beta(U)\bigr) \leq h_d\Bigl(\varepsilon-\tfrac{L}{2},\frac{t}{1+t}\Bigr).
\end{align*}

We now first consider the case that $0\leq \varepsilon \leq L/2$, and let $a = \varepsilon-\tfrac{L}{2}\in[-L/2,0]$. Since
\begin{align*}
    \log(1+t)=\log(U^\top MU)\leq \log\tau_{\max}\leq \sum_{i=1}^d\log\tau_i = L,
\end{align*}
we have $a\geq -\tfrac12\log(1+t)$, i.e., $a\in[-\tfrac12\log(1+t),0]$, and thus \Cref{prop:radial-comparison} applies:
\begin{align*}
    h_d\Bigl(a,\frac{t}{1+t}\Bigr)\leq g_d\bigl(a+\log(1+t),t\bigr).
\end{align*}

Since $g_d(\cdot,t)$ is nondecreasing in its first argument and $a+\log(1+t)\leq a+L=\varepsilon+\tfrac{L}{2}$, we get
\begin{align*}
    g_d\bigl(a+\log(1+t),t\bigr)\leq g_d\bigl(\varepsilon+\tfrac{L}{2},t\bigr) = g_d\bigl(\varepsilon+\tfrac{L}{2},\alpha(U)\bigr).
\end{align*}
Combining the last three, we have that for all $U$ and all $\varepsilon\in[0,L/2]$
\begin{align*}
    h_d\bigl(\varepsilon-\tfrac{L}{2},\beta(U)\bigr) \leq g_d\bigl(\varepsilon+\tfrac{L}{2},\alpha(U)\bigr).
\end{align*}
Taking expectation over $U$ gives $\delta_{I_d,M}(\varepsilon)\leq \delta_{M,I_d}(\varepsilon)$ for $\varepsilon\in[0,L/2]$.

We last discuss the case that $\varepsilon\geq L/2$. It is easy to see that $\delta_{I_d,M}(\varepsilon)=0$,  hence trivially $\delta_{M,I_d}(\varepsilon)\ge \delta_{I_d,M}(\varepsilon)$.

Putting all cases together establishes that, for all $\varepsilon\ge 0$,
\begin{align*}
    \delta_{M,I_d}(\varepsilon)\geq \delta_{I_d,M}(\varepsilon)
\end{align*}
and therefore
\begin{align*}
    \delta_{\rX,\rY}(\varepsilon)\geq \delta_{\rY,\rX}(\varepsilon).
\end{align*}
\end{proof}

\medskip
\noindent\textbf{Proof for~\Cref{prop: triangle for DP divergence}}
\begin{proof}[Proof of~\Cref{prop: triangle for DP divergence}]
\label{proof for prop: triangle for DP divergence}
Fix any event $A$, we have
\begin{align*}
    &\pr{X \in A} - e^{\varepsilon_1 + \varepsilon_2} \pr{Z \in A} \\
    ={}& \bigl( \pr{X \in A} - e^{\varepsilon_1} \pr{Y \in A} \bigr) + e^{\varepsilon_1} \bigl( \pr{Y \in A} - e^{\varepsilon_2} \pr{Z \in A} \bigr)\\
    \leq{}& \bigl( \pr{X \in A} - e^{\varepsilon_1} \pr{Y \in A} \bigr)_{+} + e^{\varepsilon_1} \bigl( \pr{Y \in A} - e^{\varepsilon_2} \pr{Z \in A} \bigr)_{+}\\
    ={}& \delta_{X,Y}(\varepsilon_1) + e^{\varepsilon_1} \delta_{Y,Z}(\varepsilon_2),
\end{align*}
which yields the claim.
\end{proof}

\medskip
\noindent\textbf{Proof for~\Cref{thm:ndis-analytic-form}}

\begin{proof}[Proof of \Cref{thm:ndis-analytic-form}]
\label{proof for thm:ndis-analytic-form}
Fix $\varepsilon\geq 0$ and any split $\varepsilon=\varepsilon_1+\varepsilon_2$ with $\varepsilon_1,\varepsilon_2\geq 0$. We introduce an intermediate Gaussian $\rW_1 \sim \multiNormal{\mu_2}{\Sigma_{\min}},$ which shares the \emph{target mean} $\mu_2$ and the \emph{smaller covariance} $\Sigma_{\min}$.
This induces the two-step path
\begin{align*}
    \rX=\multiNormal{\mu_1}{\Sigma_{\min}} \longrightarrow \rW_1=\multiNormal{\mu_2}{\Sigma_{\min}} \longrightarrow \rY=\multiNormal{\mu_2}{\Sigma_{\max}}.
\end{align*}
By the triangle inequality for DP divergence (\Cref{prop: triangle for DP divergence}),
\begin{align*}
    \delta_{\rX,\rY}(\varepsilon_1+\varepsilon_2) \leq{}& \delta_{\rX,\rW_1}(\varepsilon_1) + e^{\varepsilon_1}\delta_{\rW_1,\rY}(\varepsilon_2)\\
    ={}& \delta_{\multiNormal{\mu_1}{\Sigma_{\min}},\multiNormal{\mu_2}{\Sigma_{\min}}}(\varepsilon_1) + e^{\varepsilon_1}\delta_{\multiNormal{\mu_2}{\Sigma_{\min}},\multiNormal{\mu_2}{\Sigma_{\max}}}(\varepsilon_2)\\
    \leq{}&\delta_{\multiNormal{\mu_1}{\Sigma_{\min}},\multiNormal{\mu_2}{\Sigma_{\min}}}(\varepsilon_1) + e^{\varepsilon_1}\delta_{\multiNormal{\mu_2}{\Sigma_{\max}},\multiNormal{\mu_2}{\Sigma_{\min}}}(\varepsilon_2). \tag{by ~\Cref{lem: asymetric order for covariance}}
\end{align*}

Taking the infimum over all splits $\varepsilon_1+\varepsilon_2=\varepsilon$ yields
\begin{align*}
    \delta_{\rX,\rY}(\varepsilon)\leq \overline{\delta}^{(1)}(\varepsilon).
\end{align*}

For the second quantity $\overline{\delta}^{(2)}(\varepsilon)$, we could similarly create an intermediate Gaussian $\rW_2 \sim \multiNormal{\mu_1}{\Sigma_{\min}}.$ We then take the path that first change the covariance under mean $\mu_1$, then change the mean under $\Sigma_{\min}$. The rest part of proof just follow the above.

Combining the two bounds gives
\begin{align*}
    \delta_{\rX,\rY}(\varepsilon)\leq \min\{\overline{\delta}^{(1)}(\varepsilon),\overline{\delta}^{(2)}(\varepsilon)\}.
\end{align*}
Finally, the stated analytic form claims follow immediately from \Cref{prop:ndis-mean-analytic-form} (mean-only terms) and \Cref{lem:ndis-cov-analytic-form} (covariance-only terms).
\end{proof}

\section{Missing Proofs for Random-Projection DP Mechanisms}

\subsection{Privacy analysis of $\Mech_{RP}$}

\medskip
\noindent\textbf{Proof for~\Cref{lemma: rp inherent}}

\begin{proof}[Proof of~\Cref{lemma: rp inherent}]
\label{proof for lemma: rp inherent}
Let $v^\top\in\Real^{1\times d}$ be the $i$-th row of $\DB$, so that $\DB_{-i}^\top \DB_{-i} = \DB^\top\DB - v v^\top$. By \Cref{def:grp}, writing $\rX=[X^{(1)},\dots,X^{(r)}]$ and $\rY=[Y^{(1)},\dots,Y^{(r)}]$ by columns, we have i.i.d.\ columns
\begin{align*}
    &X^{(j)} \sim \multiNormal{0}{\Sigma_1}, &\qquad Y^{(j)} \sim \multiNormal{0}{\Sigma_2} \\
    &\Sigma_1 \defin \DB^\top\DB, &\qquad \Sigma_2 \defin \DB_{-i}^\top\DB_{-i}=\Sigma_1 - v v^\top.
\end{align*}
Identifying $\Real^{d\times r}$ with $\Real^{dr}$ via vectorization (a bijection), the hockey-stick divergence is invariant under this relabeling, hence
we may work with
\begin{align*}
    \widehat X \defin \mathrm{vec}(\rX)\sim \multiNormal{0}{I_r\otimes \Sigma_1}, \qquad
    \widehat Y \defin \mathrm{vec}(\rY)\sim \multiNormal{0}{I_r\otimes \Sigma_2}.
\end{align*}

\medskip
\noindent\textbf{Reduction to the NDIS quadratic form.}
Applying~\Cref{lemma: ndis} with $\mu_1=\mu_2=0$ and $\Sigma_1' \defin I_r\otimes\Sigma_1$, $\Sigma_2' \defin I_r\otimes\Sigma_2$ gives
\begin{align*}
    \Delta\mu=0, \qquad b=0, \qquad
    c(\varepsilon) = \varepsilon + \frac{r}{2}\log\frac{\det\Sigma_1}{\det\Sigma_2}.
\end{align*}
To compute $\det\Sigma_2$, use the matrix determinant lemma:
\begin{align*}
    \det(\Sigma_2)
    = \det(\Sigma_1 - v v^\top)
    = \det(\Sigma_1)\bigl(1 - v^\top \Sigma_1^{-1} v\bigr)
    = \det(\Sigma_1)(1-\leverage),
\end{align*}
recalling $\leverage$ is the leverage score of the record $v$ in $\DB$, as defined in~\Cref{def: leverage}. Thus $\det\Sigma_1/\det\Sigma_2 = 1/(1-\leverage)=\rho$, and hence
\begin{align*}
    c(\varepsilon) = \varepsilon + \frac{r}{2}\log\rho.
\end{align*}

\medskip
\noindent Next, we compute the eigenvalues of
\begin{align*}
    M \defin (\Sigma_1')^{1/2}(\Sigma_2')^{-1}(\Sigma_1')^{1/2} = I_r\otimes\Bigl(\Sigma_1^{1/2}\Sigma_2^{-1}\Sigma_1^{1/2}\Bigr).
\end{align*}
Let $u\defin \Sigma_1^{-1/2}v$, so that $u^\top u = v^\top\Sigma_1^{-1}v=\leverage$ and
\begin{align*}
    \Sigma_2 = \Sigma_1 - v v^\top = \Sigma_1^{1/2}\bigl(I_d - u u^\top\bigr)\Sigma_1^{1/2}.
\end{align*}
Therefore
\begin{align*}
    \Sigma_1^{1/2}\Sigma_2^{-1}\Sigma_1^{1/2} = (I_d-u u^\top)^{-1}.
\end{align*}
The rank-one matrix $u u^\top$ has eigenvalue $\leverage$ in direction $u$ and $0$ on $u^\perp$; hence $(I_d-u u^\top)^{-1}$ has eigenvalues $\rho=(1-\leverage)^{-1}$ in direction $u$ and $1$ on $u^\perp$. Consequently, $M$ has exactly $r$ eigenvalues equal to $\rho$ and the remaining $r(d-1)$ eigenvalues equal to $1$.

\medskip
Plugging in~\Cref{lemma: ndis}, this means the coefficients $a_k=1-\tau_k$ satisfy
\begin{align*}
    a_k = 1-\rho =-(\rho-1)\quad\text{for }k\in[r], \qquad
    a_k = 0\quad\text{for }k\in\{r+1,\dots,dr\}.
\end{align*}
Let $\rZ\sim\multiNormal{0}{I_{dr}}$ and write $T = \sum_{k=1}^r Z_k^2\sim\chi_r^2$. \Cref{lemma: ndis} yields
\begin{align*}
    &\delta_{\rX,\rY}(\varepsilon) = \Ex{1-\exp\bigl(g_\varepsilon(Z)\bigr)\Bigr)_+} \\
    & \qquad \text{where} \quad g_\varepsilon(\rZ) = \varepsilon+\frac{r}{2}\log\rho - \frac{\rho-1}{2}T.
\end{align*}

\medskip
\noindent\textbf{Expression of NDIS quadratic form in closed form.}
The integrand is positive iff $g_\varepsilon(Z)\le 0$, i.e.,
\begin{align*}
    T \geq t_0 \qquad\text{where}\qquad t_0 = \frac{2\bigl(\varepsilon+\tfrac{r}{2}\log\rho\bigr)}{\rho-1}.
\end{align*}
Hence, with $s =  r/2$ and $T\sim\chi_r^2$,
\begin{align*}
    \delta_{\rX,\rY}(\varepsilon)
    &= \Exf{T}{\Bigl(1-\exp\bigl(\varepsilon+\tfrac{r}{2}\log\rho-\tfrac{\rho-1}{2}T\bigr)\Bigr) \mathbf{1}\{T\geq t_0\}} \\
    &= \pr{T\geq t_0} - e^{\varepsilon}\rho^{s} \Exf{T}{e^{-(\rho-1)T/2}\mathbf{1}\{T\geq t_0\}}.
\end{align*}
Using the $\chi_r^2$ density $f(t)=\frac{1}{2^s\Gamma(s)}t^{s-1}e^{-t/2}$, the second term equals
\begin{align*}
       & e^{\varepsilon}\rho^s \int_{t_0}^\infty e^{-(\rho-1)t/2} f(t) \mathrm{d}t\\
    ={}& e^{\varepsilon}\rho^s \cdot \frac{1}{2^s\Gamma(s)}\int_{t_0}^\infty t^{s-1}e^{-\rho t/2} \mathrm{d}t \\
    ={}& e^{\varepsilon}\rho^s \cdot \frac{1}{2^s\Gamma(s)}\cdot \rho^{-s} \int_{\rho t_0}^\infty u^{s-1}e^{-u/2}\,du \tag{$u=\rho t$}\\
    ={}& \frac{e^{\varepsilon}}{\Gamma(s)}\Gamma\Bigl(s,\frac{\rho t_0}{2}\Bigr).
\end{align*}
Moreover,
\begin{align*}
    \pr{T \geq t_0} = \frac{1}{\Gamma(s)}\Gamma\Bigl(s,\frac{t_0}{2}\Bigr).
\end{align*}
Combining the two and recalling $s=r/2$ gives the claimed expression
\begin{align*}
    \delta_{\rX,\rY}(\varepsilon) = \frac{1}{\Gamma(\tfrac{r}{2})} \Bigl(\Gamma\Bigl(\tfrac{r}{2}, \tfrac{t_0}{2}\Bigr) - e^{\varepsilon} \Gamma\Bigl(\tfrac{r}{2}, \tfrac{\rho t_0}{2}\Bigr)\Bigr).
\end{align*}
\end{proof}

\medskip
\noindent\textbf{Proof for~\Cref{prop:grp-monotone}}

\begin{proof}[Proof is in~\Cref{prop:grp-monotone}]
\label{proof for prop:grp-monotone}
Let $s = r/2$ and recall that $\overline F(x) =  \Gamma(s,x/2)/\Gamma(s)$ is the upper tail probability of $\chi_r^2$, evaluating at point $x$. From Lemma~\ref{lemma: rp inherent} we may rewrite, for $\rho>1$,
\begin{align*}
    \delta^{\gRP}_{r}(\varepsilon;\rho) = \overline F\bigl(t_0(\rho)\bigr) - e^{\varepsilon}\overline F\bigl(\rho\,t_0(\rho)\bigr), \qquad
    t_0(\rho)\defin \frac{2(\varepsilon+s\log\rho)}{\rho-1}.
\end{align*}
Differentiating and using $\overline F'(x)=-f_{\chi_r^2}(x)$ gives
\begin{align*}
    \frac{\mathrm{d}}{\mathrm{d}\rho}\delta^{\gRP}_{r}(\varepsilon;\rho) = -f_{\chi_r^2}(t_0) t_0' + e^{\varepsilon} f_{\chi_r^2}(\rho t_0)\,(t_0+\rho t_0').
\end{align*}

\medskip
\noindent By the definition $t_0(\rho)\defin 2(\varepsilon+s\log\rho)/(\rho-1)$, we have
\begin{align*}
    \varepsilon+s\log\rho = \frac{\rho-1}{2}t_0(\rho) \qquad\Longleftrightarrow\qquad
    e^{\varepsilon}e^{-(\rho-1)t_0(\rho)/2}=\rho^{-s}.
\end{align*}
Recall that the $\chi_r^2$ density (with $s=r/2$) is
\begin{align*}
    f_{\chi_r^2}(x)=\frac{1}{2^s\Gamma(s)} x^{s-1}e^{-x/2}.
\end{align*}
Using the identity, and a substitution into the $\chi_r^2$ density gives that $ e^{\varepsilon} f_{\chi_r^2}(\rho t_0)= \frac{1}{\rho} f_{\chi_r^2}(t_0).$ Concretely,
\begin{align*}
    e^{\varepsilon} f_{\chi_r^2}\bigl(\rho t_0(\rho)\bigr)
    &= \frac{1}{2^s\Gamma(s)}e^{\varepsilon}(\rho t_0)^{s-1}e^{-\rho t_0/2} \\
    &= \frac{1}{2^s\Gamma(s)}(\rho t_0)^{s-1}e^{-t_0/2}
       \Bigl(e^{\varepsilon}e^{-(\rho-1)t_0/2}\Bigr) \\
    &= \frac{1}{2^s\Gamma(s)}(\rho t_0)^{s-1}e^{-t_0/2}\rho^{-s} \\
    &= \frac{1}{\rho}\cdot \frac{1}{2^s\Gamma(s)}t_0^{s-1}e^{-t_0/2}
     = \frac{1}{\rho} f_{\chi_r^2}\bigl(t_0(\rho)\bigr),
\end{align*}
where in the second line we used $e^{-\rho t_0/2}=e^{-t_0/2}e^{-(\rho-1)t_0/2}$ and in the third line we
substituted $e^{\varepsilon}e^{-(\rho-1)t_0/2}=\rho^{-s}$.

\medskip
\noindent Plugging this into the derivative above yields the cancellation
 \begin{align*}
     \frac{\mathrm{d}}{\mathrm{d}\rho}\delta^{\gRP}_{r}(\varepsilon;\rho) = \frac{t_0(\rho)}{\rho} f_{\chi_r^2}\!\bigl(t_0(\rho)\bigr)\geq 0.
 \end{align*}
Finally, since $\rho(p)=1/(1-p)$ and $\rho'(p)=\rho(p)^2$, the chain rule gives
\begin{align*}
    & \frac{\mathrm{d}}{\mathrm{d}p}\delta^{\gRP}_{r}(\varepsilon;p)\\
    ={}& \frac{\mathrm{d}}{\mathrm{d}\rho}\delta^{\gRP}_{r}(\varepsilon;\rho)\Big|_{\rho=\rho(p)}\cdot \rho'(p)\\
    ={}& \rho(p) t_0(p) f_{\chi_r^2}\!\bigl(t_0(p)\bigr)\geq 0.
\end{align*}
\end{proof}

\medskip
\noindent\textbf{Proof for~\Cref{thm:RP-mech-DP}}

\begin{proof}[proof of~\Cref{thm:RP-mech-DP}]
\label{proof for thm:RP-mech-DP}
Let $\DB$ and $\DB'$ be neighboring databases under the standard add/remove notion. Without loss of generality, assume $\DB$ is the larger database and $\DB'$ is obtained from $\DB$ by removing one row, i.e., $\DB'=\DB_{-i}$ for some $i\in[n]$. Let $v^\top$ be the removed row. Define the augmented databases
\begin{align*}
    \overline \DB = \begin{bmatrix}\DB\\ \sqrt{\lambda}I_d\end{bmatrix}, \qquad
    \overline \DB' = \begin{bmatrix}\DB'\\ \sqrt{\lambda}I_d\end{bmatrix}.
\end{align*}
Let the mechanism outputs be
\begin{align*}
    \rX = \defin \Mech_{RP}(\DB)=\gRP_r(\overline{\DB}), \qquad \rY = \defin \Mech_{RP}(\DB)=\gRP_r(\overline{\DB'}).
\end{align*}
By~\Cref{def:grp}, each column of $\rX$ is $\multiNormal{0}{\Sigma_1}$ and each column of $\rY$ is $\multiNormal{0}{\Sigma_2}$, where
\begin{align*}
    \Sigma_1 = \overline{\DB}^\top\overline{\DB}=\DB^\top\DB+\lambda I_d, \qquad
    \Sigma_2 = \overline{\DB}'^\top\overline{\DB}'=\DB'^\top\DB'+\lambda I_d  = \Sigma_1 - v v^\top,
\end{align*}
so $\Sigma_1\succeq \Sigma_2\succ 0$.

\medskip
\noindent\textbf{Leverage regularization under ridge.}
Let $\overline{\leverage}$ denote the leverage score of row $i$ in the augmented database $\overline{\DB}$. Then
\begin{align*}
    \overline{\leverage} = v^\top(\overline{\DB}^\top\overline{\DB})^{-1}v = v^\top(\DB^\top\DB+\lambda I_d)^{-1}v.
\end{align*}
Since $\DB^\top\DB+\lambda I_d \succeq \lambda I_d$, we have $(\DB^\top\DB+\lambda I_d)^{-1}\preceq \frac{1}{\lambda}I_d$,
and hence
\begin{align*}
    \overline{\leverage}\leq \frac{1}{\lambda}\twoNorm{v}^2 \leq \frac{l^2}{\lambda} = p^\star.
\end{align*}

Applying~\Cref{lemma: rp inherent} to the pair $(\overline{\DB},\overline{\DB}')$ yields
\begin{align*}
    \delta_{\rX,\rY}(\varepsilon) = \delta^{\gRP}_r(\varepsilon;\overline{\leverage}),
\end{align*}
where $\delta^{\gRP}_r(\varepsilon;p)$ is defined in \Cref{prop:grp-monotone}. Using Proposition~\ref{prop:grp-monotone} and $\overline{\leverage}\leq p^\star$ gives
\begin{align*}
    \delta_{\rX,\rY}(\varepsilon) \leq \delta^{\gRP}_r(\varepsilon;p^\star) \leq \delta.
\end{align*}
By the definition of $\delta_{\rX,\rY}(\varepsilon)$, this implies that for every measurable set $\cS$,
\begin{align*}
    \pr{\rX\in\cS} \leq e^\varepsilon\pr{\rY\in\cS} + \delta.
\end{align*}

\medskip
\noindent\textbf{Reverse DP inequality.}
Since $\Sigma_1\succeq \Sigma_2\succ 0$, Lemma~\ref{lem: asymetric order for covariance} implies
\begin{align*}
    \delta_{\rY,\rX}(\varepsilon) \leq \delta_{\rX,\rY}(\varepsilon) \leq \delta,
\end{align*}
equivalently, for every measurable set $\cS$,
\begin{align*}
    \pr{\rY\in\cS} \leq e^\varepsilon\pr{\rX\in\cS} + \delta.
\end{align*}

Combining the two inequalities, by definition, $\Mech_{RP}$ is $(\varepsilon,\delta)$-DP for add/remove adjacency.
\end{proof}

\subsection{Utility analysis of $\Mech_{RP}$}

\begin{proof}[Proof of~\Cref{thm:mech-rp-utility}]
\label{proof for thm:mech-rp-utility}
Recall the notations $\lambda=l^2/p^\star$ and $\bar\DB=\begin{bmatrix}\DB\\ \sqrt{\lambda}\,I_d\end{bmatrix}\in\Real^{(n+d)\times d}$ that are defined in~\Cref{alg::RP mechanism}. Write $\widetilde M = \gRP_r(\overline{\DB}) =[X^{(1)},\dots,X^{(r)}]$ by columns. Then $X^{(1)},\dots,X^{(r)}$ are i.i.d. and, by~\Cref{def:grp}, for each $j \in [r]$,
\begin{align*}
    X^{(j)} \sim \multiNormal{0}{\overline{\DB}^\top \overline{\DB}}.
\end{align*}
Since $\overline{\DB}^\top \overline{\DB}=\DB^\top\DB+\lambda I_d=\Sigma+\lambda I_d$, we have $X^{(j)}\sim \multiNormal{0}{\Sigma+\lambda I_d}.$

\medskip
\noindent\textbf{Unbiasedness.}
By the above, we have that
\begin{align*}
    \Ex{\frac{1}{r}\widetilde M\widetilde M^\top}
    = \Ex{\frac{1}{r}\sum_{j=1}^r X^{(j)}(X^{(j)})^\top}
    = \Ex{X^{(1)}(X^{(1)})^\top}
    = \Sigma+\lambda I_d.
\end{align*}
Hence
\begin{align*}
    \Ex{\widehat\Sigma} = \Ex{\frac{1}{r}\widetilde M\widetilde M^\top - \lambda I_d} = \Sigma.
\end{align*}

\medskip
\noindent\textbf{High-probability spectral-norm error.}
Define $\Sigma_{\rm reg}\defin \Sigma+\lambda I_d$ and $\widehat\Sigma_{\rm reg}\defin \frac{1}{r}\widetilde M\widetilde M^\top$. Then $\widehat\Sigma=\widehat\Sigma_{\rm reg}-\lambda I_d$, hence
\begin{align*}
    \twoNorm{\widehat\Sigma-\Sigma} = \twoNorm{\widehat\Sigma_{\rm reg}-\Sigma_{\rm reg}}.
\end{align*}

Since $X^{(1)},\dots,X^{(r)}$ are i.i.d. Gaussian with covariance $\Sigma_{\rm reg}$, $\widehat\Sigma_{\rm reg}$ then is the sample covariance of $r$ i.i.d. mean-zero Gaussian samples in $\Real^d$. We then can apply standard covariance concentration bounds for subgaussian vectors. By~\cite{vershynin2018hdp}[Remark 4.7.3], there exists an absolute constant $C>0$ such that for every $\beta\in(0,1)$, with probability at least $1-\beta$,
\begin{align*}
    \|\widehat\Sigma_{\rm reg}-\Sigma_{\rm reg}\|
    ~\le~
    C\left(\sqrt{\frac{d+\log(2/\beta)}{r}}+\frac{d+\log(2/\beta)}{r}\right)\,\|\Sigma_{\rm reg}\|.
\end{align*}
Finally,
\begin{align*}
    \twoNorm{\Sigma_{\rm reg}} = \twoNorm{\Sigma+\lambda I_d} \leq \twoNorm{\Sigma} + \lambda = \twoNorm{\DB}^2 + \lambda,
\end{align*}
which yields the claimed bound.
\end{proof}

\subsection{Privacy analysis of $\Mech_{RP}$'s variants}

\begin{proof}[Proof of~\Cref{thm:subsampling RP-mech-DP}]
\label{proof for thm:subsampling RP-mech-DP}
By \Cref{thm:RP-mech-DP}, $\Mech_{RP}$ is $(\varepsilon_0,\delta_0)$-DP under the unbounded (add/remove) neighboring relation.

The wrapper $\Mech^{\mathsf{Pois}}_{RP}$ is exactly: Poisson subsample each row with rate $q$ to form $\DB_S$, then output $\Mech_{RP}(\DB_S)$. By the standard privacy amplification theorem for Poisson subsampling (see, e.g., \cite{BalleBG18}[Thm.8]), $\Mech^{\mathsf{Pois}}_{RP}$ is $(\varepsilon',\delta')$-DP with
\begin{align*}
    \varepsilon'=\ln\bigl(1+q(e^{\varepsilon_0}-1)\bigr), \qquad \delta'=q \delta_0.
\end{align*}
With the choices in \Cref{alg:subsampled-wrapper-rp}, $\varepsilon_0=\ln\Bigl(1+\frac{e^{\varepsilon}-1}{q}\Bigr)$ gives that $e^{\varepsilon_0}=1+\frac{e^{\varepsilon}-1}{q}.$ Therefore, $\varepsilon' = \varepsilon,$ and $\delta'=q\cdot(\delta/q)=\delta$. This yields the desired claim.
\end{proof}

\begin{proof}[Proof of~\Cref{thm:ptr RP-mech-DP}]
\label{proof for thm:ptr RP-mech-DP}
Fix neighboring databases under the add/remove notion: let $\DB'\defin \DB_{-i}$ be obtained from $\DB$ by removing the $i$-th row $x_i^\top$.

\medskip
\noindent\textbf{Step 3 is $(\varepsilon_T,\delta_T)$-DP.}
Define the query $f(\DB)\defin \lambda_{\min}(\DB^\top \DB)$. We know that $f$ has global sensitivity at most $l^2$. This is because, by Weyl's inequality,
\begin{align*}
    \abs{f(\DB)-f(\DB')} = \abs{\lambda_{\min}(\DB^\top\DB)-\lambda_{\min}(\DB'^\top\DB')} \leq \twoNorm{x_ix_i^\top} \leq l^2.
\end{align*}
In Step~3, the noised query $f(\DB)+\eta$ is essentially $(\varepsilon_T,\delta_T)$-DP by the Gaussian mechanism analysis (equivalently, \Cref{prop:ndis-mean-analytic-form} specialized to a scalar shift). Moreover, $\lambda_{\mathrm{lb}}=\max\{\tilde\lambda-\alpha,0\}$ is also $(\varepsilon_T,\delta_T)$-DP by post-processing.

\medskip
We then argue that Step 4 certifies a leverage cap except with probability $\delta_{\mathrm{ptr}}$. Define the event $E \defin \{\eta \leq \alpha\}$, where $\alpha=\tau\cdot \Phi^{-1}(1-\delta_{\mathrm{ptr}}).$ Since $\eta \sim \multiNormal{0}{\tau^2}$, we have that
\begin{align*}
    \pr{E^c}=\pr{\eta>\alpha}=\delta_{\mathrm{ptr}}.
\end{align*}
On event $E$, we have $\lambda_{\mathrm{lb}}=\max\{f(\DB)+\eta-\alpha,0\}\leq f(\DB)$. Let $\lambda_{\mathrm{ridge}}\defin \max\{\frac{l^2}{p^\star}-\lambda_{\mathrm{lb}},0\}$ be the ridge chosen in Step~4. Then on $E$,
\begin{align*}
    \lambda_{\min}\bigl(\DB^\top\DB+\lambda_{\mathrm{ridge}}I_d\bigr) \geq \frac{l^2}{p^\star}.
\end{align*}
Consequently, for any row $x$ of $\DB$ , its leverage score in the augmented database $\overline \DB$ is upper bounded by $p^\star$ on event $E.$ An identical argument applies to $\DB'$ when running the mechanism on $\DB'$.

\medskip
\noindent\textbf{Step 5 is $(\varepsilon_R,\delta_R+\delta_{\mathrm{ptr}})$-DP.}
Recall $\varepsilon_R=\max\{\varepsilon-\varepsilon_T,0\}$ and $p^\star$ is chosen in Step~2 so that $\delta^{\gRP}_r(\varepsilon_R;p^\star)\leq \delta_R$.
On event $E$, the leverage cap discussed above ensures that the same privacy argument as in \Cref{thm:RP-mech-DP} (via \Cref{lemma: rp inherent} and monotonicity in \Cref{prop:grp-monotone}) on $\gRP_r(\overline \DB)$ holds and the release $\widetilde M=\gRP_r(\overline\DB)$ is $(\varepsilon_R,\delta_R)$-DP with respect to removing row $i$.

\medskip
Finally, combining the $(\varepsilon_T,\delta_T)$-DP test step (Step~3) with the $(\varepsilon_R,\delta_R)$-DP release step (Step~5) via sequential composition yields $(\varepsilon_T+\varepsilon_R,\delta_T+\delta_R)$-DP on event $E$. Finally, $E^c$ occurs with probability $\delta_{\mathrm{ptr}}$ (independent of the data), so we add $\delta_{\mathrm{ptr}}$ to $\delta$. Since $\varepsilon_T+\varepsilon_R\leq \varepsilon$ and $\delta_T+\delta_R+\delta_{\mathrm{ptr}}=\delta$, the mechanism is
$(\varepsilon,\delta)$-DP.
\end{proof}

\section{Missing Proofs for NDIS-Calibrated Gaussian Mechanism}

\subsection{Proof for~\Cref{thm:NDIS-mech-DP}}

\begin{proof}[Proof of~\Cref{thm:NDIS-mech-DP}]
\label{proof for thm:NDIS-mech-DP}
Fix an arbitrary ordered neighboring pair $\DB \sim \DB'$. For $\sigma \geq 0$, let
\begin{align*}
    X_\sigma ={}& \Mech_{\sigma I_m,q}(\DB)\sim \multiNormal{\mu(\DB)}{\Sigma(\DB)+\sigma I_m} \\
    Y_\sigma ={}& \Mech_{\sigma I_m,q}(\DB')\sim \multiNormal{\mu(\DB')}{\Sigma(\DB')+\sigma I_m}  .
\end{align*}

\medskip
\noindent\textbf{Binary search can find $\sigma^\star$.} Define the calibration objective
\begin{align*}
    F(\sigma)\defin \overline{\delta}_{\mathrm{LC}}\bigl(\varepsilon;\Delta(\sigma),\rho_\infty(\sigma),\nu(\sigma)\bigr).
\end{align*}
We first argue that we can use standard binary search on the monotone predicate $F(\sigma)\leq \delta$ to find $\sigma^\star$ (to any desired numerical tolerance). Towards that, we claim that $F(\sigma)$ is (i) nonincreasing and continuous in $\sigma$, and (ii) $F(\sigma)\to 0$ as $\sigma\to\infty$. We justify these properties from the $\sigma I_m$ regularization and the sensitivity definition.

Let $\Delta\mu =  \mu(\DB)-\mu(\DB'),$ and for each $\sigma\geq 0$ define
\begin{align*}
    m_\sigma(\DB,\DB') \defin{}& (\Sigma(\DB')+\sigma I_m)^{-1/2}\Delta\mu,\\
    R_\sigma(\DB,\DB') \defin{}& (\Sigma(\DB')+\sigma I_m)^{-1/2}(\Sigma(\DB)+\sigma I_m)(\Sigma(\DB')+\sigma I_m)^{-1/2}.
\end{align*}
Note that these are exactly the $(m_N,R_N)$ from Equation~\ref{def: m,R tuple} with $N=\sigma I_m$. Let $\tau_1(\sigma),\dots,\tau_m(\sigma)$ be the eigenvalues of $R_\sigma(\DB,\DB')$.

\medskip
\noindent\emph{Monotonicity/continuity of $\twoNorm{m_\sigma(\DB,\DB')}, \infNorm{\log\tau(\sigma)}, \abs{\sum_i\log\tau_i(\sigma)}$.} The main task is to show that $\twoNorm{m_\sigma(\DB,\DB')}, \infNorm{\log\tau(\sigma)}, \abs{\sum_i\log\tau_i(\sigma)}$

\begin{enumerate}
    \item \textbf{Mean difference decreases in $\sigma$.}
    Since $\sigma\mapsto (\Sigma(\DB')+\sigma I_m)^{-1}$ is operator monotone decreasing, then
    \begin{align*}
        \twoNorm{m_\sigma(\DB,\DB')}^2 = \Delta\mu^\top(\Sigma(\DB')+\sigma I_m)^{-1}\Delta\mu
    \end{align*}
    is nonincreasing and continuous in $\sigma.$

    \item \textbf{Loewner comparability direction is $\sigma$-independent.}
    Note that
    \begin{align*}
        R_\sigma(\DB,\DB')\succeq I_m
        \iff \Sigma(\DB)+\sigma I_m \succeq \Sigma(\DB')+\sigma I_m
        \iff \Sigma(\DB)\succeq \Sigma(\DB'),
    \end{align*}
    and similarly for $\preceq I_m$. Hence the condition ``$R_\sigma\succeq I_m$ or $R_\sigma\preceq I_m$'' does not depend on $\sigma$ and keeps for all $\sigma \geq 0.$

    \item \textbf{Covariance difference decreases in $\sigma$.}
    If $\Sigma(\DB)\succeq \Sigma(\DB')$, then $\tau_i(\sigma)\ge 1$ and $\tau_{\max}(\sigma)\defin \max_i\tau_i(\sigma)$ is nonincreasing and continuous in $\sigma$. Thus
    \begin{align*}
        \infNorm{\log\tau(\sigma)} = \log\tau_{\max}(\sigma)
    \end{align*}
    is nonincreasing and continuous in $\sigma.$ If instead $\Sigma(\DB)\preceq \Sigma(\DB')$, then $\tau_i(\sigma)\leq 1$ and $\tau_{\min}(\sigma)\defin \min_i\tau_i(\sigma)$ is nondecreasing and continuous, so
    \begin{align*}
        \infNorm{\log\tau(\sigma)} = -\log\tau_{\min}(\sigma)
    \end{align*}
    is nonincreasing and continuous in $\sigma$. Moreover,
    \begin{align*}
        \sum_{i=1}^m \log\tau_i(\sigma) = \log\det R_\sigma(\DB,\DB')
        =\log\frac{\det(\Sigma(\DB)+\sigma I_m)}{\det(\Sigma(\DB')+\sigma I_m)}
    \end{align*}
    is continuous in $\sigma$, and its absolute value is nonincreasing in $\sigma$, This is because its derivative equals $\mathrm{tr}\bigl((\Sigma(\DB)+\sigma I_m)^{-1}-(\Sigma(\DB')+\sigma I_m)^{-1}\bigr)$, whose sign is fixed by $\Sigma(\DB)\succeq \Sigma(\DB')$ or $\preceq$.
\end{enumerate}

\noindent Recall the~\Cref{def:ndis-sensitivity-loewner},
\begin{align*}
    \Delta(\sigma) \geq \sup_{\DB\sim\DB'} \twoNorm{m_\sigma(\DB,\DB')} \quad
    \rho_\infty(\sigma) \sup_{\DB\sim\DB'} \geq \infNorm{\log\tau(\sigma)} \quad
    \nu(\sigma) \geq \sup_{\DB\sim\DB'} \abs{\sum_i\log\tau_i(\sigma)},
\end{align*}
we have that $\Delta(\sigma),\rho_\infty(\sigma),\nu(\sigma)$ is also nonincreasing and continuous in $\sigma$, indeed, taking the supremum over all neighboring pairs preserves these properties.

\medskip
From \Cref{def:delta-LC}, it is easy to see that $\overline{\delta}_{\mathrm{mean}}(\varepsilon;\Delta)$ is nondecreasing in $\Delta$, and $\overline{\delta}_{\mathrm{cov}}(\varepsilon;\rho_\infty,\nu)$ is nondecreasing in $\rho_\infty$ and $\nu$. Therefore
$F(\sigma)=\overline{\delta}_{\mathrm{LC}}(\varepsilon;\Delta(\sigma),\rho_\infty(\sigma),\nu(\sigma))$
is nonincreasing and continuous in $\sigma$.

\medskip
\noindent\emph{Existence of a feasible $\sigma$ and the limit $F(\sigma)\to 0$.}
As $\sigma\to\infty$, we have
\begin{align*}
    (\Sigma+\sigma I_m)^{-1/2} = \sigma^{-1/2}(I_m+o(1)),
\end{align*}
so $\twoNorm{m_N(\DB,\DB')}=O(\sigma^{-1/2})\to 0$ and $R_\sigma(\DB,\DB')\to I_m$. Therefore $\infNorm{\log\tau(\sigma)} \to 0$, and $\abs{\sum_i\log\tau_i(\sigma)} \to 0$ for every neighbor pair. Taking the supremum over all neighboring pairs gives, as $\sigma\to\infty$,
\begin{align*}
    \Delta(\sigma)\to 0, \qquad \rho_\infty(\sigma)\to 0, \qquad \nu(\sigma)\to 0.
\end{align*}
Further, by~\Cref{def:delta-LC}, $F(\sigma)\to \overline{\delta}_{\mathrm{LC}}(\varepsilon;0,0,0)=0$.

\bigskip
\noindent\textbf{$\overline{\delta}_{\mathrm{LC}}\bigl(\varepsilon;\Delta(\sigma^\star),\rho_\infty(\sigma^\star),\nu(\sigma^\star)\bigr) \leq \delta$ implies that $\Mech_{\sigma I_m,q}$ is $\varepsilon, \delta$.}
We are left to show that if $\overline{\delta}_{\mathrm{LC}}\bigl(\varepsilon;\Delta(\sigma^\star),\rho_\infty(\sigma^\star),\nu(\sigma^\star)\bigr) \leq \delta$ then $\Mech_{\sigma I_m,q}$ is $\varepsilon, \delta$ for $\varepsilon \geq 0, \delta \in (0,1).$

\noindent By~\Cref{prop:mr-reduction}, we have that
\begin{align*}
    \delta_{X_{\sigma^\star},Y_{\sigma^\star}}(\varepsilon) = \delta_\varepsilon(m_{\sigma^\star}(\DB,\DB'),R_{\sigma^\star}(\DB,\DB')),
\end{align*}
where $\delta_\varepsilon(m,R)$ is defined in~\Cref{def: nids function}. As $q$ has $(\Delta(\sigma^\star),\rho_\infty(\sigma^\star),\nu(\sigma^\star),\sigma^\star I_m)$-Loewner-comparable NDIS sensitivity (\Cref{def:ndis-sensitivity-loewner}),
applying the Loewner-comparable decomposition bound (by \Cref{thm:ndis-analytic-form} together with \Cref{prop:ndis-mean-analytic-form}), we obtain that
\begin{align*}
    \delta_\varepsilon(m,R) \leq \overline{\delta}_{\mathrm{LC}}\bigl(\varepsilon;\Delta(\sigma^\star),\rho_\infty(\sigma^\star),\nu(\sigma^\star)\bigr).
\end{align*}

Combining all together, for any ordered neighboring pair $\DB\sim \DB'$, we have that
\begin{align*}
    \delta_{\Mech_{\sigma I_m,q}(\DB), \Mech_{\sigma I_m,q}(\DB')} ={}& \delta_\varepsilon(m_{\sigma^\star}(\DB,\DB'),R_{\sigma^\star}(\DB,\DB')) \\
    \leq{}& \overline{\delta}_{\mathrm{LC}}\bigl(\varepsilon;\Delta(\sigma^\star),\rho_\infty(\sigma^\star),\nu(\sigma^\star)\bigr) \leq \delta
\end{align*}

We then conclude the proof.
\end{proof}

\subsection{Missing Proofs for Applications}

\begin{proof}[Proof of~\Cref{prop:blr-ndis-sens}]
\label{proof for prop:blr-ndis-sens}
We verify the three function $(\Delta(\sigma), \rho_\infty(\sigma), \nu(\sigma))$ defined in \Cref{prop:blr-ndis-sens}.

\medskip
\noindent\textbf{Mean sensitivity.}
By $\lambda$-strong convexity of $\mathcal L(\cdot;\DB')$ and $\twoNorm{\nabla f_{(x,y)}(\theta)} \leq l$, the standard stability bound
gives
\begin{align*}
    \twoNorm{\hat\theta(\DB)-\hat\theta(\DB')} \leq l/\lambda.
\end{align*}

\noindent Next, we derive a uniform lower bound on $\Sigma(\widetilde{\DB})+\sigma I_d$ for $\widetilde{\DB}\in\{\DB,\DB'\}$. By the definition of the B\"ohning-style covariance,
\begin{align*}
    \Sigma(\widetilde{\DB}) = \left(\lambda I_d + \frac{1}{4}\sum_{i=1}^{|\widetilde{\DB}|} x_i x_i^\top \right)^{-1}.
\end{align*}
Since $\twoNorm{x_i}\leq l$, we have $x_i x_i^\top \preceq l^2 I_d.$ Therefore,
\begin{align*}
    \frac{1}{4}\sum_{i=1}^{|\widetilde{\DB}|} x_i x_i^\top \preceq \frac{|\widetilde{\DB}|\, l^2}{4} I_d \preceq \frac{n l^2}{4} I_d,
\end{align*}
which implies $\frac{1}{\lambda+\frac{n l^2}{4}} I_d \preceq \Sigma(\widetilde{\DB}) \preceq \frac{1}{\lambda} I_d.$ Therefore,
\begin{align*}
    \Sigma(\widetilde{\DB})+\sigma I_d
    \succeq
    \left(\sigma+\frac{1}{\lambda+\frac{n l^2}{4}}\right) I_d.
\end{align*}
Applying this bound with $\widetilde{\DB}=\DB'$, and using $N=\sigma I_d$, we obtain
\begin{align*}
    \twoNorm{m_N(\DB,\DB')} &= \twoNorm{(\Sigma(\DB')+N)^{-1/2}\bigl(\hat\theta(\DB)-\hat\theta(\DB')\bigr)} \\
    &\leq \twoNorm{(\Sigma(\DB')+\sigma I_d)^{-1/2}} \cdot \twoNorm{\hat\theta(\DB)-\hat\theta(\DB')} \\
    &\leq \frac{1}{\sqrt{\sigma+\frac{1}{\lambda+\frac{n l^2}{4}}}} \cdot \frac{l}{\lambda} = \Delta(\sigma).
\end{align*}

\medskip
\noindent\textbf{Loewner comparability.}
Let $\DB' = \DB \setminus \{(x,y)\}$ be obtained from $\DB$ by removing one record $x \in \Real^d$. By the definition of the B\"ohning-style covariance,
\begin{align*}
    \Sigma(\DB)^{-1}
    &= \lambda I_d + \frac{1}{4}\sum_{z \in \DB} zz^\top \\
    &= \lambda I_d + \frac{1}{4}\sum_{z \in \DB'} zz^\top + \frac{1}{4}xx^\top \\
    &= \Sigma(\DB')^{-1} + \frac{1}{4}xx^\top.
\end{align*}
Since $\frac{1}{4}xx^\top \succeq 0$, it follows that
\begin{align*}
    \Sigma(\DB)^{-1} \succeq \Sigma(\DB')^{-1}.
\end{align*}
Inverting both sides gives
\begin{align*}
    \Sigma(\DB) \preceq \Sigma(\DB').
\end{align*}
Finally, since $N=\sigma I_d \succeq 0$, adding $N$ preserves the order:
\begin{align*}
    \Sigma(\DB)+N \preceq \Sigma(\DB')+N.
\end{align*}

\medskip
\noindent\textbf{Covariance sensitivity.} We start by recalling an useful spectral sandwich
\begin{align*}
    m(\sigma)I_d \preceq \Sigma(\DB)+\sigma I_d \leq \Sigma(\DB')+ \sigma I_d \preceq M(\sigma)I_d,
\end{align*}
where $m(\sigma) = \sigma+\frac{1}{\lambda+\frac{n l^2}{4}}$ and $M(\sigma) = \sigma + \frac{1}{\lambda}$. This is from the one we derived above
\begin{align*}
    \frac{1}{\lambda+\frac{n l^2}{4}}I_d \preceq \Sigma(\DB)\preceq \frac{1}{\lambda}I_d.
\end{align*}

\noindent Recalling the definition of $R_{\sigma I_d}(\DB,\DB')$, and using $m(\sigma)I_d \preceq \Sigma(\DB)+\sigma I_d \leq \Sigma(\DB')+ \sigma I_d \preceq M(\sigma)I_d,$ we have that for every $i \in [d]$,
\begin{align*}
    \tau_i \geq \frac{m(\sigma)}{M(\sigma)} = \log\Bigl(\frac{\sigma+\frac{1}{\lambda}}{\sigma+\frac{1}{\lambda+\frac{n l^2}{4}}}\Bigr)
\end{align*}

\noindent Therefore,
\begin{align*}
    \infNorm{\boldsymbol{\ell}_{\sigma Id}(\DB,\DB')} \leq \log \frac{m(\sigma)}{M(\sigma)} = \log\Bigl(\frac{\sigma+\frac{1}{\lambda}}{\sigma+\frac{1}{\lambda+\frac{n l^2}{4}}}\Bigr) \qquad
    \abs{\log\det R_{\sigma I_d}(\DB,\DB')} \leq d \infNorm{\boldsymbol{\ell}_{\sigma Id}(\DB,\DB')} = \nu(\sigma).
\end{align*}
\end{proof}

\medskip
\noindent\textbf{Proof for~\Cref{prop:gpr-ndis-sens}}

\begin{proof}[Proof of~\Cref{prop:gpr-ndis-sens}]
\label{proof for prop:gpr-ndis-sens}
We verify the three function $(\Delta(\sigma),\rho_\infty(\sigma),\nu(\sigma))$ stated in \Cref{prop:gpr-ndis-sens}. Fix an ordered neighboring pair $\DB\sim \DB'$ and write $(\mu,\Sigma)\defin q_{\textnormal{GPR}}(\DB)$ and $(\mu',\Sigma')\defin q_{\textnormal{GPR}}(\DB')$. Since the GPR predictive distribution at a fixed $x_\star$ is one-dimensional, we must just write the covariance as variance (the scalar $\sigma>0$), and the induced invariants are
\begin{align*}
    m_{\sigma}(\DB,\DB') = (\Sigma'+\sigma)^{-1/2}(\mu-\mu'), \qquad
    R_{\sigma}(\DB,\DB') = \frac{\Sigma+\sigma}{\Sigma'+\sigma}.
\end{align*}

\medskip
\noindent\textbf{Mean sensitivity.}
Without lose of generality, let $\DB' = \{(x_i,y_i)\}_{i=1}^{n-1},$ and $\DB = \DB' \cup \{(x_n,y_n)\}..$
Let
\begin{align*}
    A' &\defin K' + \sigma_n^2 I_{n-1}, \\
    u &\defin \bigl(k(x_i,x_n)\bigr)_{i=1}^{n-1}, \\
    c &\defin k(x_n,x_n) + \sigma_n^2, \\
    k_\star' &\defin \bigl(k(x_\star,x_i)\bigr)_{i=1}^{n-1}, \\
    \kappa &\defin k(x_\star,x_n),
\end{align*}
so that
\begin{align*}
    A =
    \begin{pmatrix}
        A' & u \\
        u^\top & c
    \end{pmatrix},
    \qquad
    k_\star =
    \begin{pmatrix}
        k_\star' \\
        \kappa
    \end{pmatrix}.
\end{align*}
Let $\mathbf{y}' \in \Real^{n-1}$ denote the label vector of $\DB'$, and let
\begin{align*}
    \mathbf{y} =
    \begin{pmatrix}
        \mathbf{y}' \\
        y_n
    \end{pmatrix}
    \in \Real^n
\end{align*}
be the label vector of $\DB$. Define
\begin{align*}
    s \defin c - u^\top (A')^{-1} u.
\end{align*}
Since $A \succ 0$, we have $s>0$. By the block inverse formula,
\begin{align*}
    A^{-1}
    =
    \begin{pmatrix}
        (A')^{-1} + (A')^{-1} u s^{-1} u^\top (A')^{-1}
        &
        -(A')^{-1} u s^{-1}
        \\
        -s^{-1} u^\top (A')^{-1}
        &
        s^{-1}
    \end{pmatrix}.
\end{align*}
Therefore,
\begin{align*}
    \mu - \mu'
    = k_\star^\top A^{-1}\mathbf{y} - (k_\star')^\top (A')^{-1}\mathbf{y}'
    = \frac{\gamma r}{s},
\end{align*}
where
\begin{align*}
    \gamma \defin \kappa - (k_\star')^\top (A')^{-1}u,
    \qquad
    r \defin y_n - u^\top (A')^{-1}\mathbf{y}'.
\end{align*}

We next bound $\gamma$ and $r$. Consider the block matrix
\begin{align*}
    M \defin
    \begin{pmatrix}
        A' & u & k_\star' \\
        u^\top & c & \kappa \\
        (k_\star')^\top & \kappa & k(x_\star,x_\star)
    \end{pmatrix}.
\end{align*}
This matrix is positive semidefinite, since it can be written as the sum of the kernel Gram matrix on $(x_1,\ldots,x_{n-1},x_n,x_\star)$ and the positive semidefinite diagonal matrix
\begin{align*}
    \diag(\sigma_n^2 I_{n-1},\sigma_n^2,0).
\end{align*}
Taking the Schur complement with respect to $A'$ gives
\begin{align*}
    \begin{pmatrix}
        s & \gamma \\
        \gamma & \Sigma'
    \end{pmatrix}
    \succeq 0,
    \qquad
    \Sigma' = k(x_\star,x_\star) - (k_\star')^\top (A')^{-1}k_\star'.
\end{align*}
Hence,
\begin{align*}
    \gamma^2 \leq s\,\Sigma' \leq s\,l^2,
\end{align*}
where the last inequality uses $\Sigma' \leq k(x_\star,x_\star)\leq l^2$.

Next, a direct expansion using the block inverse formula yields
\begin{align*}
    \mathbf{y}^\top A^{-1}\mathbf{y}
    =
    (\mathbf{y}')^\top (A')^{-1}\mathbf{y}'
    +
    \frac{r^2}{s}.
\end{align*}
Therefore,
\begin{align*}
    \frac{r^2}{s}
    \leq
    \mathbf{y}^\top A^{-1}\mathbf{y}.
\end{align*}
Since $A = K + \sigma_n^2 I_n \succeq \sigma_n^2 I_n$, we have $\twoNorm{A^{-1}}\leq 1/\sigma_n^2$, and thus
\begin{align*}
    \mathbf{y}^\top A^{-1}\mathbf{y}
    \leq
    \twoNorm{\mathbf{y}}^2 \twoNorm{A^{-1}}
    \leq
    \frac{n B^2}{\sigma_n^2},
\end{align*}
where we used $\twoNorm{\mathbf{y}} \leq \sqrt{n}\,B$. Combining the above bounds,
\begin{align*}
    \abs{\mu-\mu'}^2
    =
    \frac{\gamma^2 r^2}{s^2}
    =
    \frac{\gamma^2}{s}\cdot \frac{r^2}{s}
    \leq
    l^2 \cdot \frac{n B^2}{\sigma_n^2}.
\end{align*}
Hence,
\begin{align*}
    \abs{\mu-\mu'}
    \leq
    \frac{l B \sqrt{n}}{\sigma_n}.
\end{align*}
Since $\Sigma' + \sigma \geq \sigma$, we obtain
\begin{align*}
    \twoNorm{m_{\sigma}(\DB,\DB')}
    =
    \abs{m_{\sigma}(\DB,\DB')}
    =
    \frac{\abs{\mu-\mu'}}{\sqrt{\Sigma'+\sigma}}
    \leq
    \frac{l B \sqrt{n}}{\sigma_n \sqrt{\sigma}}
    =
    \Delta(\sigma).
\end{align*}

\medskip
\noindent\textbf{Loewner comparability.}
Because $m=1$, the comparability between the real number trivially has.

\medskip
\noindent\textbf{Covariance sensitivity.}
We first bound $\Sigma$ and $\Sigma'$. Consider the block matrix
\begin{align*}
    \begin{pmatrix}
    A & k_\star\\
    k_\star^\top & k_{\star\star}
    \end{pmatrix}
    =
    \begin{pmatrix}
    K & k_\star\\
    k_\star^\top & k_{\star\star}
    \end{pmatrix}
    +
    \begin{pmatrix}
    \sigma_n^2 I_n & 0\\
    0 & 0
    \end{pmatrix}
    \succeq 0,
\end{align*}
where the first term is a Gram matrix and hence PSD, and the second term is PSD. Since $A\succ 0$, the Schur complement implies
\begin{align*}
    \Sigma = k_{\star\star}-k_\star^\top A^{-1}k_\star \ge 0.
\end{align*}
Moreover, by the diagonal bound $k_{\star\star}=k(x_\star,x_\star)\le l^2$ and $k_\star^\top A^{-1}k_\star\ge 0$, we also have $\Sigma \le l^2$. The same argument gives $0\leq \Sigma'\leq l^2$. Hence
\begin{align*}
    \Sigma+\sigma,\ \Sigma'+\sigma \in [\sigma,\sigma+l^2],
    \qquad\text{and therefore}\qquad
    \frac{\sigma}{\sigma+l^2} \le \frac{\Sigma+\sigma}{\Sigma'+\sigma} \le \frac{\sigma+l^2}{\sigma}.
\end{align*}
Let $\tau$ be the (unique) eigenvalue of $R_{\sigma}(\DB,\DB')$, so $\tau=(\Sigma+\sigma)/(\Sigma'+\sigma)$. Then
\begin{align*}
    |\log \tau| \le \log\Bigl(\frac{\sigma+l^2}{\sigma}\Bigr) = \rho_\infty(\sigma).
\end{align*}

Since $m=1$, combining the above proves that $q_{\textnormal{GPR}}$ has $(\Delta(\sigma),\rho_\infty(\sigma),\nu(\sigma),\sigma)$-Loewner-comparable NDIS sensitivity as stated.
\end{proof}

\section{Missing proofs for~\Cref{sec: auditing}}
\begin{proof}[Proof of \Cref{lem:ndis-upper-oracle}]
\label{proof for lem:ndis-upper-oracle}
We define a deterministic algorithm that compute an $\eta$-accurate upper bound using the CDF oracle given in~\Cref{def:genchi2-cdf-oracle}. For simplicity, let $p_0 = \pr{g_\varepsilon(\rZ)\leq 0}$ and $p_1 = \pr{g_\varepsilon(\widetilde \rZ)\leq 0}$. Since we have shown that both $g_\varepsilon(Z)$ and $g_\varepsilon(\widetilde Z)$ are generalized-$\chi^2$, then each probability $p_0,p_1$ can be evaluated up to additive error by one CDF oracle call.

Let the per-call accuracy $\eta_0 = \frac{\eta}{2(1+e^\varepsilon)}.$ Then, the deterministic algorithm is defined as follows:
\begin{enumerate}
    \item Invoke the $\eta_0$-generalized-$\chi^2$ CDF oracle on the instance corresponding to $\pr{g_\varepsilon(\rZ)\leq 0}$, and obtain an estimate $\widehat p_0$ such that $|\widehat p_0-p_0|\leq \eta_0$.
    \item Invoke the $\eta_0$-generalized-$\chi^2$ CDF oracle on the instance corresponding to $\pr{g_\varepsilon(\widetilde \rZ)\leq 0}$, and obtain an estimate $\widehat p_1$ such that $|\widehat p_1-p_1|\leq \eta_0$.
    \item Set $\widehat\delta = \widehat p_0 - e^\varepsilon \widehat p_1$ and output
    \begin{align*}
        \overline{\delta} =  \min\Bigl\{1, \max\{0,\ \widehat\delta+\eta/2\}\Bigr\}.
    \end{align*}
\end{enumerate}

\paragraph{Correctness of the Algorithm.}
By the oracle guarantee, $|\widehat p_0-p_0|\leq \eta_0, |\widehat p_1-p_1|\leq \eta_0.$ Therefore,
\begin{align*}
    |\widehat\delta-\delta_{\rX,\rY}(\varepsilon)|
    &= \bigl|(\widehat p_0-p_0) - e^\varepsilon(\widehat p_1-p_1)\bigr|\\
    &\leq |\widehat p_0-p_0| + e^\varepsilon|\widehat p_1-p_1|\\
    &\leq (1+e^\varepsilon)\eta_0 = \frac{\eta}{2}.
\end{align*}
It follows that $\widehat\delta \geq \delta_{\rX,\rY}(\varepsilon)-\eta/2$, hence
\begin{align*}
    \widehat\delta+\eta/2 \geq \delta_{\rX,\rY}(\varepsilon), \qquad \widehat\delta+\eta/2 \le \delta_{\rX,\rY}(\varepsilon)+\eta.
\end{align*}
Combining all together, we then complete the proof of \Cref{lem:ndis-upper-oracle}.

\end{proof}

\end{document}